\pdfoutput=1
\documentclass{jpp}

\usepackage{graphicx,natbib,longtable,booktabs,array}
\usepackage{amsfonts,amsmath,amssymb,bm}
\usepackage{xcolor}
\usepackage{enumitem,placeins}
\usepackage[pdftex,colorlinks,citecolor=blue,hypertexnames=false]{hyperref}
\usepackage[]{cleveref}
\crefname{section}{\S\!}{\S\S\!}
\Crefname{section}{Section}{Sections}
\crefname{appendix}{App.}{Apps.}
\Crefname{appendix}{Appendix}{Appendices}
\crefname{subappendix}{App.}{Apps.}
\Crefname{subappendix}{Appendix}{Appendices}
\crefname{subsubappendix}{App.}{Apps.}
\Crefname{subsubappendix}{Appendix}{Appendices}
\crefname{equation}{Eq.}{Eqs.}
\Crefname{equation}{Equation}{Equations}
\crefname{figure}{Fig.}{Figs.}
\Crefname{figure}{Figure}{Figures}

\graphicspath{{figures/}}

\newcommand{\Dr}{D}
\newcommand{\rpk}{r_{\rm pk}}
\newcommand{\rin}{r_{\rm in}}

\newcommand{\Ma}{{\rm Ma}}
\newcommand{\Lz}{L_z}
\newcommand{\tauE}{\tau_E}
\newcommand{\tauP}{\tau_\rho}

\title[Global self-organisation via interchange turbulence]
      {Global self-organisation via driven interchange turbulence
       in a hard-core Z-pinch}

\author[J.~Squire and others]%
{J.~Squire$^{1}$\thanks{Email address for correspondence: jonathan.squire@otago.ac.nz},
C.~Hill$^{1,2}$,
and L.~Huckle$^{1}$}

\affiliation{$^1$Physics Department, University of Otago, 730 Cumberland St., Dunedin 9016, New Zealand\\[\affilskip]
$^2$Max Planck Institute for Plasma Physics, Garching 85748, Germany\\[\affilskip]}

\pubyear{2026}
\volume{}
\pagerange{}
\date{?; revised ?; accepted ?. - To be entered by editorial office}

\begin{document}
\maketitle

\begin{abstract}
Whereas most magnetic-confinement concepts seek to suppress turbulent transport, the levitated dipole exploits it, with plasma turbulently self-organising to concentrate into the stronger-field region near the levitating 
coil. In a sufficiently large device, this can create the conditions needed for fusion, despite  stiff and vigorous transport. We study this process
in global simulations of a driven hard-core Z-pinch using  compressible
ideal magnetohydrodynamics (MHD) with self-consistent evolution of the magnetic field.  The results reveal how
plasma and field jointly settle onto a continuously turbulent, finite-$\beta$ interchange-marginal
family.  Across a wide drive scan in two-dimensional simulations, quasi-universal  pressure and
density profiles result, with the magnetic-field profile changing from a decreasing
vacuum-like form to one with an interior minimum at increasing $\beta$.  These profiles are 
robust to large changes to the heat and particle sources (including
moving both to the outside of the device) as well as to different exhaust and boundary prescriptions.  A designer
consequently has little direct control over the selected profiles, including the final separation between the pressure maximum and coil, which is
set by a  turbulent boundary layer.  
We  show that such results persist in the presence of three-dimensional
fluctuations at modest $\beta$, but that high-$\beta$ states become unstable to 
a field-parallel $m=1$ instability that is the analogue of the ballooning instability in a true dipole; this evacuates the 
core,  destroying confinement.  
Based on a near-marginal expansion and  a geometrical remapping, we argue that many features of our results should apply to any shearless closed-field-line system.
\end{abstract}

\section{Introduction}\label{sec:intro}

In a levitated-dipole  device, plasma is confined on closed field lines produced by a
current ring suspended magnetically inside a vacuum vessel.  It needs neither rotational
transform nor field-aligned plasma current, and levitation removes supports
that would otherwise intercept  field lines.  These properties have
motivated advanced-fuel reactor proposals and laboratory experiments
\citep{Hasegawa1992,Kesner2004,Garnier2009,Kesner2010}, as well as new
experimental and deuterium--tritium power-plant designs
\citep{Chisholm2026,Simpson2026}.  The confinement strategy is unusual in that, rather than 
trying to suppress anomalous turbulent transport, it can utilise continual interchange  motions to organise the plasma into the centrally
concentrated pressure and density profiles desired for fusion.

This central concentration arises because flute-like interchange motions tend to
homogenise the particle and entropy contents of closed flux tubes
\citep{hasegawa87,Hasegawa1990,kouznetsov07,boxer10,Kesner2011}.  In terms of the differential
flux-tube volume $\mathcal V(\psi)=\oint{\rm d}\ell/B$, the resulting mixed
state has $\rho\mathcal V\simeq{\rm const}$ and
$p\mathcal V^\gamma\simeq{\rm const}$ (here $p$ and $\rho$ are the plasma's pressure and density, respectively, and $\gamma$ is the adiabatic index).  Since $\mathcal V$ increases rapidly away
from a dipole coil, mixing produces an inward particle pinch and centrally
peaked density and pressure profiles.  Experiments in LDX and RT-1 have
observed this pinch, together with global convective structures and the
approach to stationary particle-content profiles
\citep{Garnier2009,Grierson2009,Grierson2010,boxer10,Saitoh2011,
Yoshida2013,Davis2014,garnier17,Kenmochi2022,Saitoh2024,Saitoh2026}.  Confinement thus relies on the properties of the  saturated
turbulence, raising the question of how this state self-organises as an operator adjusts external controls such 
as fuelling, heating, and exhaust.  The outcome of this self-organisation process is of critical importance 
for predicting the viability of 
the levitated dipole concept as a fusion reactor \citep{Kesner2004}.

Past works have studied levitated-dipole-relevant turbulence in various ways, spanning local to global domains. 
Local gyrokinetic theory and simulations allow detailed exploration of kinetic and
finite-Larmor-radius effects in these systems (particularly the entropy mode; \citealp{simakov01}), but prescribe the equilibrium
field and gradients, leaving the global profile self-organisation  to be inferred from the
calculated transport \citep{Ricci2006,kobayashi09,kobayashi10,Hallenbert2022,Hoffmann2023}.
Whole-device fluid models follow this reorganisation more
directly. The two-fluid calculations of \citet{Li2016,ou20} and flux-tube-averaged
MHD simulations of \citet{li21} evolve the plasma profiles across a fixed
dipole field, finding inward transport, nearly uniform particle and entropy
contents,  device-scale convective structures, and a near insensitivity of mean saturated profiles 
to heat or particle source location.  These models are, however,
quasi-global in their treatment of the turbulence: although the
domain captures the whole device, the magnetic field is fixed and the equations
neglect  inertial terms that are small only when the radial correlation length
is short compared to the equilibrium scale (see \cref{app:dipoleflute}).
Earlier, \citet{pastukhov05b} derived flute-averaged MHD equations for near-marginal convection in general
axisymmetric, shearless magnetic geometry that retain these global terms,  simulating the resulting turbulence in 
 cylindrical geometry (the Z-pinch) and likewise finding large-scale convective cells that hold the profile close to marginality \citep{pastukhov01,pastukhov05}.
Fully global MHD simulations, which evolve the plasma and magnetic field together to study the 
self-organisation process in its purest form, have been explored for an unforced, relaxing
plasma \citep{Makhin2005}, but not (to our knowledge) in the presence of sustained heating, fuelling and exhaust. 
This is of particular interest in the high-$\beta$ regime where the plasma starts 
strongly influencing the magnetic field, a limit to which the levitated dipole concept is particularly well suited \citep{Krasheninnikov1999,garnier99,simakov00,Garnier2006}.

In this paper, we present a detailed study of self-consistent interchange
reorganisation in the simplest levitated-dipole analogue: a cylindrical annulus (hard-core
Z-pinch) with compressible MHD.  Global simulations of the driven system evolve the
density, pressure, flow, and confining field together into a saturated turbulent steady state.  Across a broad range of
forcing and loss conditions, we follow the formation of the finite-$\beta$
pinched plasma and measure its dependence on idealised prescriptions for 
heating, fuelling, and exhaust.  We also explore
the plasma's transport and confinement properties in steady state and test how the self-organised plasma survives in
three dimensions.

We find that the plasma and magnetic field self-organise together onto the
finite-$\beta$ interchange-marginal family 
$p\mathcal V^\gamma\simeq{\rm const}$  and $\rho\mathcal V\simeq{\rm const}$.  
At low $\beta$ the family reduces
to the usual vacuum-field power laws with $\mathcal V\propto r^{2}$ for the simple
cylindrical field of the Z-pinch.  As the heating rate is raised, pressure
modifies the magnetic field, flattening the marginal profiles by flattening $\mathcal V$; a single member of this family, fixed by the
measured $\beta$ at one point, describes
both pressure and density through the source-free bulk over a range of several 
thousand in heating rate.  Most dramatically, the profile, along with the location of the pressure and density maxima,  remains very similar 
when the heating or fuelling sources are moved from the inside to the outside of the cylinder \citep{ou20,li21}, as 
well as across different exhaust prescriptions and numerical controls. Such changes do, however, change the
time evolution towards the saturated state, as well as the total stored energy, turbulence strength, and particle flux direction (inwards versus outwards).

Although the MHD hard-core Z-pinch  is a highly idealised model, it
retains a surprisingly wide range of the relevant physics, suggesting that
many of our results should carry over to a levitated dipole.  When its  axial length is taken to be approximately $2\pi$ times its radius, 
the Z-pinch is
the close, wide-coil limit of a physical dipole.  
More generally, we
show explicitly, using the near-marginal reduced equations derived by
\citet{pastukhov05b} for a general axisymmetric, shearless, closed-field-line geometry, that the same particle and
entropy contents are advected in both systems, and that the local nonlinear flute dynamics differ
only through a rescaling of time, axial length, and interchange drive strength.  The pinch therefore retains very similar interchange physics at that level of approximation, while also evolving
the finite-amplitude compressible flow and magnetic response of its own geometry.  This
supports the idea that the profile-selection picture and turbulence
properties should transfer qualitatively to dipole geometry, even in the presence of  a 
more complex two-dimensional magnetic equilibrium at finite $\beta$.  Likewise, the
appearance of the same inward pinch, global convection,  and peaked profiles in
MHD, two-fluid and gyrokinetic descriptions  supports the relevance of MHD turbulent dynamics to 
the full kinetic problem \citep{kobayashi10,ou20,li21}, although there
are various important effects that are kinetic in nature.

Our simulations also reveal several subsidiary results.  The first relates to the ballooning limit, a 
field-line bending instability that arises above a critical 
$\beta$.  While 
previous dipole studies have considered linear ballooning
stability in detail \citep{garnier99,pastukhov01,simakov00,Kesner2001,Testa2026}, to our knowledge, our
three-dimensional calculations give the first nonlinear study of its
outcome.  We find that a low-$\beta$ state remains
effectively 2-D, 
although weak non-axisymmetric fluctuations, 
at a level consistent with critical balance \citep{goldreich95}, alter the mean profiles slightly compared to 2-D.
 At higher $\beta$, an
$m=1$ deformation evacuates the core catastrophically, overwhelming
the inward turbulent pinch.   This transition occurs at $\beta\approx0.5$ in the Z-pinch, but is 
in general significantly less stringent in dipole geometry. 

The second subsidiary result is that the saturated state becomes sensitive to the axial domain length when the latter falls below
the energy-containing scale.  The global overturning motions can no longer
fit, energy accumulates in zonal flows that suppress transport, and the pressure profile steepens.  Our 
shortest boxes confine better than the converged long boxes by producing 
this zonally regulated state. Because these short 
boxes are less physically justified for the dipole --- they are shorter than the circumference of a circular coil at the inner radius --- this result suggests  caution is needed, including in local simulations, in extrapolating the confinement benefits of zonal flows from restricted domains to a physical device.

Third, we show that turbulence also smooths non-axisymmetric heating and fuelling (as might occur from an antenna located
at one toroidal location), leaving
similar centrally peaked profiles 
(\cref{app:zmod}).  This is promising for a dipole reactor supplied through
localised ports: its poloidal field provides no parallel route for toroidal
mixing, yet interchange motions substantially reduce any imposed source asymmetry.

\subsection{Outline}

\Cref{sec:model} defines the full-field model, forcing and simulation campaign,
providing the common set-up for all subsequent comparisons.  Numerical results
are presented in Sections~\ref{sec:state}--\ref{sec:3d}.  \Cref{sec:state}
establishes the jointly selected plasma and magnetic state and derives the
finite-$\beta$ marginal family used to interpret it.  \Cref{sec:sources} moves
the sources and changes the exhaust to explore the near invariance of  the profile selected by the
turbulence, even though different transport routes are selected by the forcing.  \Cref{sec:operating}
studies how the turbulence amplitude is regulated, then connects this
 to stored energy and particle residence.  \Cref{sec:peak} shows how the saturated
position of the pressure maximum is determined via the excursion of turbulent eddies onto the hard core and
uses it to assess the heat load to the coil.  Finally, \cref{sec:3d} releases the flute
symmetry to determine which selected states survive field-line bending.

\Cref{app:runs} catalogues the simulations and their averaging intervals, while 
other appendices catalogue subsidiary results and tests.
\Cref{app:zonal} isolates the short-box regime, showing how zonal flows steepen
the profiles and increase confinement.  
\Cref{app:zmod} tests axially nonuniform heating and fuelling, whose
imprint is partly removed by turbulent mixing.  
\Cref{app:dipole} explores analytically the relationship of the pinch to the true dipole 
geometry, constructing an effective 
radial map between a pinch and a dipole flute dynamics, then comparing their 3-D stability.  Finally, \cref{app:numerics}
gives the  linear eigenvalue problem for interchange and bending
modes, using the former to test the nonlinear solver. 

\section{Model and set-up}\label{sec:model}

We model a hard-core Z-pinch: a plasma annulus surrounding a rigid,
current-carrying central rod.  The rod produces a purely azimuthal magnetic
field, whose field lines are circles about the axis.  Their curvature is
therefore unfavourable everywhere for a plasma whose pressure falls outward.
The configuration also has no rotational transform, magnetic shear or
field-aligned current.  In this section, we describe the equations and numerical method in
\cref{sec:equations}, the heating, fuelling and exhaust model in
\cref{sec:forcing}, and the simulation campaign in \cref{sec:diagnostics}.

\subsection{Equations, geometry and numerics}\label{sec:equations}

We solve the equations of ideal, compressible, adiabatic MHD in cylindrical
coordinates $(r,\theta,z)$,
\begin{subequations}\label{eq:mhd}
\begin{align}
 \frac{\partial\rho}{\partial t}
 +\boldsymbol{\nabla}\bcdot\left(\rho\bm u\right)&=S_\rho,\label{eq:mass}\\
 \frac{\partial\left(\rho\bm u\right)}{\partial t}
 +\boldsymbol{\nabla}\bcdot\left[\rho\bm u\bm u-\frac{\bm B\bm B}{4\pi}
   +\left(p+\frac{B^2}{8\pi}\right)\mathsf{I}\right]&=\bm uS_\rho,
   \label{eq:momentum}\\
 \frac{\partial\bm B}{\partial t}
 -\boldsymbol{\nabla}\mathbin{\times}\left(\bm u\mathbin{\times}\bm B\right)&=0,
   \label{eq:induction}\\
 \frac{\partial\mathcal{E}}{\partial t}
 +\boldsymbol{\nabla}\bcdot\left[
   \left(\mathcal{E}+p+\frac{B^2}{8\pi}\right)\bm u
   -\frac{\bm B\left(\bm B\bcdot\bm u\right)}{4\pi}\right]
   &=S_\mathcal{E},\label{eq:energy}\\
 \frac{\partial\left(\rho C_i\right)}{\partial t}
 +\boldsymbol{\nabla}\bcdot\left(\rho C_i\bm u\right)&=S_i.\label{eq:scalar}
\end{align}
\end{subequations}
\noindent Here $\rho$ is the mass density, $\bm u$ is the flow velocity, $p$ is
the thermal pressure and $\bm B$ is the magnetic field.  The total energy
density is $\mathcal{E}=p/(\gamma-1)+\rho u^2/2+B^2/8\pi$, with adiabatic
index $\gamma=5/3$, and $\mathsf{I}$ is the unit tensor.  We write
${\rm D}/{\rm D}t\equiv\partial/\partial t+\bm u\bcdot\bnabla$ for the Lagrangian derivative
following the flow.  The quantity $C_i$
is the concentration of the $i$th of two passive labels.  The terms on the
right-hand sides describe the particle, heat and label sources and the energy
exhaust, which are defined in \cref{sec:forcing}.  Replacement particles are
added co-moving with the local flow, which implies the momentum source $\bm
uS_\rho$ and a kinetic energy contribution contained in $S_\mathcal{E}$.  We do
not include explicit viscosity, resistivity or thermal conduction; 
dissipation occurs at the grid scale, and we test its effect by
scanning the resolution.  Ideal MHD likewise omits much other relevant physics including finite-Larmor-radius effects,
kinetic entropy modes, pressure anisotropy, radiation, and the sheath or
recycling physics of a material surface.  We
discuss these omissions in \cref{sec:conclusions}.

For a purely azimuthal equilibrium field, $\bm
B=B_\theta(r)\hat{\bm\theta}$, radial force balance is
\begin{equation}
 \frac{{\rm d}}{{\rm d}r}\left(p+\frac{B_\theta^2}{8\pi}\right)
 +\frac{B_\theta^2}{4\pi r}=0,
 \label{eq:forcebalance}
\end{equation}
where the final term is the inward hoop force of the curved field.  Except in
\cref{sec:3d}, we set $\partial/\partial\theta=0$ and solve \cref{eq:mhd} on a
two-dimensional $(r,z)$ grid.  The equilibrium field strength, curvature and
pressure are constant around each field line, so this restriction removes
field-line bending exactly.  It retains all three vector components,
compression and the self-consistent evolution of $B_\theta$, and hence the
full flute dynamics of the pinch.  

As shown in  \cref{app:dipole}, it transpires that these cylindrical dynamics involve the same conserved
flux-tube contents in any closed-field-line configuration \citep{pastukhov05b}, and,
near marginality,  MHD dynamics in a dipole can be mapped onto those of the pinch up to a
local rescaling. The same is not true of 3-D dynamics (\cref{app:dipoleballooning}).

\subsubsection{Normalisation and averaging}\label{sec:units}

Ideal MHD without explicit dissipation has no intrinsic length, density or
magnetic-field scale.  The physical parameters are therefore dimensionless
ratios, rather than the absolute size of the annulus or strength of its field.
We measure lengths in units of the hard-core radius $\rin$ and magnetic fields
in units of the field $B_{\rm in}$ at that surface.  Pressure is measured in
$B_{\rm in}^2/8\pi$, so the plotted pressure is directly a local
$\beta$-like quantity; as usual, $\beta\equiv8\pi p/B^2$ is the ratio of
thermal to magnetic pressure.  We use the initial density $\rho_{\rm IC}$ at
the pressure maximum as the density unit and
$v_{A,{\rm IC}}=B_{\rm in}/(4\pi\rho_{\rm IC})^{1/2}$ as the velocity unit.
The unit of the stored (volume-integrated) thermal energy,
$W=\int p/(\gamma-1)\,{\rm d}V$, is
$(B_{\rm in}^2/8\pi)V_{\rm dom}$, using the cylindrical volume of each
run, $V_{\rm dom}=\pi\Lz(r_{\rm out}^2-\rin^2)$.  The radial geometry is set by
$r_{\rm out}/\rin$, together with the dimensionless axial period $L_{z}/\rin$ and the
positions and widths of the sources.  Fixed input time constants, including
those of the selective sinks, are given in
$\tau_{A0}\equiv\rin/v_{A,{\rm IC}}$.

We use two further reference values for the drive and axial domain.  The first
is the total heat input $Q_{\rm ref}$ of the reference run defined in
\cref{sec:defaults}.  The second is the default axial period
$L_{z0}\equiv2\pi r_{\rm mid}=6\pi\rin$, based on a circular ring circumference near the middle of the domain ($r_{\rm mid}=3\rin$). 
For the reference geometry, with
$r_{\rm out}=6\rin$ and $\Lz=L_{z0}$, the domain volume is
$V_{\rm dom} = V_{\rm ref} = 210\pi^2\rin^3$.  
The reference heat input is
$Q_{\rm ref}\simeq2.79\times10^{-5}
(B_{\rm in}^2/8\pi)V_{\rm ref}/\tau_{A0}$, which is chosen to give a useful modest-$\beta$ default simulation against which to compare others.

An overbar denotes an average over the periodic directions ($z$, and
$\theta$ where it is resolved).  For mean profiles and other averaged
diagnostics it also includes an average over a saturated time interval,
except where a quantity is shown as a function of time (e.g.\ $\bar f(r,t)$,
time histories, or profiles at given times); $\langle f\rangle_t$ denotes a
time average alone.  We choose
this interval separately for each run from the stationarity of its mass and
energy budgets.  The interval $(t_{\rm av},t_{\rm end})$ is listed for every
run in \cref{tab:usedruns}.  A fluctuation $\delta f$ is the departure of field $f$ from
the instantaneous average over $z$ and $\theta$.  For dynamical
times and measured rates, we use the state-dependent, saturated peak
Alfv\'en curvature time
\begin{equation}
 \tau_{A,{\rm pk}}=\frac{\rpk}{v_{A,{\rm pk}}},\qquad
 v_{A,{\rm pk}}=\frac{\bar B_\theta\left(\rpk\right)}
  {\left[4\pi\bar\rho\left(\rpk\right)\right]^{1/2}},
 \label{eq:tauapk}
\end{equation}
where $\rpk$ is the radius of the mean pressure maximum in the saturated state and a subscript
``pk'' always denotes a quantity evaluated there.  When written with
arguments, $\rpk(t)$ and $\rpk(z,t)$ denote the maxima of the instantaneous
axially averaged and local pressure profiles, respectively.  This is a natural
timescale for a curvature-driven instability, although the interchange is slower at low $\beta$.  We do not compare raw
simulation times between runs because their sound and Alfv\'en speeds differ significantly.

\subsubsection{Numerical method and initial state}\label{sec:numerics}

We run the simulations with \textsc{Athena++} \citep{stone20}, using the HLLD
Riemann solver, the second-order Van Leer unsplit integrator and
piecewise-linear reconstruction.  Constrained transport maintains
$\boldsymbol{\nabla}\bcdot\bm B=0$, and the CFL number is $0.3$.  We do not
use mesh refinement.  Density and pressure floors act only as positivity
guards and have no material effect on the saturated two-dimensional states
reported here.  The late phases of  strongly disrupted
three-dimensional continuations can become floor affected (see \cref{sec:3d}).

Most runs start from a smooth, magnetostatic equilibrium.  The pressure and density rise
from their low inner-boundary values to a broad maximum at $2\rin$, which coincides with
the default heating location $r_Q=2\rin$; outward of
the maximum, the density declines smoothly while the pressure levels onto a
small outer shelf.  The magnetic field is chosen to
satisfy radial force balance, and the initial profile is interchange stable,
so the turbulence is produced by the subsequent forcing rather than inherited
from the initial condition. More precisely, writing
$\tilde\rho\equiv\rho/\rho_{\rm IC}$, 
$\tilde p\equiv8\pi p/B_{\rm in}^2$, and $x\equiv r/\rin$, outside the maximum we set
\begin{equation}
 \tilde\rho=W_1(x),\qquad
 \tilde p=\tilde p_0\left[W_2(x)W_1(x)+\left(1-W_2(x)\right)W_{1}(x_c)\right],
 \label{eq:initial}
\end{equation}
where $W_1(x)=[1+(x-x_{\rm pk})^2/4]^{-1}$,
$W_2(x)=\{1+\exp[-2.5(x_c-x)]\}^{-1}$, $x_{\rm pk}=2$, $x_c=4$, and $\tilde p_0=0.032$.
Inward of the maximum, a raised cosine joins both quantities to the values in
the inner boundary conditions.  We obtain the field by integrating
\cref{eq:forcebalance},
\begin{equation}
 \frac{B_\theta}{B_{\rm in}}=\frac{1}{x}\left[1
  -\int_1^{x}x'^2\frac{{\rm d}\tilde p}{{\rm d}x'}\,{\rm d}x'\right]^{1/2}.
 \label{eq:binit}
\end{equation}
This gives $\beta\simeq0.13$ at the maximum and a field within a few per cent
of the vacuum $r^{-1}$ profile.  Direct evaluation of the criterion derived in
\cref{sec:kadomtsev} confirms that the profile is stable everywhere.  Heating
and fuelling subsequently push it beyond marginality.

\subsection{Heating, particle replacement and exhaust}\label{sec:forcing}

Interchange turbulence flattens the profiles that drive it.  A sustained
turbulent state therefore requires continuous heating and fuelling, together
with a way to remove energy and particles.  We represent these processes with
 simple sources and sinks, varying their locations and forms to test how strongly different
 features  determine the final state. 

\subsubsection{Heating and energy exhaust}
\label{sec:sourceterms}

Thermal energy is added through a Gaussian centred at $r_Q$ with standard
deviation $\sigma_Q$,
\begin{equation}
 S_\mathcal{E}=Q\,g_Q(r),\qquad
 g_Q(r)\propto\exp\left[-\frac{\left(r-r_Q\right)^2}{2\sigma_Q^2}\right],
 \qquad \int g_Q\,{\rm d}V=1,
 \label{eq:heatsource}
\end{equation}
where the normalisation of $g_Q$ makes $Q$ the total heat input.  This source
changes neither mass nor momentum.  Throughout the paper, $Q$ is the measured
rate from the run history rather than the nominal input value.  By default,
$r_Q=2\rin$ and $\sigma_Q=0.4\rin$.  For the displaced sources in
\cref{sec:sourceprofiles}, we keep both the total power and the shape of
${\rm d}Q/{\rm d}r$ fixed, accounting for the cylindrical volume element.
When $\Lz$ is changed, $Q$ and the other extensive rates are scaled with it,
so these runs differ only in their axial geometry.

Energy must also leave the domain.  There is no unique prescription for this
loss because the appropriate outer condition depends on device edge physics omitted
from the model.  We therefore use two different forms and test which results
are common to both.  Each removes only thermal energy and acts only in the
outermost part of the annulus.
\begin{description}
\item[Fractional sink.]\hspace{0.3em} Thermal energy is removed in proportion
to itself,
\begin{equation}
 \frac{\partial}{\partial t}\frac{p}{\gamma-1}
 =-\Gamma_E(r)\,\frac{p}{\gamma-1},\qquad
 \Gamma_E(r)=\Gamma_0\exp\left(\frac{r-r_{\rm out}}{w_s}\right),
 \label{eq:fractionalsink}
\end{equation}
with $w_s=0.2\rin$ and
$\Gamma_0\tau_{A,{\rm pk}}\simeq1.5$ at the reference drive.  Here
$\Gamma_0$ is the coefficient at $r=r_{\rm out}$, and the local removal rate
decreases exponentially inward.  This prescription does not fix the edge
temperature, so both the edge temperature and the profile amplitude rise with
the drive.  That freedom lets us follow the marginal family to high $\beta$,
although it is likely less physical than fixing an edge temperature.  We scan
$\Gamma_0$ over a factor of $16$ at fixed $Q$ to test the dependence on this
arbitrary coefficient.
\item[Edge thermostat.]\hspace{0.3em} The temperature excess above a set point
is relaxed away in a one-sided way,
\begin{equation}
 \frac{\partial}{\partial t}\frac{p}{\gamma-1}
 =-\frac{\Theta(r)}{\tau_T}\,
  \frac{\rho\max\left(T-T_{\rm set},0\right)}{\gamma-1},
 \label{eq:thermostat}
\end{equation}
where $T=p/\rho$.  The function $\Theta$ rises smoothly from zero to unity
over a width $0.1\rin$ about $r=5.5\rin$, restricting the thermostat to the
outermost layer.  The relaxation time is $\tau_T=0.05\,\tau_{A0}$, meaning the outer layer is  effectively clamped, and the temperature set point is $T_{\rm set}=0.9\,T_{\rm BC}$, where $T_{\rm BC}$ is the prescribed hard-core boundary temperature (see below). This
provides a simple limiter analogue: because the edge temperature is fixed, a
stronger drive can be accommodated only by a steeper profile.   At strong drive, a
sharp, several-cell-wide shoulder develops just inside the clamped layer; 
no analysis includes this region.
\end{description}

\subsubsection{Mass fuelling}

Most simulations replace particles using a global proportional controller.
The controller compares the domain mass $M$ with a target
$M_t=1.2\,M_{\rm IC}$, and mass is added at the rate
\begin{equation}
 \dot M_{\rm c}=\min\left[\dot M_{\max},
  \max\left(0,\frac{M_t-M}{\tau_M}\right)\right],
 \label{eq:controller}
\end{equation}
which is distributed over a separately normalised Gaussian at
$(r_\rho,\sigma_\rho)$ with the same form as \eqref{eq:heatsource}.  The response is slow compared with an eddy turnover, with 
$\tau_M\simeq100\,\tau_{A,{\rm pk}}$ by default, which means that the controller maintains a mass below $M_t$ due to continual loss through the boundaries (see below).  The cap
$\dot M_{\max}$ is reached only at the strongest drives.  Added material is
cold and locally co-moving, meaning we scale density and momentum together to leave
the velocity unchanged, and add no thermal energy directly.  This avoids the
drag on turbulent and zonal flows that would result from adding momentum at a
fixed rate.  The default source is at the outer edge,
$r_\rho=5.2\rin$ with $\sigma_\rho=0.2\rin$.  In this configuration,
transport must build a centrally peaked density profile rather than having
one imposed by the source.  

This choice of fuelling lets the boundary conditions set the mass-loss rate, which 
can influence the
density structure and residence times. We test the boundary's influence by also running cases 
fuelled at a fixed, prescribed rate $\dot M_{\rm fix}$, as well as cases with an  outer particle sink similar to 
the energy exhaust described above, which removes density, co-moving momentum,
thermal and kinetic energy, and passive labels in the same local proportion.

\subsubsection{Passive tracers}

We use two passive scalars as simple composition markers.  Label $0$
represents fusion ash and is injected with the heat source; label $1$
represents impurities and is injected at the edge with the replacement particles.  They
are labels rather than separate species, and have no charge state, radiation
or mass difference.
They allow us to ask whether material released in the hot interior accumulates
there and whether an edge extractor can remove it.  By default, both labels leave via
the boundary conditions, like the mass, although we also consider cases
where label $0$ is removed by an edge sink with the same shape as the thermal sink (see
\cref{sec:confinement}).  The labels take longer than the plasma to reach a steady
inventory, and in many runs they are still filling during the averaging interval
(label-$0$ loss below 90 per cent of its source); this does not affect the results
reported here unless noted.

\subsubsection{Radial boundaries}\label{sec:boundaries}

The domain is periodic in $z$ and $\theta$.  We treat the inner boundary as
the levitating coil.  Its ghost cells contain a fixed reservoir with prescribed
density and temperature $(\rho_{\rm BC},T_{\rm BC})$.  The default reservoir
is cold, with $\beta\simeq6\times10^{-3}$; it has zero velocity and continues
the field as $B_\theta=B_{\rm in}\rin/r$.  This represents an absorbing hard surface, without recycling,
sheath physics or finite thermal conductance.

At the outer boundary, the ghost zones have zero velocity and continue the
low-$\beta$ marginal power laws $\rho\propto r^{-2}$,
$p\propto r^{-10/3}$ and $B_\theta\propto r^{-1}$.  At the strongest drives,
the field  flattens well away from $r^{-1}$
(\cref{sec:marginal}) and piles up against this boundary condition.  We do not
use the outer region for quantitative conclusions in those runs.

We record every source and loss channel, together with the radial fluxes at
both boundaries, and check the global mass and energy budgets in every run.

\subsection{The simulation campaign}\label{sec:diagnostics}

\subsubsection{Default configuration}\label{sec:defaults}

Unless stated otherwise, runs use the geometry and source prescriptions
specified above: $r_{\rm out}=6\rin$, $\Lz=L_{z0}=6\pi\rin$,
heating at $2\rin$ and edge fuelling at $5.2\rin$.
The axial period represents one circuit at $r_{\rm mid}=3\rin$, allowing
motions on the scale of the machine.  The reference run uses the fractional
sink and $Q=Q_{\rm ref}$, reaching $\beta_{\rm pk}\simeq0.13$.

The default $(r,z)$ grid has $384\times1440$ nearly square cells, with
resolution tests from $N_r=64$ to $1024$ ($1024\times3840$).
Two-dimensional runs last  hundreds to thousands of
$\tau_{A,{\rm pk}}$.  As described in \cref{sec:units}, averages are taken
over stationary intervals, defined by checking the mass and thermal-energy budgets
and the pressure-maximum radius.  \Cref{tab:usedruns} lists these intervals,
$Q/Q_{\rm ref}$, and the attained $\beta_{\rm pk}$ for each run, with the
stationarity criterion given in \cref{app:runs}.

For the three-dimensional continuations in \cref{sec:3d}, we extrude relaxed
two-dimensional states in $\theta$ to avoid repeating the initial transient.
The six-case survey uses $N_r\times N_z\times N_\theta=192\times720\times192$,
and we run  two cases bracketing the transition at $384\times1440\times384$.
These continuations mostly cover shorter time frames than the 2-D set.

\subsubsection{Scans and controls}\label{sec:campaign}

With the fractional sink and edge fuelling, we vary
$Q/Q_{\rm ref}$ from $0.04$ to $200$ in nine steps, which takes
$\beta_{\rm pk}$ from $0.02$ to $4$.  With the edge thermostat, we scan the
same geometry over the lower part of this range, using
$Q/Q_{\rm ref}=0.04$ to $10$ in eight steps.  The two sinks produce
different responses to increasing drive: the fractional sink allows the
plasma to heat up, leaving the turbulence and transport comparable relative to
the state itself; the thermostat fixes the edge temperature, so a stronger
drive requires much stronger transport.

The source and boundary controls test whether the relaxed state remembers
how it is forced.  At matched total power, we move the heat source from
$r_Q=2\rin$ to $4\rin$.  These tests use both exhaust
prescriptions and more than one drive, and several are paired with runs fuelled
at the pressure maximum, $r_\rho=2\rin$, instead of the edge.  
We
also replace the mass feedback controller with fixed-rate fuelling, taking the
rate from the late-time outcome of the corresponding run, as well as testing $0.25$ and
$4$ times that rate.  In addition to letting the passive scalars escape from 
the boundaries, we test extraction of label $0$ via a sink and scan over a
factor of about $300$ in sink strength.
We do the same test for the mass with an edge particle-sink scan, 
but the sink's strength is limited by the large Alfv\'en speeds that arise  due to the low density 
if mass is evacuated too rapidly.
We scan the
prescribed inner core boundary over a factor of $32$ in $\rho_{\rm BC}$ and $10$
in $T_{\rm BC}$.  One run starts from a different initial equilibrium whose pressure maximum sits
on the hard core.

The remaining controls test numerical and domain effects.  They include the
resolution ladder and axial lengths from $\Lz=L_{z0}/12$ to $2L_{z0}$, as well as two wider annuli with $r_{\rm out}=9\rin$ and $12\rin$.
\Cref{tab:usedruns} names and lists every simulation, its parameters and its averaging
interval.

\section{The self-organised marginal state}\label{sec:state}
\subsection{A representative turbulent state}\label{sec:representative}

\begin{figure}
  \centerline{\includegraphics[width=\textwidth]{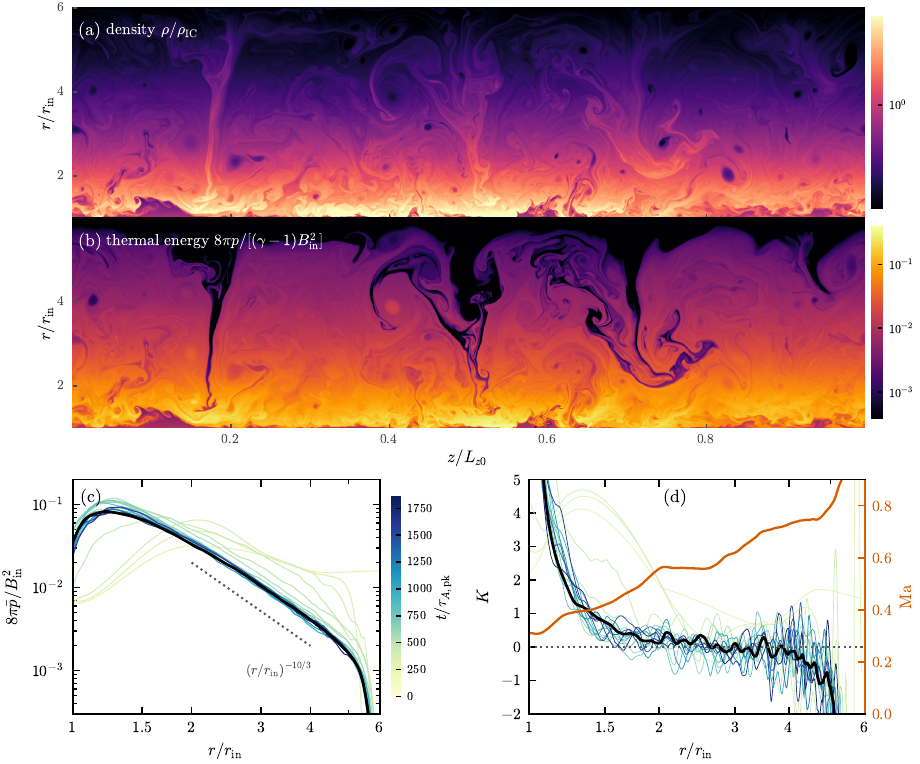}}
  \caption{The self-organised state at the reference drive $Q=Q_{\rm ref}$,
  with edge fuelling.
  (\textit{a},\textit{b}) Density and thermal energy, normalised as described in
  \cref{sec:units}, in the $(r/\rin)$--$(z/L_{z0})$ plane at
  $N_r=1024$ ($1024\times3840$ cells), across the full axial extent and on a
  common radial scale.
  (\textit{c}) Pressure profiles through the relaxation, pale early and dark
  late.  The relaxed mean is black and the low-$\beta$ marginal slope
  $(r/\rin)^{-10/3}$ is included for reference; time is measured in the
  saturated peak Alfv\'en curvature time.
  (\textit{d}) Evolution of the marginality $K$ \eqref{eq:kdef} in the same time colours, with
  its relaxed mean in black; the
  dotted line marks $K=0$. Red, on the right-hand axis, shows the Mach number $\Ma\equiv(\overline{\delta u^{2}})^{1/2}/c_s$
  profile in the saturated state.}
  \label{fig:state}
\end{figure}

\Cref{fig:state}(\textit{a},\textit{b}) show the density and thermal energy in
the $r$--$z$ plane of the default configuration at $N_r=1024$ long after saturation.  Motions fill much of the
annulus.  Dense plumes launched near $r\simeq1.3\rin$ reach the outer boundary,
while cold, low-pressure tongues from the outer edge penetrate inward past
$r\simeq2\rin$; the overturning cells that carry both have radial and axial
sizes comparable to the annulus width.  At a fixed radius the density varies
by more than an order of magnitude, and the thermal energy by more.  

Reaching this state is slow (\cref{fig:state}\textit{c}).  The pressure maximum
starts at $r=2\rin$, migrates inward and settles near $1.25\rin$.  Meanwhile,
the initially flat outer shelf fills in and steepens over
$2\lesssim r/\rin\lesssim4$.  During the final three hundred
$\tau_{A,{\rm pk}}$, the profiles fluctuate about a reproducible mean (black
curve) with no residual trend.  Relaxation takes of order $10^{3}$
peak-Alfv\'en curvature times in this run.  Displacing the sources can extend
it by several times (\cref{sec:sources}), so profiles should be compared only
after a similarly long interval.

This rearrangement is the manifestation of the classic turbulent pinch familiar from
many past studies  in  laboratory plasmas
\citep{Garnier2009,boxer10,Kesner2010,Saitoh2011,Davis2014,Saitoh2026} and 
models ranging from flux-tube-averaged MHD to gyrokinetics
\citep{pastukhov01,pastukhov05,kobayashi10,ou20,li21}.  When two flux tubes interchange, they carry
their mass and entropy with them.  Mixing therefore homogenises the associated
tube contents, while the local density and pressure retain radial gradients.
Since flux-tube volume $\mathcal V$ decreases inward (\cref{eq:tubevolume} below), uniform
particle content produces a centrally peaked mean density.  At low $\beta$ and for $\gamma=5/3$, the
ideal point-dipole scaling $\mathcal V\propto R_{\rm eq}^4$, with $R_{\rm eq}$
the outer-equatorial radius of the field line, gives
$\bar p\propto R_{\rm eq}^{-20/3}$ and $\bar\rho\propto R_{\rm eq}^{-4}$, while the Z-pinch scaling $\mathcal V\propto r^2$ (as studied here) gives
$\bar p\propto r^{-10/3}$ and
$\bar\rho\propto r^{-2}$ \citep{hasegawa87,Hasegawa1990,kadomtsev66,kouznetsov07,Kesner2011}.
This confines the plasma even in the presence of sustained turbulence and provides the physics motivation for the  levitated-dipole concept.   

\subsection{Equilibrium and Kadomtsev marginality}\label{sec:kadomtsev}

Although turbulent transport determines the relaxed profiles, the turbulent
Reynolds and magnetic stresses are small compared with the magnetic pressure
over most of the drive scan, so the mean profiles $\bar p$ and $\bar B_\theta$
remain in approximate radial force balance, \cref{eq:forcebalance}.  We
combine this with the conservation of the tube contents to predict the relaxed
profiles.

Since the field is purely azimuthal and $\partial/\partial\theta=0$, a flux tube is a
ring of plasma at fixed $(r,z)$, whose differential volume per unit flux is
\begin{equation}
 \mathcal{V}\equiv\oint\frac{{\rm d}\ell}{B}=\frac{2\pi r}{B_\theta}.
 \label{eq:tubevolume}
\end{equation}
The conservation laws are particularly simple in this geometry.  In the
two-dimensional flute sector, ideal flux freezing gives
${\rm D}\ln(B_\theta/r)/{\rm D}t=-\bnabla\bcdot\bm u$, and hence
${\rm D}\ln\mathcal{V}/{\rm D}t=\bnabla\bcdot\bm u$.  Away from sources, sinks and
irreversible dissipation, continuity and adiabaticity give
${\rm D}\ln\rho/{\rm D}t=-\bnabla\bcdot\bm u$ and
${\rm D}\ln p/{\rm D}t=-\gamma\bnabla\bcdot\bm u$.  Therefore
\begin{equation}
 \frac{{\rm D}}{{\rm D}t}(\rho\mathcal{V})=0,\qquad
 \frac{{\rm D}}{{\rm D}t}\left(p^{1/\gamma}\mathcal{V}\right)=0,
 \label{eq:tubeinvariants}
\end{equation}
so that the tube mass $\rho\mathcal{V}$ and tube entropy content
$p\mathcal{V}^{\gamma}$, or equivalently $G\equiv p^{1/\gamma}\mathcal V$, are materially conserved by a flute displacement
\citep{freidberg14}.  Turbulent mixing drives both toward uniformity across
the mixed region.  To turn this into a statement about the mean profiles we
evaluate the tube contents on $\bar p$, $\bar\rho$ and
$\bar{\mathcal V}\equiv2\pi r/\bar B_\theta$, effectively neglecting the contribution of 
the turbulent fluctuations to the mean of the tube contents.  As
discussed below, it transpires that uniform $p\mathcal{V}^{\gamma}$ is also the condition for marginality to
linear interchange modes; uniform $\rho\mathcal{V}$ is often termed
turbulent equipartition \citep{yankov94,isichenko95}, but the density profile does not itself affect the plasma's stability.

The low-$\beta$ power laws mentioned above ($\bar p\propto r^{-10/3}$, $\bar\rho\propto r^{-2}$) assume that the pressure does not modify the
field.  At finite $\beta$, force balance must be included when deriving the
marginal profile.  Dividing \cref{eq:forcebalance} for the mean profiles by $\bar B_\theta^2/8\pi$ gives
\begin{equation}
 \frac{{\rm d}\ln \bar B_\theta^2}{{\rm d}\ln r}
  =-2-\beta\frac{{\rm d}\ln \bar p}{{\rm d}\ln r},
 \label{eq:blogslope}
\end{equation}
so that the flux-tube volume varies as
\begin{equation}
 \frac{{\rm d}\ln\bar{\mathcal{V}}}{{\rm d}\ln r}
  =1-\frac{{\rm d}\ln \bar B_\theta}{{\rm d}\ln r}
  =2+\frac{\beta}{2}\frac{{\rm d}\ln \bar p}{{\rm d}\ln r},
 \label{eq:vlogslope}
\end{equation}
and ${\rm d}(\bar p\bar{\mathcal{V}}^{\gamma})/{\rm d}r=0$ becomes
\begin{equation}
 K\equiv\frac{1}{\gamma}\frac{{\rm d}\ln \bar p}{{\rm d}\ln r}
  +\frac{4}{2+\gamma\beta}=0 .
 \label{eq:kdef}
\end{equation}
This is the classic Kadomtsev stability criterion for a Z-pinch
\citep{kadomtsev66,Makhin2005,Crews2024}, usually derived as the criterion for linear stability, rather than via this  
entropy argument (see \cref{app:numerics}).  We use $K$ as
the marginality diagnostic throughout, with $K>0$ stable and $K<0$ unstable.
Linearising about a static equilibrium with these profiles gives the flux-tube
analogue of a squared buoyancy frequency, $N^{2}=2c_s^{2}K/r^{2}$ with
$c_s^{2}=\gamma\bar p/\bar\rho$, and, where $K<0$, an interchange growth rate
$(c_s/r)\sqrt{-2K}$ at large axial wavenumber (\cref{app:numerics}).  At low
$\beta$, \cref{eq:kdef} reduces to ${\rm d}\ln\bar p/{\rm d}\ln r=-2\gamma$,
recovering $\bar p\propto r^{-10/3}$ for $\gamma=5/3$.  At finite $\beta$ the
marginal slope is shallower.  Plasma pressure makes $\bar B_\theta$ fall off
more slowly than the vacuum $r^{-1}$ (\cref{eq:blogslope}), so
$\bar{\mathcal{V}}$ grows more slowly with radius and the same entropy content
gives a flatter pressure profile.

\Cref{fig:state}(\textit{d}) follows $K$ toward its relaxed mean and also
shows the saturated local  Mach number profile, 
$\Ma(r)\equiv[\overline{\delta u^{2}}(r)]^{1/2}/c_s(r)$.  In the source-free bulk, where the turbulence is only mildly subsonic, $K$ eventually oscillates
within a few tenths of zero.  It rises steeply inside the pressure maximum and
exceeds $5$ at the hard core.  Beyond $r\simeq4\rin$, it becomes negative
because the edge heat sink steepens the profile faster than the turbulence can
relax it.  The inner layer is stable in the mean \citep{pastukhov01,pastukhov05}, yet eddies continue to cross
it, and $\Ma$ remains $0.35$--$0.45$ there, smoothly dropping from $0.6$--$0.8$ in the bulk, with these overshooting eddies  driving 
relatively strong heat throughput to the coil
(see \cref{sec:peak,sec:coil}).

\begin{figure}
  \centerline{\includegraphics[width=\textwidth]{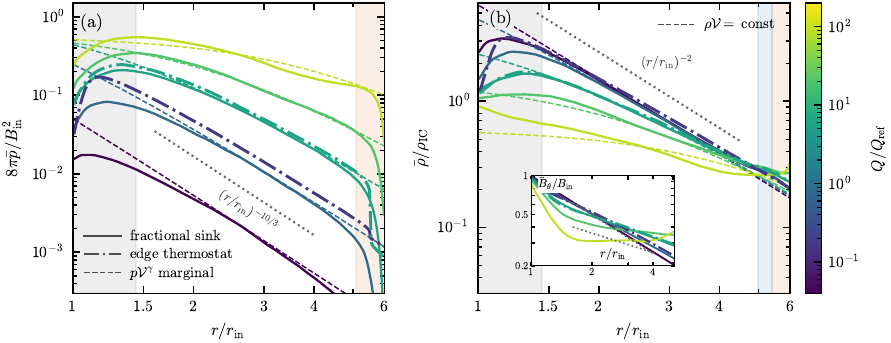}}
  \caption{Both invariants in the relaxed bulk follow the finite-$\beta$
  interchange-marginal family \eqref{eq:family}.
  (\textit{a}) Pressure $8\pi\bar p/B_{\rm in}^2$ for five heating rates under
  the fractional sink (solid) and two under
  the edge thermostat (dot-dashed); dashed curves, drawn only for the
  fractional-sink sequence to keep the comparison legible, are the marginal family
  $\bar p\bar{\mathcal{V}}^\gamma=$ const, anchored at $r/\rin=2.6$ in each run and
  integrated with no
  fitted parameters. The grey dotted line is the low-$\beta$ $(r/\rin)^{-10/3}$
  limit.
  (\textit{b}) Density $\bar\rho/\rho_{\rm IC}$ for the same runs, compared
  with the particle counterpart $\bar\rho\bar{\mathcal{V}}=$ const using the {same}
  finite-$\beta$ field.  The family  thus tests both invariants, with the
  dashed curves for the fractional-sink sequence showing $\bar\rho\propto\bar B_\theta/r$
  anchored at the same radius (the grey dotted $r^{-2}$ line  is the
  corresponding low-$\beta$ profile).  Inset: the field
  profiles $\bar B_\theta$ that set $\bar{\mathcal{V}}=2\pi r/\bar B_\theta$, with $r^{-1}$ (vacuum) and
  $r^{-1/2}$
  dotted references.  Light shading marks the
  boundary-contaminated inner region and the source/sink layers.  Line colour
  encodes the heating rate, as shown by the colour bar, here and in later figures.}
  \label{fig:marginal}
\end{figure}

\subsubsection{The finite-$\beta$ marginal family}\label{sec:marginal}
Marginality alone does not determine the profile, since \cref{eq:kdef}
contains $\beta$, which is itself part of that profile.  Differentiating $\beta=8\pi\bar p/\bar B_\theta^2$ and substituting
\cref{eq:blogslope,eq:kdef} gives
\begin{equation}
 \frac{{\rm d}\beta}{{\rm d}\ln r}
  =-\frac{2\beta\left[2(\gamma-1)+\gamma\beta\right]}{2+\gamma\beta},
 \label{eq:family}
\end{equation}
whose solutions form a one-parameter family labelled by $\beta$ at any one
radius, with $\bar p$ and $\bar B_\theta$ determined up to an overall amplitude \citep{pastukhov01}.  This
fixes the \emph{shape} of the profiles through the mixed region.  Their
amplitude, the structure of the stable inner layer and the radius of the
maximum still depend on the sources and losses
(\cref{sec:operating,sec:peak}).  At low $\beta$, \cref{eq:family}
gives $\beta\propto r^{-2(\gamma-1)}$, $\bar p\propto r^{-2\gamma}$ and
$\bar B_\theta\propto r^{-1}$, the vacuum field and the marginal power law of
\citet{kadomtsev66}, to which quasilinear treatments of this geometry also
relax \citep{kouznetsov07}.  As $\beta$ increases the profile flattens, and for
$\beta>2/\gamma$ the marginal field increases outward.  Joining such a region
to the decreasing field near the pressure maximum can produce an interior
minimum in $B_{\theta}$ \citep{Guazzotto2007}.

\Cref{fig:marginal}(\textit{a}) compares the relaxed pressure profiles with
this family across the drive scan.  For each run, we read $\beta$ from the
measured profile at the single radius $r=2.6\rin$, well inside the source-free
interval, and integrate \cref{eq:family} both inward and outward.  There are no
fitted shape parameters after anchoring the amplitudes at this radius.  The family follows the measured profiles through the bulk
over the whole scan.  The weakest lies close to the low-$\beta$
$(r/\rin)^{-10/3}$ limit, and the profiles flatten successively as $\beta$
rises, as required by \cref{eq:family}.

Panel (\textit{b}) shows the corresponding densities.  Uniform
$\bar\rho\bar{\mathcal{V}}$ requires $\bar\rho\propto\bar B_\theta/r$, so its shape follows from
the same solution; fuelling and the retained inventory set its amplitude
separately.  When anchored at the same radius, this relation reproduces the
measured bulk density well over most of the scan, with
larger departures at the strongest drive and inside the maximum, where the
family has no interior turnover.  The inset shows the field
profiles responsible for the finite-$\beta$ correction.  As the drive rises,
$\bar B_\theta$ departs from the vacuum $r^{-1}$ profile and, at the strongest
drives, flattens and turns over in the interior.  The $r^{-1/2}$ reference line
marks the slope at which the local $m=1$ bending threshold is crossed.  This
eventually limits the two-dimensional state (see \cref{sec:3dtheory}).

The five fractional-sink states shown span $\beta_{\rm pk}=0.02$ to $4.2$, and
the two thermostat states span $0.29$ to $0.62$.  The prescriptions reach the
same $\beta$ at very different drives because the thermostat pins the edge
temperature; even its weakest saturated run has $\beta_{\rm pk}\simeq0.28$.  Across this
range the amplitudes differ by more than an order of magnitude, yet the
normalised shapes change little.  The same distinction recurs throughout the
paper.  Turbulence selects the profile \emph{shape}, while the sources and
losses select its \emph{operating point} \citep[see also][]{pastukhov01,pastukhov05}.  Agreement with the marginal family
is closest at intermediate drive.  At the top of the scan, the extreme
throughput requires a larger departure from marginality (\cref{sec:closures}).
The outer-field pile-up discussed in \cref{sec:boundaries} also contaminates
the shaded band at the right of each panel, where the comparison is only
qualitative.

\section{Robustness of the marginal state}\label{sec:sources}
\subsection{Displacing heat and particles}\label{sec:sourceprofiles}

We test the robustness of the marginal family to changes in the source geometry
 by holding the total power and all other parameters
fixed while moving the heating location from $r_Q=2\rin$ (\texttt{S-Q1}, \texttt{T-Q1}) to $3\rin$ (\texttt{S-Q1-H3}, not shown)
or $4\rin$ (\texttt{S-Q1-H4}, \texttt{T-Q1-H4}).  We also move the particle fuelling from the
outer edge, $r_\rho=5.2\rin$, to the core, $r_\rho=2\rin$
(\texttt{S-Q1-CF}, \texttt{T-Q1-CF}), with tests repeated using both exhaust prescriptions.
\Cref{tab:usedruns} lists the full simulation set.

\begin{figure}
  \centerline{\includegraphics[width=\textwidth]{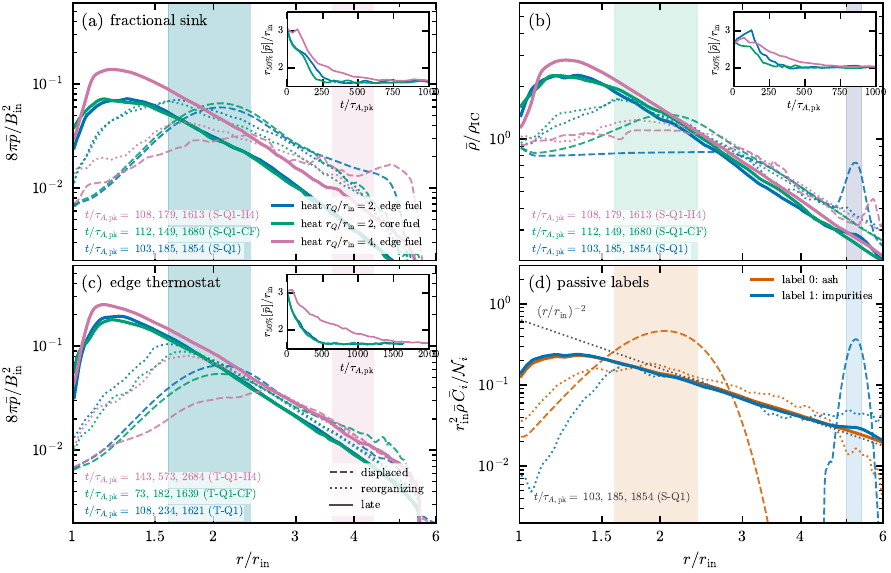}}
  \caption{Comparison of the effect of moving the heating and fuelling  
  locations in different runs. 
  Different colours show different heat and particle
  source locations, while line style shows three stages of each simulation in time (early,
  reorganising and late).  Because the runs evolve at different rates, these
  stages are chosen separately rather than at common times.  The times, followed by the run name, are
  printed in the case colour and use each run's saturated
  $\tau_{A,{\rm pk}}$.  Translucent bands mark one Gaussian standard deviation
  of the relevant source: heat in the pressure panels and particle replacement
  in the density panel.
  (\textit{a}) Pressure and (\textit{b}) density under the fractional sink;
  (\textit{c}) pressure under the edge thermostat.  Insets follow
  $r_{50\%}$, the radius dividing the area under the mean profile in half, on a
  common $1.5$--$3.5$ scale in $r/\rin$.  (\textit{d}) The two passive-label
  densities in \texttt{S-Q1}.  Label $0$, the ash
  analogue, enters with the heat; label $1$, the impurity analogue, enters with
  the replacement particles.  Each is normalised only by its own inventory
  (the ordinate is $\rin^2\bar\rho\,\bar C_i/\mathcal N_i$, with
  $\mathcal N_i=\int\bar\rho\,\bar C_i\,r\,{\rm d}r$), so their convergence onto the
  $(r/\rin)^{-2}$ guide concerns the labels rather than the bulk density.
  Early profiles track their different sources; the late profiles converge in
  position but retain different amplitudes.}
  \label{fig:sources}
\end{figure}

\Cref{fig:sources} compares three stages of each run, with the times shown chosen separately
because the different source configurations reorganise at very different
rates.  Blue is the default heat and edge-fuelling case, green moves the fuelling
inward,  and pink moves the heat outward; dashed, dotted and solid curves show different stages in time
(early, intermediate, and late, respectively).  The early profiles differ due to 
their different sources, but the relaxed states are nearly identical in shape: source
placement changes only the relaxation timescale and the final amplitude.

The insets quantify this migration using $r_{50\%}[\bar f]$, defined by
$\int_{\rin}^{r_{50\%}}\bar f\,{\rm d}r
=\tfrac12\int_{\rin}^{r_{\rm out}}\bar f\,{\rm d}r$, which is much steadier
than the position of the maximum itself.  Heating at $r_Q=4\rin$ first pulls
the maximum \emph{outward}, to $2.4\rin$ under the fractional sink and
$2.6\rin$ under the thermostat, and builds a broad shoulder over its
deposition layer.  At that stage the maximum is around half the peak pressure of
the two reference runs, which remain near their initial $2\rin$.  All six
maxima then move inward and converge between $1.19$ and $1.27\rin$.  This
spread is smaller than the widths of the maxima themselves, while $r_{50\%}$
converges to within several per cent: $1.63$--$1.66\rin$ for $\bar p$ and
$2.02$--$2.06\rin$ for $\bar\rho$.  The relaxation time scales approximately 
with the energy confinement time $\tauE=W/Q$, which is $220$--$530\,\tau_{A,{\rm pk}}$ for these  runs and will 
be explored in more detail below.

The passive labels provide a stricter test of the mixing
(\cref{fig:sources}\textit{d}).  Label $0$ is released with the heat at
$2\rin$ and label $1$ with the replacement particles at $5.2\rin$.  Each is
normalised by its own inventory, so the comparison concerns shape rather than
amount.  Although the labels begin as images of their source layers, both end
on the bulk-density profile, close to the low-$\beta$ equipartition form
$(r/\rin)^{-2}$, and their shapes agree closely across the bulk.  Both
concentrations therefore become flat \citep{pastukhov05}.  The 
reason is simply that the same density-mixing dynamics that flatten $\bar\rho \bar{\mathcal V}$ 
also apply to any passive scalar that satisfies \cref{eq:scalar}.

Surprisingly, moving the heating outward at fixed power \emph{improves} confinement for the same profile shape.  At
$Q=Q_{\rm ref}$, the stored energy and peak pressure rise by ${\simeq}40$ per
cent under the thermostat (panel \emph{c}) and ${\simeq}70$ per cent under the fractional sink (panel \emph{a}).
The increases are similar at $4Q_{\rm ref}$ (\texttt{T-Q4-H4} and \texttt{S-Q4-H4}, not shown;
see \cref{tab:usedruns}), while $\rpk$ is nearly unchanged across all tests.  The reason is that an outwards-displaced source
drives the interior less strongly: between the maximum and the source, $K$
averages ${\simeq}{+}0.2$, compared with ${\simeq}0$ in the reference runs
(not shown).  Transport through the interior is therefore weaker and less
power reaches the coil (\cref{sec:coil}).  The relaxed profiles are also slightly
shallower with outboard heating.

\subsection{Fluxes and  transport routes}\label{sec:sourceroutes}

The system must organise its mass, entropy and energy fluxes so that the same
quasi-universal marginal state can form under different heating and fuelling scenarios.
Although turbulence-driven fluxes can in principle have either sign
\citep{KesnerGarnier2000,Kesner2003,Kesner2011}, as we now show, the tube-content laws suggest that
the entropy flux is constrained to point outwards, while mass flux is not.  A standard local diffusive closure for either
materially conserved quantity in \cref{eq:tubeinvariants} is
\begin{equation}
 \overline{\delta u_r\,\delta f}\simeq
 -\frac{\chi_f}{\bar{\mathcal{V}}}
  \frac{{\rm d}(\bar{\mathcal{V}}\bar f)}{{\rm d}r},
 \qquad \chi_f>0,\qquad f=\rho\ \hbox{or}\ p^{1/\gamma}.
 \label{eq:tubecontentflux}
\end{equation}
Here $\chi_f$ is an effective radial diffusion coefficient, with the usual
mixing-length estimate $\chi_f\sim u_{r,{\rm rms}}\ell_r$ for rms turbulent
radial speed $u_{r,{\rm rms}}\equiv(\overline{\delta u_r^2})^{1/2}$ and eddy extent $\ell_r$; of course
a local closure is not accurate for domain-scale eddies  (\cref{fig:state}) but nonetheless
provides a helpful intuition.  For entropy, the gradient in
\cref{eq:tubecontentflux} is related directly to stability.  Rearranging
\cref{eq:vlogslope,eq:kdef}, which rely on the radial equilibrium, gives
\begin{equation}
 \frac{{\rm d}\ln(\bar p^{1/\gamma}\bar{\mathcal{V}})}{{\rm d}\ln r}
 =\frac{2+\gamma\beta}{2}K.
 \label{eq:kentropygradient}
\end{equation}
In a locally unstable region, $K<0$ both drives the turbulence and, through
\cref{eq:tubecontentflux}, gives an outward entropy flux. In contrast,  an inward entropy
flux would (under this closure) require a stable gradient $K>0$, which  would not be able to  drive the eddies
that carry it unless forced in another way.  We therefore expect entropy to flow mainly outward through
unstable regions.\footnote{Kinetic physics could change this conclusion.  The entropy mode of
closed-field-line systems can be unstable where the ideal interchange is stable ($K>0$), because it is
driven by the relative density and temperature gradients rather than by $K$ \citep{kobayashi10}.  Eddies
driven in this way could carry entropy through ideally stable regions without the constraint above.} 

In contrast, there is no stability constraint on
${\rm d}(\rho\mathcal{V})/{\rm d}r$.  Pressure-driven eddies can mix particle
content inward or outward, depending on the gradient produced by the source
and exhaust.  Mass transport should therefore follow the imposed fuelling and
losses rather than the sign of $K$.

\begin{figure}
  \centerline{\includegraphics[width=\textwidth]{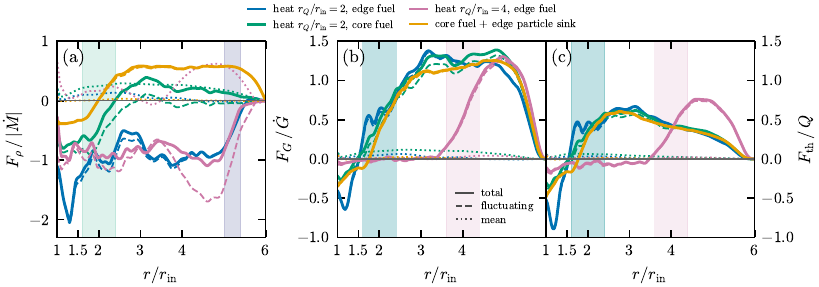}}
  \caption{The energy and mass fluxes illustrate how the system sets up different transport routes to the same marginal state.  Relaxed
  shell-integrated (\textit{a}) mass, (\textit{b}) entropy and
  (\textit{c}) internal-energy fluxes, \cref{eq:fluxes}, split into total,
  fluctuating and mean-advective parts and normalised by the rate at which
  each source supplies the transported quantity.  The blue, green and pink
  curves are the three fractional-sink cases of
  \cref{fig:sources}(\textit{a},\textit{b}).  Orange in all three panels is
  the strong core-fuelled edge-particle-sink control \texttt{S-Q1-CF-PSs}, in which the outer sink extracts $58$ per cent of the supplied
  particles, and which has the same fractional energy sink.  Panels (\textit{b}) and
  (\textit{c}) are aligned and share a scale as two views of thermal transport.
  The adiabatically conserved $G$ content is mathematically the most appropriate quantity to consider; its outer plateau
  above unity occurs because entropy is produced by shocks and dissipation. 
  The non-conserved $F_{\rm th}$ instead measures the total internal energy advected
  to a material surface, which is of more interest physically.  Translucent bands mark
  one standard deviation of the mass source in (\textit{a}) and the heat
  source in (\textit{b},\textit{c}).  We see that particle flux follows the imposed source and loss
  geometry and can have either sign. In contrast, entropy is transported only outward through most of the domain, other than a small inward overshoot through 
  the stable ($K>0$) region.}
  \label{fig:fluxes}
\end{figure}

\Cref{fig:fluxes} tests this expectation using the relaxed, shell-integrated
radial fluxes of mass, entropy, and internal energy,
\begin{equation}
 F_\rho=2\pi r\Lz\,\overline{\rho u_r},\qquad
 F_G=2\pi r\Lz\,\overline{p^{1/\gamma}u_r},\qquad
 F_{\rm th}=\frac{2\pi r\Lz}{\gamma-1}\,\overline{p\,u_r},
 \label{eq:fluxes}
\end{equation}
for the three fractional-sink cases in
\cref{fig:sources}(\textit{a},\textit{b}).  Blue is the default edge-fuelled run (\texttt{S-Q1}), green
moves the fuel to the core (\texttt{S-Q1-CF}), and pink moves the heat outward (\texttt{S-Q1-H4}), matching \cref{fig:sources}.  In orange, we also include
a simulation with an outer mass sink similar to the energy sink (\texttt{S-Q1-CF-PSs}) to test how redirecting
the mass throughput changes the fluxes (the marginal pressure and density profiles of this state are very similar but not shown).  Solid,
dashed, and dotted curves are the total, fluctuating and mean-advective fluxes,
respectively; negative values are inward.  The entropy flux $F_G$ can be read in the same way as $F_\rho$,
with  heating supplying it at the volume-integrated rate
$\dot G=[(\gamma-1)/\gamma]\int p^{1/\gamma-1}Q g_Q\,{\rm d}V$,
with no mass-source contribution.
The $G$ budget also contains imposed sink terms and irreversible production
by shocks and grid-scale viscous and resistive dissipation, the latter implying that  unlike mass, its
flux need not be perfectly flat between the source and exhaust.  The thermal energy (heat) flux, $F_{\rm th}$, is 
more relevant physically, but $p$ itself is not a conserved quantity --- the pressure work $p\bnabla\bcdot\bm u$ enters
its evolution --- meaning that $F_{\rm th}$ need not be flat away from sources. 

The fluxes differ markedly between these simulations, even though the profiles themselves
 do not (\cref{fig:sources}).  
The mass flux follows the fuelling and exhaust and can be either positive or negative
(\cref{fig:fluxes}\textit{a}).  Edge fuelling with the ordinary exhaust gives
an inward throughput to the coil, while core fuelling drives only a weak
outward flux outside the source, since more of the added mass flux 
still goes inwards into the coil for these boundary conditions.  Adding an outer particle sink (orange) creates a 
robust outward mass flux
between the source and extractor.  Overall, we see that the mass-flux route is strongly affected by the choice of boundary condition or sinks.
The fixed inner density and temperature make the hard core an
efficient absorbing reservoir, whereas the outer amplitudes, which  follow the instantaneous
edge state and continue as power laws (\cref{sec:boundaries}), are less inclined to absorb significant
mass flux. 

In contrast, entropy and energy, shown in \cref{fig:fluxes}\textit{b}--\textit{c}, follow the stability argument above,  
 exhibiting only outwards fluxes except in the (stable) near-wall overshoot region.  Moving the heat source outward 
reduces the entropy flux to zero through the inner bulk,  while  still allowing the strong pressure peak shown in \cref{fig:sources}\textit{a} 
to grow and sustain itself (a slight inwards internal energy flux is visible in panel 
\textit{c}
 because compression and pressure work add energy on the
way to the coil).  Most importantly, we see that the inward migration of the pressure peak seen in \cref{fig:sources}
is not explained by a self-sustained inward entropy pinch.  Instead,
the pressure maximum moves as turbulence erodes the stable side of the
marginality front (\cref{sec:peakexcursion}), and there is little heat transport 
inward from the source. 

\section{Turbulence and confinement}\label{sec:operating}

The energy and particle confinement times, $\tauE=W/Q$ and
$\tauP=M/\dot M_{\rm loss}$, depend on the turbulence that carries heat and
particles through the marginal profile.  We first determine what sets its
amplitude and temporal response (\cref{sec:closures,sec:cycle}) and study the turbulence properties (\cref{sec:globality}), then 
measure stored energy and mass residence times
(\cref{sec:storedenergy,sec:confinement}).

Various diagnostics below use the bulk band
$\mathcal B:2.5<r/\rin<5$, between the default heating layer and the outer
source and sink.  The displaced-heating controls retain this band for comparison,
although it contains their heat source.  We define the following turbulent-weighted average for any radial profile $X(r)$,
\begin{equation}
 \langle X\rangle_{\mathcal B}
 =\frac{\int_{\mathcal B}X(r)E_K(r)\,{\rm d}V}
        {\int_{\mathcal B}E_K(r)\,{\rm d}V}
 =\frac{\int_{\mathcal B}X(r)rE_K(r)\,{\rm d}r}
        {\int_{\mathcal B}rE_K(r)\,{\rm d}r},
 \qquad
 E_K=\tfrac12\bar\rho u_{r,{\rm rms}}^2,
 \label{eq:bulkaverage}
\end{equation}
which thus weights the radii that contain radial
turbulent kinetic energy rather than quiet parts of the band.  The $E_K$ weight is
formed from the time-averaged profile over the interval being analysed and is
kept fixed as a weight, even in time-dependent calculations.  

\subsection{How the turbulence regulates its amplitude}\label{sec:closures}

We test the hypothesis that the saturated amplitude is set by a balance between interchange growth and
nonlinear turnover, with the profile  remaining sufficiently supercritical  to carry the
imposed heating rate. A locally supermarginal state,
$\Delta K\equiv-K>0$, grows at
$\gamma_{\rm int}=(c_s/r)\sqrt{2\Delta K}$ (\cref{sec:kadomtsev} and \cref{app:numerics}), while an eddy of radial extent $\ell_r$
turns over at $\omega_{\rm nl}=u_{r,{\rm rms}}/\ell_r$.  Saturation should make these rates
comparable, which, with global-scale 
energy-containing eddies $\ell_r\sim r$
(see \cref{sec:globality}),  reduces to the simple estimate
$\Ma\sim\sqrt{\Delta K}$.  

\begin{figure}
  \centerline{\includegraphics[width=\textwidth]{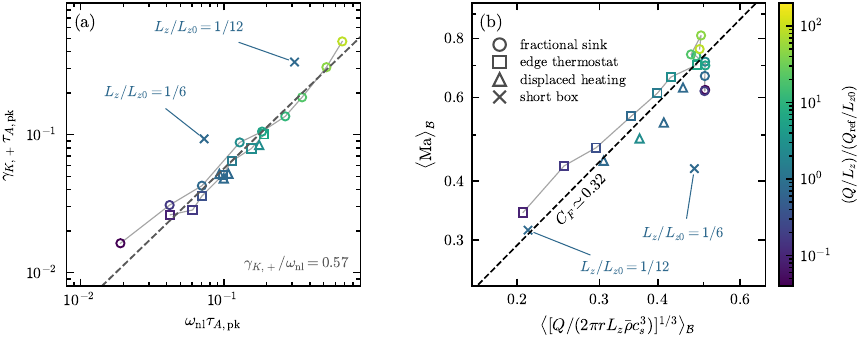}}
  \caption{Amplitude of the saturated turbulence across many runs. 
  (\textit{a}) Turbulence-weighted  interchange rate $\gamma_{K,+}$
  against the nonlinear rate $\omega_{\rm nl}$, \cref{eq:rates}, both in
  saturated peak-Alfv\'en units; the dashed line is the fixed-slope best fit
  $\gamma_{K,+}/\omega_{\rm nl}=0.57$ to the non-short-box cases.  Circles and
  squares are the fractional-sink and
  thermostat drive scans, respectively, triangles the displaced-heating runs \texttt{S-Q1-H3},
  \texttt{S-Q1-H4}, \texttt{T-Q1-H4} and \texttt{T-Q4-H4} (the source-placement class of
  \cref{tab:usedruns}), and crosses the short boxes \texttt{S-Lz/6} and \texttt{S-Lz/12}, which
  are labelled by $L_z/L_{z0}$ and excluded from the fit and quoted range.
  (\textit{b}) Mach number against the mixing-length combination of
  \cref{eq:cf}; both coordinates use the bulk average in
  \cref{eq:bulkaverage}, and the dashed guide is $C_F\simeq0.32$.  The same symbols apply.  Colour encodes the heating per unit axial length,
  $(Q/L_z)/(Q_{\rm ref}/L_{z0})$, which equals $Q/Q_{\rm ref}$ at the standard box length. 
  The measures support the physical picture that the system saturates to balance interchange driving and nonlinear mixing against the imposed heating.}
  \label{fig:closures}
\end{figure}

We test this in the simulations by evaluating $\gamma_{\rm int}$ and $\omega_{\rm nl}$ using
\cref{eq:bulkaverage}:
\begin{equation}
 \gamma_{K,+}^{2}
  =\left\langle\frac{2c_s^{2}}{r^{2}}\max(-K,0)\right\rangle_{\mathcal B},
 \qquad
 \omega_{\rm nl}^{2}
  =\left\langle\left(\frac{u_{r,{\rm rms}}}{\ell_r}\right)^{2}
   \right\rangle_{\mathcal B},
 \label{eq:rates}
\end{equation}
where $\ell_r(r)$ is the measured radial e-folding correlation length.  This is the mean of the inward and outward
separations $|r'-r|$ at which the normalised correlation of radial-velocity fluctuations,
\begin{equation}
 C_{u}(r,r')=\frac{\overline{\delta u_r(r)\,\delta u_r(r')}}{u_{r,{\rm rms}}(r)\,u_{r,{\rm rms}}(r')},
 \label{eq:urcorr}
\end{equation}
first falls to $e^{-1}$.  The average in \cref{eq:urcorr} is over $z$ and time, and
\cref{fig:globality}(\textit{a}) shows examples.  We clip $-K$ at zero to isolate the locally unstable contribution, since stable
radii will  provide restoring buoyancy to a global displacement but not necessarily a damping. Thus,  rather than a global eigenmode
growth rate,
$\gamma_{K,+}$ is a crude measure of the available drive, weighted to favour the contribution of  the most unstable regions
with stronger turbulence.
\Cref{fig:closures}(\textit{a}) shows the results across the drive scans
(circles and squares), simulations with displaced heating (triangles) and short boxes (crosses).
The growth and turnover rates track one another from
$\omega_{\rm nl}\tau_{A,{\rm pk}}\simeq0.02$ to $0.5$, with
$\gamma_{K,+}/\omega_{\rm nl}\simeq0.5$--$0.9$ and a fixed-slope best fit of
${\simeq}0.6$.   The short-box crosses lie well above
the trend because their transport becomes limited by zonal flows (\cref{app:zonal}).

Panel (\textit{b}) asks what turbulence amplitude is reached at a given heating rate.
Mixing-length reasoning supposes that eddies of speed
$u_{r,{\rm rms}}$ carry heat across a shell at a rate set by their own kinetic
energy flux, ${\sim}\bar\rho u_{r,{\rm rms}}^{3}$ per unit area, so that in
steady state $Q\simeq C_F\,\bar\rho u_{r,{\rm rms}}^{3}\,2\pi r\Lz$ with $C_F$
an order-unity efficiency.  Written in terms of the Mach number, this gives
\begin{equation}
 \Ma=\left[\frac{Q}{C_F\,2\pi r\Lz\,\bar\rho c_s^{3}}\right]^{1/3}.
 \label{eq:cf}
\end{equation}
The
measured $\langle\Ma\rangle_{\mathcal B}$ follows \eqref{eq:cf} reasonably well with a single
$C_F\simeq0.32$ across the full drive range, both
exhaust prescriptions, and
displaced heating, although it is worth noting that with the standard fractional heat sink boundary, changing $Q$ does not strongly 
affect $\Ma$, with the plasma responding by adjusting $c_{s}$ instead.   The two
short boxes are once again shown separately because their zonal-flow states change both
the local thermodynamics and the kinetic-energy weight.\footnote{The fact that the $\Lz=L_{z0}/12$ case 
lies closely on the fit line appears to be a coincidence, as its saturated turbulence is the most strongly zonal-flow affected 
of all simulations; see \cref{fig:zonal}.}
While the numerical value of 
$C_F$ is presumably specific to this geometry and model,
the reasonable agreement of both measures (panels \emph{a} and \emph{b}) provides good evidence that the system sits 
in a quasi-steady balance between modestly supercritical interchange driving and turbulent mixing. 

\subsection{The regulation cycle}\label{sec:cycle}

The time-averaged balances in \cref{fig:closures} do not show how marginality
is maintained and would be consistent with either a state at fixed supercriticality or a relaxation cycle
in which the gradient repeatedly overshoots and decays.  These pictures imply
different regulation physics; we explore them here by diagnosing the lag between the
drive and the turbulent response.

\begin{figure}
  \centerline{\includegraphics[width=\textwidth]{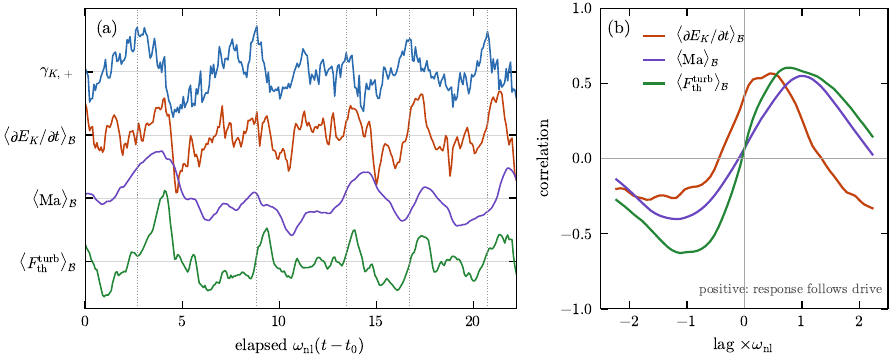}}
  \caption{The regulation cycle, measured in the reference state \texttt{S-Q1} at high
  output cadence.  (\textit{a}) Standardised traces of the
  drive $\gamma_{K,+}$, the growth rate of the radial turbulent kinetic energy density
  $\partial E_K/\partial t$, the Mach number, and the fluctuating advected
  internal-energy flux, each evaluated via the $\mathcal B$ average.  
  The drive has a mean of $0.65\,\omega_{\rm nl}$ and a standard deviation of $0.13\,\omega_{\rm nl}$, so it
  oscillates about a positive mean rather than through zero.
  Dotted guides mark five
  prominent, well-separated maxima of the drive $\gamma_{K,+}$.  (\textit{b})
  Lagged correlation of each response with the drive.  Elapsed time
  in (\textit{a}) and lag in (\textit{b}) are in units of the median
  $\omega_{\rm nl}^{-1}$ over the interval, and positive lag means the
  response follows the drive, illustrating how the system sets up a cycle whereby pressure gradients build in time, thereby 
  driving more turbulence that increases the transport, flattening the gradients again.}
  \label{fig:cycle}
\end{figure}

\Cref{fig:cycle} follows the reference state \texttt{S-Q1} at high cadence, using three
hundred samples over twenty-two turnover times.  We show four diagnostics, the drive $\gamma_{K,+}$, 
the rate of change of the radial turbulent kinetic energy density $\partial E_K/\partial t$, the Mach number $\Ma$, and
the time variation of the internal-energy flux in \cref{eq:fluxes},
\begin{equation}
 F_{\rm th}^{\rm turb}(r,t)
 =\frac{2\pi r\Lz}{\gamma-1}
   \overline{\delta p\,\delta u_r}.
 \label{eq:turbheatflux}
\end{equation}
For each, we average over
$\mathcal B$ via 
$\langle\Ma(r,t)\rangle_{\mathcal B}$ and plot each time trace on a normalised axis (labels omitted)
to show their time variation and correlation.

We mark five prominent drive maxima in \cref{fig:cycle}(\textit{a}) via vertical dotted lines. These recur
roughly every four to five turnovers, and --- as seen in the time traces of $\partial E_K/\partial t$, whose maxima align with those of $\gamma_{K,+}$ --- drive 
the growth of turbulence. The response of $\Ma$ then follows, bringing with it an internal 
energy flux $F_{\rm th}^{\rm turb}$ that acts to flatten $p$, thereby decreasing $\gamma_{K,+}$.  The correlations in panel (\textit{b}) show this
more precisely: we see that
$\partial E_K/\partial t$ responds within about half a turnover, whereas $\Ma$ and
$F_{\rm th}^{\rm turb}$ follow after about one turnover.

The marginal state of \cref{sec:state} is thus maintained by a regulation
cycle by which  an excess gradient builds periodically, driving  turbulence that drives enhanced transport that then relaxes the gradient. 
This process keeps the profile close to marginality, oscillating about a slightly supercritical state ($K\lesssim0$), whose residual
drive is measured by the always-positive mean $\gamma_{K,+}$ in \cref{fig:closures}(\textit{a}).

\subsection{Global structure of the turbulence}\label{sec:globality}

\begin{figure}
  \centerline{\includegraphics[width=\textwidth]{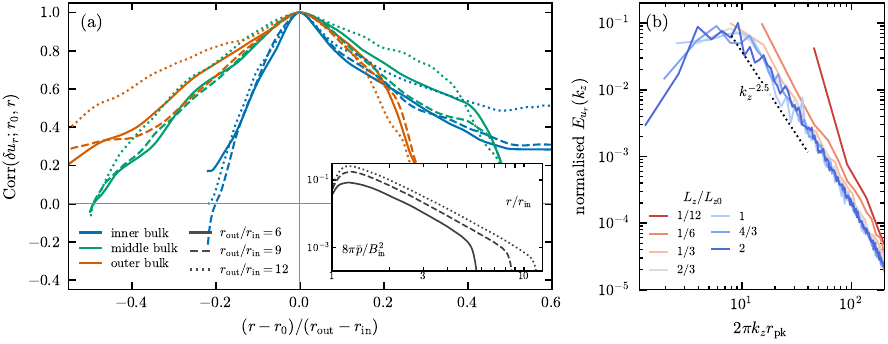}}
  \caption{Turbulent correlation functions and spectra.
  (\textit{a}) Correlation of $\delta u_r$ about three anchor radii $r_0$,
  placed at fractions $0.22$ (inner bulk), $0.50$ (middle) and $0.72$ (outer)
  of the annulus width from the
  hard core, for the reference simulation \texttt{S-Q1} and the wider-domain
  simulations \texttt{S-Q1-R9} and \texttt{S-Q1-R12}, which have $r_{\rm out}/\rin=6$, $9$, and $12$, respectively.
  The radial coordinate is normalised by the annulus width, so a fixed
  physical eddy size would appear as curves narrowing in proportion as the
  domain widens; instead they show similar functional forms for all cases.  The inset shows the corresponding
  relaxed pressure profiles, which lie on the same marginal power law in each
  case, so the wider correlations are obtained at similar logarithmic
  pressure slopes.
  (\textit{b}) Source-free axial spectra of $u_r$ against $2\pi k_z\rpk$ for
  $\Lz=L_{z0}/12$ to $2L_{z0}$, each normalised to unit integrated power.  The
  shortest boxes are constrained by their lowest admissible mode, while the
  longer ones share a common resolved outer energy-containing scale.  The dotted guide is a purely 
  empirical $k_z^{-2.5}$ fit to the slope of smaller-scale turbulence.}
  \label{fig:globality}
\end{figure}

The turbulent motions are radially global, with one or two long-lived vortex pairs
spanning  the full annulus (see also
\citealt{pastukhov01,pastukhov05}).  To test the dependence on the
available radial extent, we compare \texttt{S-Q1} with two additional simulations that
move the outer boundary from $r_{\rm out}/\rin=6$ to $9$ and $12$ (\texttt{S-Q1-R9}
and \texttt{S-Q1-R12}) while retaining the reference drive and approximately the same
resolution per unit radius.  Individual plumes in \cref{fig:state} span much of the reference
annulus, and the velocity correlations in
\cref{fig:globality}(\textit{a}) show that their size grows in the wider runs.
The correlations are measured about three anchor radii, at fractions $0.22$,
$0.50$, and $0.72$ of each annulus.  After separation is normalised by
$r_{\rm out}-\rin$, the curves collapse, whereas a fixed  eddy size
would make these correlations narrow as the domain widened.  The eddies therefore occupy a
fixed, order-unity fraction of the entire annulus.  For \texttt{S-Q1}, the e-folding length is
$\ell_r\simeq1.7\rin$, about one third of its radial width; this measured value
sets $\omega_{\rm nl}$ in \cref{eq:rates}.

The inset also illustrates the relaxed pressure profiles in these wider simulations, demonstrating
the continued power law across
the full domain.  Widening the annulus therefore extends both the
marginal profile and the global motions that sustain it.

The axial spectra in \cref{fig:globality}(\textit{b}) illustrate the scale at which
the turbulence holds its energy when left to develop freely.  We see that in the long boxes, with
a spectral peak at $2\pi\rpk k_z\simeq5$--$10$, the energy-containing eddies have an axial wavelength of
$5$--$8\rin$, meaning a single overturning cell is of size several $\rin$, comparable to
its radial extent of roughly $2\ell_r$.  Thus, the turbulence is approximately isotropic in the $r$--$z$ plane (see also
\cref{fig:state}\textit{a},\textit{b}).
The same spectra tell us the box length needed to accommodate the turbulence without causing 
artificial restrictions.  The shortest
boxes place most of their power in the lowest admissible mode (red lines), whereas the
spectra overlap over the resolved range once $\Lz\gtrsim2L_{z0}/3$.  We use
this comparison to set the  length of the box for production simulations and treat the dynamics of shorter boxes
in \cref{app:zonal}, showing how they become strongly affected by zonal flows.

Over the resolved range, the axial spectrum falls approximately as
$k_z^{-2.5}$.  We do not have an explanation for this slope, and it is not
converged: the fitted value changes from about $-3$ at $N_r\leq256$ to $-2.4$
at $N_r=1024$.  These spectra are steeper than the $-3/2$ or $-5/3$ spectra of
homogeneous two-dimensional MHD \citep{biskampwelter89,ishizawa98,jha26} and,
unlike that system, the flute sector has $B_r=B_z=0$ and no in-plane magnetic
cascade.  Its broad radial eddies also sample a background that changes across
their own extent, which may limit the relevance of single-scale cascade
arguments.

\subsection{Stored energy}\label{sec:storedenergy}

\begin{figure}
  \centerline{\includegraphics[width=\textwidth]{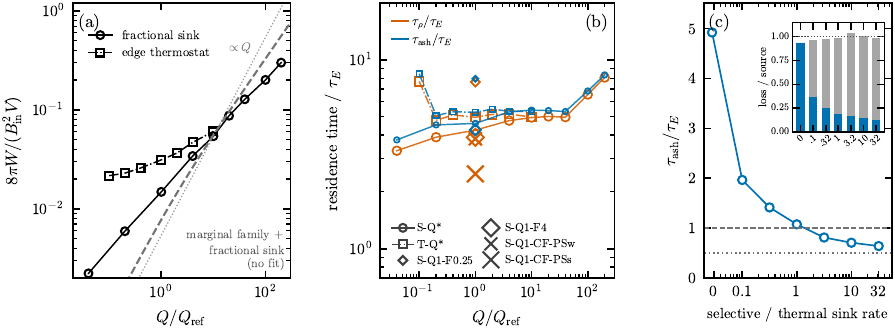}}
  \caption{Steady-state stored energy and residence times.
  (\textit{a}) Stored thermal energy, normalised so that it is the 
  volume-averaged ratio of thermal to inner-edge magnetic pressure,
  against $Q/Q_{\rm ref}$ for the two exhaust prescriptions in the edge-fuelled
  scans \texttt{S-Q*} and \texttt{T-Q*}.  The dashed curve is the finite-$\beta$ marginal family
  closed with the fractional-sink rule (see text), drawn with no fitted constants, and the
  thin dotted line is a linear reference, $W\propto Q$.
  (\textit{b}) Bulk-particle (red) and heat-released ash-label (blue)
  inventory-to-loss times divided by $\tauE$.  These equal the steady-state residence
  time once injection and loss balance, which in some runs the ash label has not yet reached.  Circles and squares denote the \texttt{S-Q*} fractional-sink
  and \texttt{T-Q*} thermostat scans as in panel (\textit{a}).  Small and large diamonds are the fixed-fuelling
  controls \texttt{S-Q1-F0.25} and \texttt{S-Q1-F4} in which the mass controller is turned off; small and large red crosses are the weak and
  strong core-fuelled particle-extraction controls \texttt{S-Q1-CF-PSw} and
  \texttt{S-Q1-CF-PSs}, which use an edge particle sink similar to the heat sink (only $\tau_\rho$ is shown).   We see that $\tau_{\rm ash}$ closely
  follows $\tau_\rho$, as expected.
  (\textit{c}) Ash-label residence time against the strength of an
  edge sink that is similar to the energy sink but acts on the ash,  for the seven runs \texttt{S-Q1-Anone} to \texttt{S-Q1-Asuperstrong} (see  \cref{tab:usedruns}); the dashed line is
  $\tau_{\rm ash}=\tauE$ and the dotted line the geometrical estimate
  $\tau_{\rm ash}/\tauE\simeq1/2$ of \cref{eq:kesnerratio}.  The inset shows the partition of the injected label between
  hard-core loss (blue) and sink extraction (grey), and checks
  loss/source closure; the outer boundary removes less than $0.1$ per cent of
  the label in every run.}
  \label{fig:confinement}
\end{figure}

The fractional energy sink has an unusual (and unphysical) analytical
advantage for building a model of the global structure.  Because its local
removal rate is linear in $p$, the heating rate it can balance follows directly
from the pressure in the sink layer, and the marginal profile then predicts
the stored energy $W$ with no free constants.  We take the pressure maximum to be at
$\rpk=1.25\rin$ (see \cref{fig:peak}), continuing the stable interior inward at constant
$p_{\rm pk}$, then
outside $\rpk$ follow the marginal family of \cref{eq:family}. This leaves
$\beta_{\rm pk}$ as the only  parameter determining the profile.  The profile is not marginal inside the
sink layer itself, so we evaluate the family at its inner edge,
$r_s=r_{\rm out}-w_s$, and take that pressure $p_s$ to fill the layer (the model is relatively insensitive to this choice).
Neglecting boundary losses, steady state then requires
$Q\simeq p_s\int\Gamma_E\,{\rm d}V/(\gamma-1)$ for the linear sink of
\cref{eq:fractionalsink}.  Varying $\beta_{\rm pk}$ traces the dashed curve in
\cref{fig:confinement}(\textit{a}), which is drawn without free parameters other than 
the choice of $\rpk$.  The curve is sublinear (cf.\ thin dotted line) due to the finite-$\beta$ flattening
of the profile, whereby  the edge pressure rises
faster than the peak pressure, making the fractional sink progressively more
effective.

The measured branch is more sublinear still, with $W\sim Q^{0.6}$, while the 
model slope never falls below $W\propto Q^{0.75}$. This is because the closure's  main
omissions both change with drive $Q$.  Most importantly, the boundaries, mostly the inner core,
remove approximately three-quarters of the heat at $Q_{\rm ref}$ but only two-fifths
at $200\,Q_{\rm ref}$, and such losses are neglected in the model. Other
effects include the inner boundary condition setting a floor on $W$ that affects very low drive cases, 
and modest departures of profiles from
marginality (see \cref{fig:marginal}).  The agreement to within a factor of three
over four decades in $Q$ is  partly a cancellation between these
effects.

The edge-thermostat branch, shown with squares, is flatter again, and we do not have a predictive
model for it.  Pinning the edge temperature constrains the profile's rescaling,
but extra power creates a steep boundary layer only several cells wide (see \cref{fig:marginal}).
The comparison demonstrates how the exhaust and boundary
prescriptions control the stored energy in the device,  therefore highlighting  the crucial  importance 
of realistic edge physics for making confinement predictions about levitated dipoles.

\subsection{Particle and passive-label confinement times}\label{sec:confinement}

Whether a fusion device confines energy longer than particles is important for
its fuel cycle, since a short particle confinement  can allow
it to rapidly expel fusion ash while maintaining a hot core.  \citet{Kesner2010} argued
that the concept is uniquely well suited to such a regime, with dipole geometry allowing energy to particle confinement ratios of up to ${\simeq}50$. 
The argument is simple: if interchange mixing holds the profiles at marginality and
both energy and particles leave through the edge, each unit of mass lost
carries away only the thermal energy of the edge, whereas the confined plasma
holds the much larger energy of the hot interior.  Written for a general
profile, the ratio of confinement times is then geometrical,
\begin{equation}
 \frac{\tauE}{\tauP}\simeq\frac{\langle T\rangle_M}{\gamma T_s},\qquad
 \langle T\rangle_M\equiv\frac{\int\bar p\,{\rm d}V}{\int\bar\rho\,{\rm d}V},
 \label{eq:kesnerratio}
\end{equation}
where $\tauE=W/Q$ and $\tauP=M/\dot M_{\rm loss}$ are the energy and particle
confinement times defined at the start of \cref{sec:operating},
$T_s$ is the value of $\bar p/\bar\rho$ where the plasma leaves (the sink in our simulations), and the
factor $\gamma$ accounts for the enthalpy carried by a convective outflow.
In a dipole, the rapid growth of $\mathcal V$ makes the interior far hotter
than the edge such that  this ratio can be large.  The pinch, with $\mathcal V\propto r^2$
at low $\beta$, is much less favourable.  Evaluating \eqref{eq:kesnerratio} on
the marginal profiles of \cref{sec:state}, from $\rpk$ to the inner edge of
the sink layer, gives $\tauE/\tauP\simeq2$ at the reference drive, falling to
$1.5$ for the flatter profiles at $\beta_{\rm pk}\simeq1$.

\Cref{fig:confinement}(\textit{b}) shows the native result of the
simulations, which exhibits the opposite ordering.  The bulk-particle residence
time $\tauP$ (red) is always approximately $5\tauE$ under both
exhaust prescriptions, although $\tauE$ itself varies by more than two orders
of magnitude across the scans. The residence time $\tau_{\rm ash}$ of the ash label, which is released in the
hot interior (blue), follows that of the bulk particles closely.  Replacing
the mass controller by fixed fuelling rates (diamonds) leaves this unchanged,
except in \texttt{S-Q1-F0.25} (small markers), which has a much lower mass fuelling rate and reaches the regime where the inner reservoir supplies mass instead of
absorbing it,  thereby increasing $\tau_{\rho}$.  These results show that the ratio $\tau_{\rho}/\tauE$ is set by where the plasma leaves the system, which 
is a function of the boundary conditions in our simulations. Nearly all of
the particles are absorbed at the coil, whereas a reasonable fraction ($\gtrsim 25\%$) of the
energy leaves through the edge sink, so the premise of
\cref{eq:kesnerratio} fails.  The red crosses test this dependence directly by adding an
edge particle sink, similar to the energy sink, to the core-fuelled run
\texttt{S-Q1-CF}; for the stronger sink (\texttt{S-Q1-CF-PSs}) this diverts  $58$ per cent of the throughput outwards, away from the coil,
 causing $\tauP/\tauE$ to fall to $2.5$ while $\tauE$ changes only modestly.  Overall, we thus see that the native $\tau_{\rho}/\tauE$
ratio  measures the boundary prescription in most simulations, and we do not regard it as
a useful prediction for a dipole.

Panel (\textit{c}) extends this sink  treatment of particles  by giving the ash label
an edge sink with the same envelope as the thermal one and scanning its rate.
This is a crude model of the selective extraction proposed for dipole reactors 
\citep{Kesner1997,Kesner2004} while also avoiding the severe
timestep constraint of a  strong density sink, which evacuates the
outer layer, raising the Alfv\'en speed (panel \emph{b} suggests a density sink will behave similarly).  As the sink
strengthens, the label stops reaching the coil (inset), and
$\tau_{\rm ash}/\tauE$ falls from ${\simeq}5$ to  ${\simeq}1$ at equal label and
thermal sink rates, before levelling off near ${\simeq}2/3$ at stronger sinks.  In this limit, where the label leaves through the edge as heat
does, $\tauE/\tau_{\rm ash}\simeq1.5$, comparable to the estimate of
\cref{eq:kesnerratio} (dotted line).  The comparison is not
exact, as the bulk particles and much of the heat still reach the coil in these
runs, with the floor also partly reflecting how quickly the turbulence can carry plasma
from the interior to the edge.

\section{The pressure maximum and hard-core load}\label{sec:peak}

\subsection{A consistently selected inward maximum}\label{sec:peakfacts}

The radius at which the pressure maximum settles is of crucial importance for dipole confinement.  It sets the good-curvature and fusion-weighted volumes,
the maximum pressure and temperature at a given coil current, and the distance from the hottest
plasma to the coil.  Yet, as shown above, moving either source does not move the
maximum substantially.   While physics beyond our model may do so --- for example, kinetic
instabilities can grow in the good-curvature region where interchange modes
cannot \citep{ivanov25} ---  in these simulations the turbulence and boundary conditions alone
select the final steady-state profile.

We interpret the maximum as the inward limit of turbulent transport.  By successive overshoot of turbulent eddies, the
turbulence carries the profile towards the hard core until it encounters the
cold boundary imposed at the wall.  The time-averaged maximum then sits in an
overshoot layer, close enough to the wall for eddies to reach it but far enough
away that they do not erase the profile behind it.  Its radius is therefore
 set by the radial extent of the eddies, rather than by either source.

\begin{figure}
  \centerline{\includegraphics[width=\textwidth]{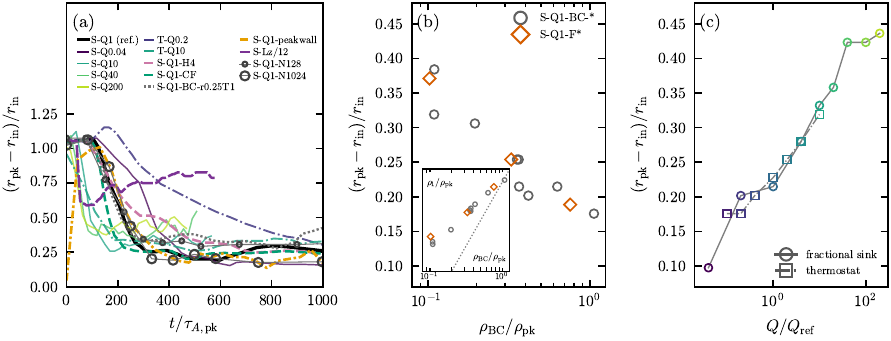}}
  \caption{The turbulence selects the radius of the maximum.
  (\textit{a}) Histories of the normalised standoff $(\rpk-\rin)/\rin$ of the pressure maximum, with every curve labelled by
  its name in \cref{tab:usedruns}, showing how most runs settle into a similar close-to-the-wall peak 
  location despite quite different physical and numerical parameters.  (\textit{b}) Saturated standoff against the prescribed
  inner-reservoir density normalised by the saturated peak density, for the
  inner-boundary scan (circles) and the fixed-rate-fuelling runs (diamonds); see \cref{tab:usedruns}.
  The inset compares the density of the first active cell $\rho_{1}$ with the prescribed
  boundary density $\rho_{\rm BC}$, illustrating how a thin boundary layer develops;
  the dotted line is equality.  (\textit{c}) Residual  dependence of the standoff on $Q$ at fixed
  boundary state, comparing the two exhaust prescriptions and illustrating a weak increase with $Q$.
  The two prescriptions coincide at equal $Q$ even though their peak-to-boundary
  pressure ratios differ by up to a factor of five, showing how the standoff is weakly affected by the
  heat throughput to the coil.}
  \label{fig:peak}
\end{figure}

\Cref{fig:peak}(\textit{a}) shows the time history of the maximum, making clear
that the transients differ much more than the final positions.  The reference
state settles by ${\simeq}300\,\tau_{A,{\rm pk}}$, whereas \texttt{S-Q1-H4}, which 
heats at $r=4\rin$,
takes $2$--$3$ times longer (this is shown in \cref{fig:sources}\emph{a} also).  \texttt{S-Q1-peakwall} starts with its maximum on the
hard core, moves outwards to near the heating location $\rpk\approx2\rin$, and then returns to
the same band as the other runs, demonstrating the independence from initial conditions.  Likewise, the resolution ladder 
from $N_{r}=128$ to $1024$ (excluding the coarsest member, $N_r=64$) exhibits only modest dependence
($\rpk=1.26$--$1.19\rin$), showing that this is not a numerical effect. Overall, the ranges between 
different simulations are
narrower than the maximum itself, unless the system is artificially restricted such as for \texttt{S-Lz/12} (see \cref{app:zonal}).

Panels (\textit{b}) and (\textit{c}) isolate  two weak dependencies that
remain.  The standoff $\rpk-\rin$ is always a small fraction of $\rin$, and no
control changes it significantly.  Panel
(\textit{b}) shows the effect of the inner reservoir, which indirectly (combined with the fuelling rate) sets the contrast
between the boundary and the peak plasma.  As the prescribed density
approaches the peak density, the maximum moves inward, and we see that either 
changing the boundary conditions themselves (scan \texttt{S-Q1-BC-*}) or
changing the fixed-fuelling rate (scan \texttt{S-Q1-F*}) follows the same trend against the reservoir-to-peak contrast
$\rho_{\rm BC}/\rho_{\rm pk}$.  The inset shows that the first active cell is
systematically denser than the prescribed reservoir, so a thin boundary layer
prevents the plasma from simply matching $\rho_{\rm BC}$.  Panel (\textit{c})
shows the remaining drive dependence.  A factor of $5\times10^{3}$ in $Q$
increases the normalised standoff $(\rpk-\rin)/\rin$ by only a factor of about four (from $0.10$ to $0.44$), and both
exhaust prescriptions follow the same curve.  The trend is not simply a result of the
pressure contrast with the boundary; for example, at fixed $Q$ the thermostat runs hold a
peak-to-reservoir pressure ratio two to five times larger than the
fractional-sink runs, yet have the same standoff.  The
standoff instead appears to change as a result of the heat throughput that the overshoot layer
must carry to the wall.

\subsection{The maximum as the inward reach of the turbulent front}
\label{sec:peakexcursion}

\begin{figure}
  \centerline{\includegraphics[width=\textwidth]{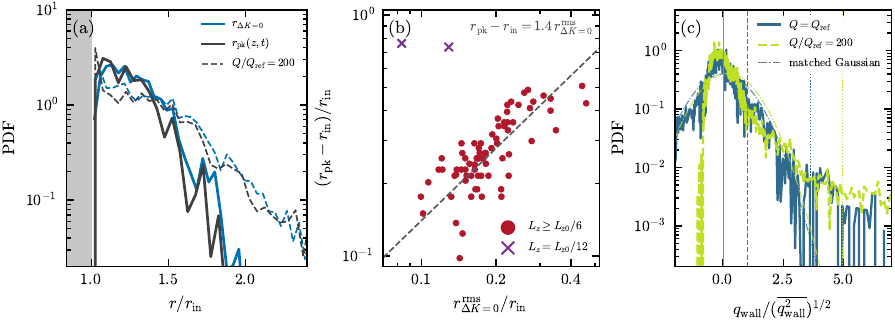}}
  \caption{(\textit{a}) Probability densities of the instantaneous, $z$-resolved
  marginality front $r_{\Delta K=0}$ (blue) and local pressure maximum $\rpk(z,t)$ (black)
  at the reference drive (solid; \texttt{S-Q1}) and strongest drive (dashed; \texttt{S-Q200}); the shaded strip
  is the coil, against which both distributions are peaked.
  (\textit{b}) Mean peak-to-wall distance $\rpk-\rin$ against the root-mean-square deviation
  $r_{\Delta K=0}^{\rm rms}$ of $r_{\Delta K=0}$, over nearly all of the simulations.  Only the two
  shortest boxes, with $\Lz=L_{z0}/12$ (\texttt{S-Lz/12} and its core-fuelled partner),
  depart from the trend and are drawn as crosses ($\Lz=L_{z0}/6$ and
  $L_{z0}/3$ behave like the long boxes for this diagnostic and are grouped
  with them).  The dashed line shows
  $\rpk-\rin=1.4\,r_{\Delta K=0}^{\rm rms}$ as a guide.
  (\textit{c}) PDFs of the signed inner-boundary energy-load proxy
  $q_{\rm wall}=-F_{\mathcal E,r}(r_2,z,t)$ \eqref{eq:qwall}, normalised by its rms; positive
  values point into the hard core.  The common dashed line marks one rms and
  dotted vertical lines mark the 99th percentiles.  Dot-dashed curves are
  Gaussian distributions with the same mean and variance as each measured
  PDF.}
  \label{fig:front}
\end{figure}

The axial average, in which the peak is visible, hides the instantaneous structure visible in
\cref{fig:state}: at a given axial position, the pressure maximum does not sit
at a fixed radius but is dominated by strong turbulent fluctuations across a wide range in radius.  
To understand this, at each $(z,t)$ we first locate the radius $\rpk(z,t)$ of
the largest pressure, and then search outward for the first zero of
$\Delta K=-K$, evaluated with the instantaneous $p$ and $\beta$ at that
height.  We denote this radius by $r_{\Delta K=0}(z,t)$ and use it to define
the marginality front.  The search must start from the maximum because the
front cannot lie inside it: here $\partial p/\partial r\geq0$,
\cref{eq:kdef} gives $K>0$, so a
supercritical region can exist only on the falling side of the maximum.\footnote{Beginning the search for $K=0$ at the hard core instead picks up  noisy secondary maxima, and produces a misleadingly
narrow distribution.}
\Cref{fig:front}(\textit{a}) collects the distributions of both $\rpk$ and $r_{\Delta K=0}(z,t)$ over $z$ and the saturated
interval.  Their broad distributions almost completely overlap and both distributions peak against
the coil: at a given height and time, the wall is the most likely position of
the local pressure maximum.  The separated maximum in the axially averaged state
 is therefore a property of the mean profile. 

\Cref{fig:front}(\textit{b}) tests this interpretation by comparing the
peak-to-wall distance with $r_{\Delta K=0}^{\rm rms}$ for the relaxed controls
listed in the caption.  The two are proportional within the scatter, with the
mean maximum lying one to two front widths from the wall.  Thus, the maximum
sits about one excursion width outside the coil, the closest position the
wandering front can repeatedly reach without eroding the profile further.  Further comparisons 
with the  turbulence correlation length in the inner region (not shown) show that the standoff 
is a turbulent correlation length, which is also consistent with the
idea of profile erosion into the wall. The modest dependence on reservoir density and  heat throughput results
from associated changes to the overshooting eddies. 
The shortest boxes are the key exceptions, because
the system there drives axially averaged zonal flows (\cref{app:zonal}, \cref{fig:zonal}).  These 
flows inhibit the inward overshoot by breaking up eddies, and stop the 
approach of the front to the wall. 

The distributions in panel (\textit{a}) show that fronts and eddies repeatedly
enter the stable inner layer, making the energy load on the coil intermittent.
From the cell-centred primitive variables, we reconstruct the radial component
of the ideal-MHD total-energy flux and define
\begin{equation}
 q_{\rm wall}(z,t)\equiv-F_{\mathcal{E},r}(r_2,z,t),\qquad
 F_{\mathcal{E},r}=\left(\mathcal{E}+p+\frac{B^2}{8\pi}\right)u_r
 -\frac{B_r(\bm B\bcdot\bm u)}{4\pi},
 \label{eq:qwall}
\end{equation}
where $r_2$ is the centre of the second active radial cell and positive
$q_{\rm wall}$ points into the hard core.  \Cref{fig:front}(\textit{c}) shows
the PDF over $z$ and the saturated interval, normalised by
$(\overline{q_{\rm wall}^2})^{1/2}$.  The distribution is modestly
skewed towards positive loads, with frequent local reversals and a $99$th
percentile four to five times the rms.  The corresponding Gaussian (dot-dashed) has a much
shorter positive tail.  Because $q_{\rm wall}$ is reconstructed at $r_2$ rather
than measured at the surface, and the model contains no sheath physics, it is
an inner-boundary energy-load proxy rather than a material heat flux.

\subsection{Consequences for a dipole}\label{sec:coilconsequence}%
\label{sec:coil}

These results have two direct consequences for a levitated dipole.  On the positive side, broad
heating and edge fuelling  still produce a hot, dense, centrally peaked
state, as required by the concept, allowing considerable flexibility in source placement.  
More harmfully, the sources provide almost no
control over the pressure-maximum radius, which itself determines the coil heat load,
shielding requirement and fusion-volume weighting.  In contrast, the power-plant study of
 \citet{Simpson2026} assumes a fixed and controlled peak location, constraining it to lie at
least two $3.5$-MeV $\alpha$-particle gyro-orbits from the first closed flux
surface. A design must also allow
for the intermittent heat load seen in \cref{fig:front}(\textit{c}).  

The
fractions and peak dynamics described above come from an ideal bath with no recycling, no sheath, and infinite
heat capacity, rather than a quantitative prediction for a material surface.  Losses
are therefore limited by transport rather than by the surface, and calculations
with realistic sheath and recycling physics are needed to determine the load
received by the coil.  
Likewise, a transport barrier on
the coil surface or an imposed sheared flow could suppress the local excursion
and help shield the coil, a possibility that is supported by our short-box zonally dominated runs. Active regulation of interchange
fluctuations would be another possibility \citep{Roberts2015}.

\section{Three-dimensional limit of the selected state}\label{sec:3d}

The preceding results were obtained in the flute sector with $\partial/\partial\theta=0$ for all fluctuations.  We now ask whether
the selected state persists in three dimensions.  As discussed in
\cref{sec:equations}, $\partial/\partial\theta=0$ is an exact symmetry of the hard-core
pinch, so the two-dimensional calculations solve the $m=0$ problem exactly
while retaining the compressional and magnetic responses omitted from reduced
dipole models.  But, this same symmetry removes all $m\geq1$ modes, which include the
 field-line-bending instability of a finite-$\beta$ pinch, the
counterpart of the ballooning instability that sets the $\beta$ limit of a
levitated dipole \citep{garnier99,simakov00}.  We show here that a sufficiently strong $m=1$
instability can erode the self-organised core, limiting the
range of $\beta$ over which the two-dimensional profiles persist.    

\subsection{Bending stability of the marginal family}\label{sec:3dtheory}

Consider a perturbation $\propto\exp({\rm i}m\theta+{\rm i}k_zz)$ of a relaxed
state.  We measure its drive via
\begin{equation}
 \Dr\equiv-\beta\frac{{\rm d}\ln \bar p}{{\rm d}\ln r},
 \qquad\hbox{or}\qquad
 \frac{{\rm d}\ln \bar B_\theta^2}{{\rm d}\ln r}=-2+\Dr, 
 \label{eq:bendingdrive}
\end{equation}
where the second form follows  from \cref{eq:blogslope} and shows how $\Dr$ measures the
departure from a vacuum field $B_{\theta}\propto r^{-1}$.  In the low-frequency reduction used in \cref{sec:kadomtsev}, a mode
with radial scale $k_r^{-1}$ is unstable when
\begin{equation}
 \Dr>m^{2}\left(1+\frac{k_r^{2}}{k_z^{2}}\right),
 \label{eq:bendingcriterion}
\end{equation}
which is derived in \cref{app:numerics}. Here $k_r$ is a local radial wavenumber of the perturbation within the WKB approximation.  For $k_r\ll k_z$, the $m=1$ condition
becomes $\Dr>1$, 
equivalently, when $B_\theta$ falls more slowly than $r^{-1/2}$ (the reference slope
shown in the inset of \cref{fig:marginal}\textit{b}).

Along the marginal family of \cref{eq:family},
$\Dr=4\gamma\beta/(2+\gamma\beta)$ increases monotonically with $\beta$ and
approaches 4.  The family therefore crosses a fixed sequence of bending
thresholds as the drive increases.  Local $m=1$ onset occurs at $\Dr=1$, or
$\beta=2/(3\gamma)=0.4$ \citep{pastukhov01}.  At $\Dr=2$, or $\beta=2/\gamma$, the slope of
$B_\theta$ becomes positive, creating an
interior minimum, as  seen in the
highest-drive profile in the inset of \cref{fig:marginal}(\textit{b}).  Modes
with $m\geq2$ require $\Dr>4$ and remain stable for all $\beta$ on the interchange-marginal family.  

The turbulence may also restrict which bending modes can grow, which (we will see below) appears
to slow the instability onset over intermediate times.  An $m=1$ mode driven by a
supercritical layer of width $\Delta r$ must fit inside it, so
$k_r\gtrsim\pi/\Delta r$, and \cref{eq:bendingcriterion} then requires
$k_z\gtrsim\pi/(\Delta r\sqrt{\Dr-1})$, with $\Dr$ loosely averaged over the
layer.  A quiescent equilibrium can always meet this with a short enough axial
wavelength, but in the turbulent state, structure much finer in $z$ than the
eddies will be mixed away before it can grow.  Since the axial spectrum in
\cref{fig:globality}(\textit{b}) peaks at $2\pi k_z\rpk\simeq10$, effective
growth should then require $(\Delta r/\rpk)\sqrt{\Dr-1}\gtrsim2\pi^{2}/10\approx2$.

A final aspect worth mentioning is that field-line bending can generate  departures from the flute state even when the system is stable. 
Eddies must maintain their flute structure by communicating along the field
via magnetic tension.  Comparing Alfv\'enic communication at rate
$\omega_{\rm A}=v_{\rm A}/r$ with turbulent decorrelation at
$\omega_{\rm nl}\sim u_{\rm turb}/\ell_r$, where $u_{\rm turb}\equiv(\overline{\delta u^2})^{1/2}=\Ma\,c_s$ includes all velocity components, in the spirit of critical balance
\citep{goldreich95}, suggests that the relative amplitude of the $m=1$
departure from axisymmetry is
\begin{equation}
 \frac{\delta_1}{\delta_0}\sim\frac{\omega_{\rm nl}}{\omega_{\rm A}}
 \sim\frac{r}{\ell_r}\frac{u_{\rm turb}}{v_{\rm A}}
 =\frac{r}{\ell_r}\Ma\left(\frac{\gamma\beta}{2}\right)^{1/2},
 \label{eq:cbresponse}
\end{equation}
where $\delta_m$ is a normalised  perturbation amplitude of  azimuthal harmonic $m$ 
(see \cref{sec:3dresults}).  A weak field-line-dependent response can
therefore coexist with a subcritical mean profile, becoming of order unity
when the two rates are comparable.  

\subsection{Extruded-state test}\label{sec:3dtest}

We test these predictions by extruding six relaxed two-dimensional states in
$\theta$ and continuing them with the same sources, sinks and controller, but in 3-D.  The
parents have $\beta_{\rm pk}=0.13$, $0.33$, $0.51$, $0.89$, $1.37$ and $2.38$, at
$Q/Q_{\rm ref}=1$, $4$, $10$, $20$, $40$ and $100$, respectively.  This sequence
crosses the thresholds of \cref{sec:3dtheory} while remaining at drives of
fusion interest.  We impose no non-axisymmetric perturbation: numerical noise supplies the
seed, which can be amplified by the inherited turbulence.  We  survey cases at resolution
$N_r\times N_z\times N_\theta=192\times720\times192$ and repeat the two cases
that bracket the change in behaviour, $\beta_{\rm pk}=0.33$ and $0.89$, at
$384\times1440\times384$.  Depending on drive, the survey continuations run for
$168$--$564\tau_{A,{\rm pk}}$; see \cref{tab:usedruns}.

\begin{figure}
  \centerline{\includegraphics[width=\textwidth]{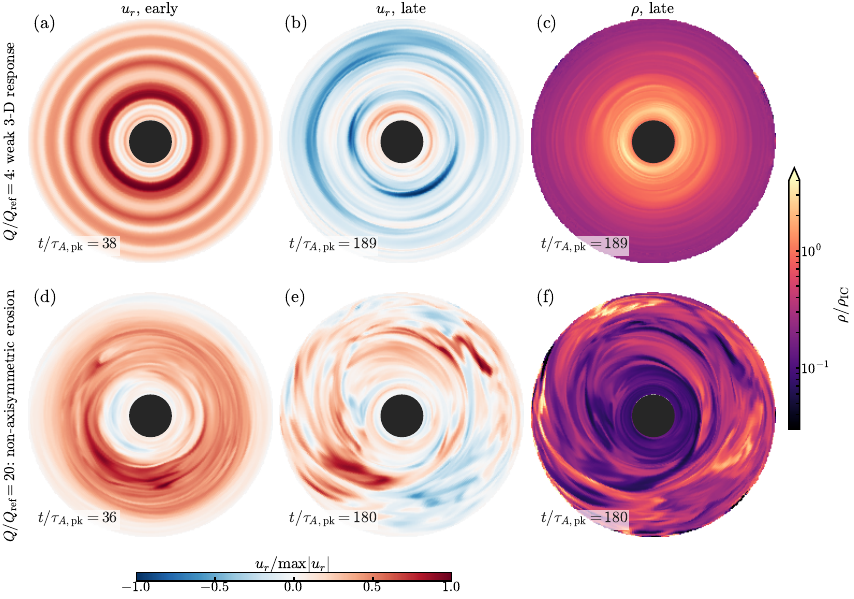}}
  \caption{Annular sections of two three-dimensional continuations.  Top row:
  $Q=4Q_{\rm ref}$ ($\beta_{\rm pk}=0.33$ in the parent), for which the
  non-axisymmetric response remains relatively weak through
  $t/\tau_{A,{\rm pk}}\simeq189$.  Bottom row: $Q=20Q_{\rm ref}$
  ($\beta_{\rm pk}=0.89$), for which a low-$m$ deformation grows and removes much of the
  mass from the core.  Left and centre panels show radial velocity early and late in
  each run; right panels show density at the later time.  Radial velocity is
  normalised to the maximum $|u_r|$ of each snapshot, and density is shown as
  $\rho/\rho_{\rm IC}$ on a common scale, so the emptied core in
  (\textit{f}) can be compared directly with (\textit{c}).  Times use the
  saturated peak-Alfv\'en clock of the corresponding 2-D parent.}
  \label{fig:render}
\end{figure}

\Cref{fig:render} compares annular sections of the two high-resolution runs
early and late in their evolution.  The weaker case retains its concentric
rings and peaked density core, with only a modest late-time non-axisymmetric distortion.  The stronger case develops a large
low-$m$ deformation that removes significant mass from its core.

\subsection{Weak-drive persistence and high-drive erosion}\label{sec:3dresults}

\begin{figure}
  \centerline{\includegraphics[width=\textwidth]{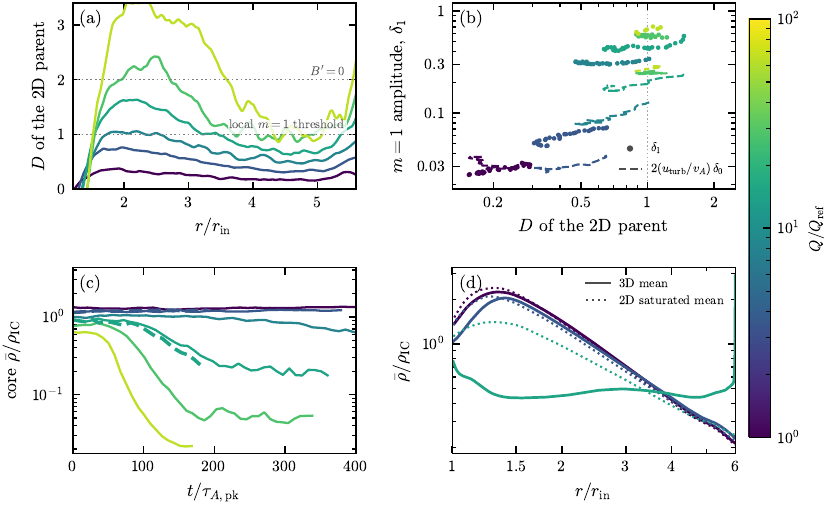}}
  \caption{(\textit{a}) The bending drive $\Dr$ \eqref{eq:bendingdrive} of
  each 2-D parent, with the
  local $m=1$ threshold and zero-field-gradient limits shown as dashed lines; colour is the heating rate, as in
  \cref{fig:marginal}.  In increasing heating-rate order the parents' $\beta_{\rm pk}$
  are $0.13$, $0.33$, $0.51$, $0.89$, $1.37$ and $2.38$.
  (\textit{b}) Fractional $m=1$ density amplitude $\delta_1$ (points), compared
  with the critical-balance estimate $2(u_{\rm turb}/v_{\rm A})\delta_0$ (dashed),
  against the parent drive at the same radius.  Each colour uses the final
  quarter of its survey continuation and $2.5<r/\rin<5$.  The vertical line
  marks $\Dr=1$, and we omit radii in which $>1\% $ of the cells are affected by the density floor.  (\textit{c}) Core density, normalised to $\rho_{\rm IC}$,
  against time.
  Solid curves show the standard resolution cases while dashed curves show the high-resolution pair at
  $Q/Q_{\rm ref}=4$ and $20$.
  (\textit{d}) Time-averaged density profiles from the continuations (solid), compared with saturated-state time averages
  from 2-D runs at matching radial and axial resolution (dotted;
  \texttt{S-Q1-N192} at $Q/Q_{\rm ref}=1$). The $Q/Q_{\rm ref}=4$ and $20$ 3-D cases are averaged over their second half; the $Q/Q_{\rm ref}=1$ case is averaged over $100<t/\tau_{{\rm A,pk}}<400$.}
  \label{fig:3d}
\end{figure}

\Cref{fig:3d} shows the three-dimensional evacuation and collapse of
confinement more quantitatively.

Panel~(\textit{a}) shows the drive $\Dr$ for the six 2-D parents.  Each
profile rises away from the hard core, reaches a maximum in the inner bulk, and
then decays outward, with a peak $\Dr$ that increases from about $0.4$ (stable) to $3.6$ (very unstable) across the scan.
The two weakest parents are subcritical everywhere.  The $Q/Q_{\rm ref}=10$
parent is only marginally supercritical, with $\Dr-1\lesssim0.06$ across a
layer $0.6\rin$ wide, so $(\Delta r/\rpk)\sqrt{\Dr-1}\simeq0.1$, suggesting 
its unstable modes will be unable to grow due to mixing from the interchange turbulence (see \cref{sec:3dtheory}).  At $\beta_{\rm pk}=0.89$, $\Dr-1\simeq0.6$ across a wider layer, so the
same combination rises to ${\approx}1$; it reaches $1.8$ and $3.2$ in the two
strongest parents, which also cross $\Dr=2$.  The transition between the
$Q/Q_{\rm ref}=10$ and $20$ parents therefore reflects an increase in both the
strength and the width of the supercritical layer.

\Cref{fig:3d}(\textit{b}) shows how the non-axisymmetric response increases
across this sequence.  We take $\delta_m(r)$ to be the time-averaged rms
amplitude of azimuthal harmonic $m$ of $\rho/\bar\rho(r,t)-1$, including all axial
wavenumbers.  The points show $\delta_1$ at radii in the bulk; each dashed
curve gives $2(u_{\rm turb}/v_{\rm A})\delta_0$ from the same three-dimensional
state, using the shell turbulent kinetic energy for $u_{\rm turb}$ and the
mean azimuthal field for $v_{\rm A}$.  This is \cref{eq:cbresponse} with
$\ell_r\sim r/2$, using the same coefficient for every run.  The weakest run ($Q/Q_{\rm ref}=1$)
follows its guide across the bulk, with $\delta_1/\delta_0\simeq0.26$ showing how 
 the Alfv\'enic tension maintains the turbulence in a flute-like state.
 At $Q/Q_{\rm ref}=4$, still subcritical
but closer to the threshold, the response is above the estimate, and the unstable runs
whose parents exceed $\Dr=1$ lie somewhat further above it, as expected once
the $m=1$ mode is linearly driven. 

Panel (\textit{c}) shows the volume-weighted mean of $\bar\rho$ inside
$3\rin$, the region containing the maximum.  At $Q/Q_{\rm ref}=1$ and $Q/Q_{\rm ref}=4$, the 
system maintains its peaked core over long times, showing 
the applicability of the 2-D simulations to the more realistic 3-D setting.
In contrast, at $\beta_{\rm pk}=0.89$ ($Q/Q_{\rm ref}=20$) and above, a low-$m$, low-$k_z$ deformation grows and
the core empties.  The core density falls by a factor of 5 at this drive and by
even larger factors at higher $Q$.  The effect is driven by outward mean flow
that grows during the continuation and
eventually exceeds the turbulent particle flux, which, interestingly, remains inward with 
approximately the magnitude of the pinch in the 2-D runs (see
\cref{sec:sourceroutes}).  
The intermediate case ($Q/Q_{\rm ref}=10$) behaves as expected for a barely supercritical layer, with a much slower onset;
its local growth rate is ${\simeq}3$ times smaller than  the $Q/Q_{\rm ref}=20$ case, and its thin supercritical layer admits
only short axial wavelengths that are easily mixed by the 2-D turbulence.   The strongest runs become affected by the controller
and a numerical density floor after the core has emptied, so the final states 
of these cases should not be analysed in detail.

Panel~(\textit{d}) compares late-time mean profiles from the 3-D continuations
with saturated-state averages, at matching two-dimensional resolution (using \texttt{S-Q1-N192} for $Q/Q_{\rm ref}=1$).  The disrupted case has lost its
maximum and instead increases outward through the bulk, a dramatically different profile.  The two weaker cases
keep the two-dimensional shape through the bulk, but their larger
non-axisymmetric fluctuations move the maximum modestly away from the wall,
by an amount that shrinks for decreasing $D$.  At $Q/Q_{\rm ref}=4$, where
$\delta_1/\delta_0\simeq0.8$, the pressure maximum moves from $1.28$ to
$1.37\rin$ along with a decrease in wall density.  At
$Q/Q_{\rm ref}=1$, where $\delta_1/\delta_0\simeq0.26$, it moves only from
$1.27$ to $1.33\rin$, comparable to the spread between 2-D simulations in the saturated state (see \cref{sec:peakfacts}), and the bulk density agrees to ${\simeq}3.5\%$
across the domain.  Other diagnostics, such as the inward turbulent mass flux, also match the control
to within $10\%$ (not shown).
Below the bending threshold, the 3-D results therefore support
the persistence of the peaked flute state, with mean profiles that approach
the 2-D ones at decreasing $\Dr$.

While this 3-D transition occurs for 
$\beta_{\rm pk}\gtrsim 0.5$  in the
3-D pinch, the corresponding dipole limit is likely to be less
restrictive:  variation along the field line changes the balance between
curvature drive and bending, stabilising 3-D modes \citep{garnier99,simakov00,Kesner2001}.  In \cref{app:dipoleballooning} we provide
an illustrative calculation for the dipole, showing how the critical
$\beta$ rises rapidly
above the uniform-line value of $0.4$ even for field lines close to the coil
(\cref{fig:dipole}\textit{a}), reaching  ${\simeq}2.1$ even directly adjacent 
to the LDX coil  (this value is indicative only because
the equilibrium geometry is not self-consistent at this $\beta$).  More realistic calculations
seem to  give yet higher limits: the LDX equilibrium of \citet{garnier99} remained
ballooning stable at $\beta_{\max}\simeq10$ (see also \citealt{Testa2026}), while the finite-pressure
point-dipole equilibrium of \citet{Krasheninnikov1999} was shown to be stable
for arbitrary $\beta$ \citep{simakov00}.  Further study is needed to understand
if other non-flute effects become important for confinement before reaching these stability limits. 

\section{Discussion and conclusions}
\label{sec:conclusions}

In this paper, we use global ideal-MHD simulations of a driven hard-core
Z-pinch as an analogue of a levitated dipole to study interchange
self-organisation on closed magnetic field lines.  Building on previous studies \citep{Makhin2005,pastukhov05,ou20}, we determine the state selected under
sustained heating, fuelling and exhaust in full MHD, across a broad range of $\beta$.
The pressure, density and magnetic field relax onto a one-parameter family
obtained by combining the Kadomtsev marginality criterion
with radial force balance \citep{kadomtsev66,pastukhov05}.  This state, which 
is the ``turbulent pinch'' seen experimentally  \citep{boxer10,Saitoh2026}, 
strongly concentrates density and temperature in the core of the device. 
Measuring $\beta$ at
one radius then fixes the bulk profile shapes, with their amplitudes anchored
separately and no fitted shape parameters.
The prediction follows both pressure and density through the source-free bulk
of every run in the drive scan, from $\beta_{\rm pk}\simeq0.02$ to $4$
(\cref{fig:marginal}).  As $\beta$ increases, the profiles flatten and
$B_\theta$ develops an interior minimum \citep{Guazzotto2007}, organising the plasma and its confining field as a single system.

The selected bulk profiles are only weakly sensitive to how the state is forced.  Moving the heat source
from $2\rin$ to $4\rin$, moving particle replacement between $2\rin$ and
$5.2\rin$, changing the exhaust prescription, or refining the grid leaves the
normalised profiles and pressure-maximum radius nearly unchanged
(\cref{fig:sources}).  These changes do, however, affect the amplitude of the state by changing the transport efficiency;  
likewise, the timescale 
required to reach the common profile, which is approximately the energy-confinement time, 
depends on the source and sink physics.  Entropy and particles
also take different routes during this relaxation in how they 
adjust to the system's forcing.  The entropy, whose 
flux is proportional to the marginality $K$
(\cref{eq:kentropygradient}) within a simple diffusive model, is only transported outwards through  supercritical
regions, and its inward migration is driven by successive overshoot of eddies 
into an inner stable region that continually shrinks with time. Particle content
has no corresponding constraint, and its flux follows the imposed sources and
sinks with either inwards or outwards transport (\cref{fig:fluxes}).  Passive labels likewise follow
sources and sinks without constraint and always settle into a final profile that follows the bulk density 
(\cref{fig:sources}\textit{d}).
This profile insensitivity even extends to axially nonuniform sources:
with a source that varies from zero to twice its mean along $z$,
turbulence drives  similar peaked radial profiles with only modest axial imprint of the source 
variation and some degradation in confinement  (\cref{app:zmod}).
It does not, however, extend to unphysically short axial domains, where zonal flows 
develop and modify the transport, improving confinement and modifying the final state (\cref{app:zonal}).

A consequence of this self-organisation is that the location of the pressure maximum, where
fusion would occur, is not easily controlled by external drivers.
Across the full set of standard simulations, the  maximum in the relaxed state 
always lies in a small range of radii very close to the coil (\cref{fig:peak}).  We argue its location is set by the mean inward reach of a marginality
front whose excursions are comparable to its distance from the wall
(\cref{fig:front}) --- effectively a turbulent boundary layer 
that connects the fixed-condition boundary to the core plasma by continually overshooting eddies.  This explains why
the heating and fuelling locations, as well as the numerical resolution, barely change the position of the maximum, although we
cannot yet predict the size of the excursion from external parameters. 

The interchange turbulence that maintains this state is radially global, with a correlation
length that grows with the annulus size and remains about one third of the domain width (\cref{fig:globality}).  
This turbulence  maintains itself via
a regulation cycle lasting several turnover times: heating induces an excess pressure gradient that   increases
the {growth} of fluctuations, then  the fluctuation amplitude and transport
follow about one turnover later, flattening the gradient to restart the cycle (\cref{fig:cycle}).  The residual of this repeated overshoot and relaxation is a small mean
supercriticality $K<0$ across much of the domain.

The conclusions above are derived from simulations in the flute sector ($\partial/\partial\theta=0$), which is an exact 
symmetry of the equations but excludes field-line bending and related instabilities.  To test such physics, 
we continue some cases into three dimensions, which gives,  to our knowledge, the first tests of the nonlinear effects of the $m=1$
instability and the accuracy of the flute assumption. At weak drive, peaked states persist robustly, with a small non-axisymmetric
response whose amplitude relative to the 2-D fluctuations is approximately
 consistent with the standard critical
balance condition of Alfv\'enic turbulence (\cref{fig:3d}; \citealp{goldreich95}).  At
stronger drive, for $\beta_{\rm pk}\gtrsim 0.5$, the 2-D marginal state develops an $m=1$ unstable layer,
which drives an outward mean flow that overwhelms the inward turbulent
pinch, evacuating the peaked density core (\cref{fig:render,fig:3d}). This 
instability is the analogue of the ballooning mode in the dipole context \citep{garnier99,simakov00} 
and must be avoided to maintain confinement. We show in \cref{app:dipole} how these pinch estimates
are highly pessimistic compared to the corresponding dipole geometry, since the latter remains stable at much higher $\beta$.

\subsection{Implications for levitated-dipole confinement}\label{sec:dipoleimplications}

A core conclusion, already well appreciated from reduced models
\citep{pastukhov01,pastukhov05,ou20,li21}, is the robustness of interchange
self-organisation.  Within the limitations of the Z-pinch geometry, we show how the
turbulent pinch survives fully self-consistent global evolution and remains
effective at high $\beta$.
The hard-core pinch is, of course, only an approximation to a dipole, but its physics is closer than its
different geometry might suggest.  Physically, it is the limit of a dipole with a very
high-aspect-ratio coil, describing the region close to the coil (although such a coil would
be difficult to engineer; \citealp{Simpson2026}).  More generally, within the near-marginal reduced  ordering of \citet{pastukhov05b}, the
local nonlinear flute equations in any axisymmetric, shearless,
closed-field-line geometry can be mapped onto those of the pinch by rescaling
angle, time and the interchange drive strength, and even models that capture global variation remain very similar (\cref{app:dipole}).  Both systems advect the same particle and entropy
tube contents, $\rho\mathcal V$ and $p\mathcal V^\gamma$, with the same form of interchange drive (albeit with a different scaling), so the pinch contains the same interchange physics at that level of
approximation. The finite-amplitude compressibility, some long-correlation-length effects,  and magnetic evolution
are specific to the geometry, but not of particular importance to our main results.

With this caveat, our results carry several direct implications for a dipole.  Interchange turbulence builds a hot,
dense, centrally peaked plasma, even with broad heat deposition and edge
fuelling, while the field evolves self-consistently with it.  The axial-source tests (\cref{app:zmod}) 
further suggest that heating and fuelling need not
be distributed uniformly around the torus, because turbulence can smooth localised
input even without an imposed toroidal flow, although quantitative details of the response
will certainly depend on the geometry and omitted kinetic effects.

This profile robustness  limits an operator's  control over where the hot plasma sits, with good and bad consequences.
Detailed control of the radial heating and fuelling locations seems unnecessary, because the turbulence builds the
same peaked state wherever they are placed. In fact, in our simulations, heating on the outside of the device yields better confinement,
with a hotter core for the same heating rate. However, the same turbulence carries the plasma inward until it meets
the coil, and in our simulations a large fraction of the injected power, and most of the particles, reach the coil,
with an intermittent load (\cref{fig:front}\textit{c}).  These loading fractions depend on our highly idealised boundaries (a cold,
absorbing inner wall and a simple outer exhaust) and should not be taken as predictions.  The ratio of particle to
energy confinement, which matters for ash removal and advanced fuel cycles \citep{Kesner2004,Kesner2010}, also illustrates 
the importance of the boundaries, changing by almost an order of magnitude when an edge sink removes the ash before it
reaches the coil (\cref{fig:confinement}).  Predictions of coil loading and confinement in a dipole will therefore
depend on realistic physics at both the coil and the edge, at least as much as on the interchange turbulence itself.

By contrast, the three-dimensional erosion that limits the pinch at high $\beta$ (\cref{sec:3d}) is probably of
little relevance to an actual dipole, because variation along the field line raises the bending threshold well
above the pinch value, even close to the coil (\cref{app:dipoleballooning}).  Our 
results do, however,  suggest that  if a dipole is driven past its own threshold, the likely outcome will 
be strong evacuation of the core plasma. Clearly, this should be avoided.  Also
relevant may be the weak-drive critical-balance result, which suggests that residual field-line-dependent
fluctuations persist well below the ballooning limit, with a relative amplitude (compared to the flute turbulence) proportional
to the strength of said turbulence via critical balance.  Such fluctuations can modestly alter transport and coil loading even
when the mean profile is stable to field-line bending, and their manifestation in the dipole geometry will be interesting for further study.

\subsection{Omitted physics and future directions}

Three omissions are likely to affect our results.  The first is kinetic physics, since ideal MHD
contains no finite-Larmor-radius effects, heat fluxes, pressure anisotropy, or hot electrons. 
One clear  consequence is that the model misses the drift-frequency entropy mode \citep{simakov01}.  Gyrokinetic calculations of
this mode in the pinch and dipole
\citep{kobayashi09,kobayashi10}, together with related kinetic treatments
\citep{Ricci2006PRL,Hallenbert2022,Hoffmann2023,Numata2025}, find inwards heat transport in some regimes, which  is
absent from ideal MHD (\cref{sec:sourceroutes}). Nonetheless, both interchange and  entropy 
modes drive the system towards the same marginal profiles, and given the interchange 
is more virulent, more study is needed to understand whether there exist relevant regimes in which entropy-mode-driven turbulence 
can be of global relevance. Other, electromagnetic instabilities in the good-curvature region could also
play a role
\citep{ivanov25}; most interestingly, such effects could enhance transport in the inner region, and thereby 
push the saturated pressure-peak location outwards, but further study is needed. 
Overall, however, we speculate that MHD results are likely to be 
more relevant to levitated-dipole design than in tokamaks or stellarators: 
tokamak transport is usually set by drift-wave turbulence in a plasma that is stable to ideal MHD modes,
whereas a driven dipole carries its transport in large, MHD-unstable
eddies, with the fluid description  capturing more of these dynamics.

The second omission is the global dipole geometry, as discussed in \cref{sec:dipoleimplications}.  The equivalence given 
in  \cref{app:dipole} shows how to map a pinch onto a dipole by defining an
$r_{\rm eff}$ that preserves the differential tube volume, replacing the
corresponding pinch field by
$B_{\rm eff}(r_{\rm eff})=2\pi r_{\rm eff}/\mathcal V$.  A vacuum Z-pinch has
$B_{\rm eff}\propto r_{\rm eff}^{-1}$, whereas a point dipole maps to the
steeper $B_{\rm eff}\propto r_{\rm eff}^{-5/3}$.   Compared with the pinch, the steeper
effective field makes the flux-tube volume grow faster and gives more strongly
peaked marginal profiles, likely increasing $\tauE/\tau_{\rho}$  and thereby improving the overall performance \citep{Kesner2010}.
Likewise, if pushing to $\beta\gtrsim1$, the dipole is more stable to three-dimensional perturbations that seem very damaging for confinement. 

The third omission is the boundary treatment.  Our inner surface is a cold,
fixed-density and fixed-temperature reservoir that absorbs plasma without
recycling, sheath physics, or finite thermal conductance.  It arrests the inward
migration, sets the distance between the pressure maximum and the wall, and determines how energy
and particles divide between  loss routes.  The outer boundary continues the
low-$\beta$ marginal gradients, so is also idealised. A true  dipole will have distinct inner and outer
scrape-off layers, with fluctuations in the edge playing important roles \citep{Saitoh2026}; clearly, more realistic density boundaries and surface physics are needed
before the present coil-load and residence-time estimates can be treated as
useful predictions.

\section*{Acknowledgements}
The authors thank T.~Adkins, J.~Ball, D.~Garnier, J.~Guterl, and P.~Ivanov for helpful discussions.
JS and CH acknowledge the support of the Royal Society Te Ap\=arangi, through Marsden-Fund grant MFP-UOO2221 and Rutherford Discovery Fellowship RDF-U001804. LH was supported
by a University of Otago PhD Scholarship.
Computational support was provided by the New Zealand
 eScience Infrastructure (NeSI) high performance computing facilities, funded jointly by NeSI’s
 collaborator institutions and through the NZ MBIE, as well as through the Aoraki system at the University of Otago.

Generative AI tools were used extensively in this work: Claude Opus and Fable (Anthropic) and GPT Sol and Astra (OpenAI), accessed through agentic coding interfaces. They were used to edit existing \textsc{Athena++} simulation set-ups and analysis scripts, to launch and monitor simulations, to analyse simulation output and produce figures, to check analytic calculations, and for assistance in drafting and editing the manuscript text. All of this work was directed by the authors, and its outputs were reviewed and checked by them.

\appendix
\crefalias{section}{appendix}\crefalias{subsection}{subappendix}\crefalias{subsubsection}{subsubappendix}

\section{Simulations used in the manuscript}\label{app:runs}

This appendix provides a brief description and the parameters of all simulations presented in the manuscript.
Names should be interpreted as follows: first the exhaust prescription,  \texttt{S} for the fractional sink
of \cref{eq:fractionalsink} and \texttt{T} for the edge thermostat of
\cref{eq:thermostat}; then the heat input as a multiple of $Q_{\rm ref}$; then
changes compared to the default configuration of
\cref{sec:defaults}.  For example, \texttt{S-Q20-H4} is a fractional-sink run at
$20\,Q_{\rm ref}$ heated at $r_Q=4\rin$, and \texttt{S-Q1-CF-PSw} is the
reference drive, core fuelled, with the weakest edge particle sink; see \cref{tab:usedruns}.  Lengths in
names are in units of $\rin$, and the axial suite is named by $\Lz/L_{z0}$ ($L_{z0}=6\pi\rin$
being the circumference at the reference radius $r_{\rm mid}=3\rin$).  Edge fuelling,
$r_Q=2\rin$, $r_{\rm out}=6\rin$ and $\Lz=L_{z0}$ are defaults and therefore not listed.

Other configuration parameters are common to every run unless the table says
otherwise, with time  given in units of the fixed input clock
$\tau_{A0}=\rin/v_{A,{\rm IC}}$.  The default
heat source has $\sigma_Q=0.4\rin$, edge fuelling is at $r_\rho=5.2\rin$ with
$\sigma_\rho=0.2\rin$, and core fuelling at $2\rin$ with $0.4\rin$.  The mass
controller of \cref{eq:controller} uses $\tau_M=250\,\tau_{A0}$ and a cap
$\dot M_{\max}$ that would replace the initial inventory in
$6.7\times10^{2}\,\tau_{A0}$, while the fixed-rate runs use
$\dot M_{\rm ref}$, the late-time controller rate of the corresponding feedback
run, which replaces the same inventory in $2.2\times10^{3}\,\tau_{A0}$.  The
fractional sink has $w_s=0.2\rin$ and e-folds the local thermal energy in
$1.6\,\tau_{A0}$ at $r=r_{\rm out}$; the thermostat uses
$\tau_T=0.05\,\tau_{A0}$ and a set point at $0.9\,T_{\rm BC}$.  The inner
reservoir holds $\rho_{\rm BC}=0.9\,\rho_{\rm IC}$ at
$\beta_{\rm BC}=6.4\times10^{-3}$, about a fifth of the initial temperature at
the maximum, and the suffix \texttt{BC-r$x$T$y$} multiplies the $\rho_{\rm BC}$ and $T_{\rm BC}$ by $x$ and
$y$ respectively.  The edge particle sink has the same radial
shape as the energy sink, stops removing below $0.03\,\rho_{\rm IC}$ and takes
at most a tenth of the local density in a step, while the label-extraction runs
place the label-$0$ source at $1.5\rin$ with width $0.4\rin$; the rates at
which the two labels are injected are passive normalisations and cancel from
every diagnostic reported in the main text.  Floors are
$10^{-8}\rho_{\rm IC}$ in density and $3\times10^{-9}B_{\rm in}^2/8\pi$ in
pressure, except in the three-dimensional continuations, which raise the
density floor to $10^{-2}\rho_{\rm IC}$; note that the highest-$\beta$ of those
become floor affected once their cores empty (\cref{sec:3dresults}).  Two-dimensional runs are seeded initially with a random radial velocity of amplitude
$4\times10^{-4}v_{A,{\rm IC}}$ at axial mode number 3 (or one in the two
shortest boxes).  In the axial suite the heat input, the sink coefficient, and
the controller cap are all scaled with $\Lz$, so that only the axial geometry
differs.

The averaging interval begins at $t_{\rm av}$, the point at which the mass and
stored thermal energy have become statistically steady.  One run,
\texttt{T-Q0.04}, never reaches that state, and is listed for completeness but
never plotted.  Here $t_{\rm end}$ is the last output of the run, at which
every averaging window closes, so the second entry of the pair is also the
duration simulated.

\begingroup
\scriptsize
\setlength{\tabcolsep}{3.0pt}
\renewcommand{\arraystretch}{1.08}
\begin{longtable}{@{}p{0.192\textwidth}l r r r r l>{\raggedright\arraybackslash}p{0.225\textwidth}@{}}
\caption{Simulations used in, or directly relevant to, the manuscript, grouped by the comparison each belongs to.  The grid column is $N_r\times N_z$ for two-dimensional runs and $N_r\times N_z\times N_\theta$ for the three-dimensional continuations, which also carry the $Q$ and $\beta_{\rm pk}$ of the two-dimensional parent they were extruded from.  Radii are in units of $\rin$.  $\Ma$ is the turbulent Mach number defined in \cref{sec:kadomtsev}, as a median over $2.5<r/\rin<5$ of the averaged state, and is not quoted for the evolving three-dimensional runs.  Every mean reported in the paper is formed between $t_{\rm av}$ and $t_{\rm end}$, both measured from the start of the run in the fixed input clock $\tau_{A0}=\rin/v_{A,{\rm IC}}$; for the three-dimensional continuations the pair is instead the whole duration simulated, none of which is a saturated state.  Names, defaults and the parameters shared by every run are set out in the text.\label{tab:usedruns}}\\
\hline
Name & grid & $Q/Q_{\rm ref}$ & $\beta_{\rm pk}$ & $\Ma$ & $r_{\rm out}$ & $(t_{\rm av},t_{\rm end})/\tau_{A0}$ & Notes\\
\hline\endfirsthead
\multicolumn{8}{c}{\tablename\ \thetable\ (continued)}\\\hline
Name & grid & $Q/Q_{\rm ref}$ & $\beta_{\rm pk}$ & $\Ma$ & $r_{\rm out}$ & $(t_{\rm av},t_{\rm end})/\tau_{A0}$ & Notes\\
\hline\endhead
\hline\endfoot
\hline
\multicolumn{8}{@{}l}{{\itshape fractional-sink drive}: main drive scan, loss proportional to local thermal energy}\\
\nolinkurl{S-Q0.04} & $384\!\times\!1440$ & 0.04 & 0.0212 & 0.65 & 6 & 3509, 4512 & \\
\nolinkurl{S-Q0.2} & $384\!\times\!1440$ & 0.2 & 0.0551 & 0.64 & 6 & 2005, 3008 & \\
\nolinkurl{S-Q1} & $384\!\times\!1440$ & 1 & 0.135 & 0.71 & 6 & 3760, 4512 & \\
\nolinkurl{S-Q4} & $384\!\times\!1440$ & 4 & 0.331 & 0.75 & 6 & 501, 1504 & \\
\nolinkurl{S-Q10} & $384\!\times\!1440$ & 10 & 0.51 & 0.71 & 6 & 501, 1504 & \\
\nolinkurl{S-Q20} & $384\!\times\!1440$ & 20 & 0.893 & 0.77 & 6 & 752, 1504 & \\
\nolinkurl{S-Q40} & $384\!\times\!1440$ & 40 & 1.37 & 0.75 & 6 & 1003, 1504 & \\
\nolinkurl{S-Q100} & $384\!\times\!1440$ & 100 & 2.38 & 0.84 & 6 & 1003, 1504 & \\
\nolinkurl{S-Q200} & $384\!\times\!1440$ & 200 & 4.2 & 0.77 & 6 & 752, 1504 & \\
\hline
\multicolumn{8}{@{}l}{{\itshape thermostat drive}: the same scan with the edge temperature clamped}\\
\nolinkurl{T-Q0.04} & $384\!\times\!1440$ & 0.04 & 0.238 & 0.31 & 6 & 6016, 7520 & not saturated\\
\nolinkurl{T-Q0.1} & $384\!\times\!1440$ & 0.1 & 0.282 & 0.35 & 6 & 5013, 7520 & \\
\nolinkurl{T-Q0.2} & $384\!\times\!1440$ & 0.2 & 0.289 & 0.45 & 6 & 5013, 7520 & \\
\nolinkurl{T-Q0.4} & $384\!\times\!1440$ & 0.4 & 0.311 & 0.48 & 6 & 5013, 7520 & \\
\nolinkurl{T-Q1} & $384\!\times\!1440$ & 1 & 0.36 & 0.57 & 6 & 2005, 4512 & \\
\nolinkurl{T-Q2} & $384\!\times\!1440$ & 2 & 0.388 & 0.64 & 6 & 2005, 4512 & \\
\nolinkurl{T-Q4} & $384\!\times\!1440$ & 4 & 0.486 & 0.70 & 6 & 501, 3008 & \\
\nolinkurl{T-Q10} & $384\!\times\!1440$ & 10 & 0.622 & 0.73 & 6 & 2005, 3008 & \\
\hline
\multicolumn{8}{@{}l}{{\itshape source placement}: heat and particle sources moved at fixed total rates}\\
\nolinkurl{S-Q1-CF} & $384\!\times\!1440$ & 1 & 0.156 & 0.77 & 6 & 2005, 4512 & fuelled at $2\rin$\\
\nolinkurl{T-Q1-CF} & $384\!\times\!1440$ & 1 & 0.346 & 0.56 & 6 & 2005, 4512 & fuelled at $2\rin$\\
\nolinkurl{S-Q1-H3} & $384\!\times\!1440$ & 1 & 0.162 & 0.64 & 6 & 1003, 3008 & heating at $r_Q=3\rin$\\
\nolinkurl{S-Q1-H4} & $384\!\times\!1440$ & 1 & 0.234 & 0.54 & 6 & 2005, 4512 & heating at $r_Q=4\rin$\\
\nolinkurl{T-Q1-H4} & $384\!\times\!1440$ & 1 & 0.492 & 0.45 & 6 & 5013, 7520 & heating at $r_Q=4\rin$\\
\nolinkurl{S-Q4-H4} & $384\!\times\!1440$ & 4 & 0.555 & 0.58 & 6 & 1755, 3008 & heating at $r_Q=4\rin$\\
\nolinkurl{T-Q4-H4} & $384\!\times\!1440$ & 4 & 0.724 & 0.49 & 6 & 3509, 6016 & heating at $r_Q=4\rin$\\
\nolinkurl{S-Q20-H4} & $384\!\times\!1440$ & 20 & 1.39 & 0.55 & 6 & 1003, 3008 & heating at $r_Q=4\rin$\\
\hline
\multicolumn{8}{@{}l}{{\itshape radial resolution}: convergence in grid spacing at fixed geometry and drive}\\
\nolinkurl{S-Q1-N64} & $64\!\times\!240$ & 1 & 0.187 & 0.55 & 6 & 2005, 4512 & \\
\nolinkurl{S-Q1-N128} & $128\!\times\!480$ & 1 & 0.151 & 0.60 & 6 & 2005, 4512 & \\
\nolinkurl{S-Q1-N192} & $192\!\times\!720$ & 1 & 0.147 & 0.64 & 6 & 2005, 4512 & \\
\nolinkurl{S-Q1-N256} & $256\!\times\!960$ & 1 & 0.148 & 0.70 & 6 & 3259, 4512 & \\
\nolinkurl{S-Q1-N384} & $384\!\times\!1440$ & 1 & 0.157 & 0.68 & 6 & 2005, 4512 & \\
\nolinkurl{S-Q1-N512} & $512\!\times\!1920$ & 1 & 0.143 & 0.76 & 6 & 2005, 4512 & \\
\nolinkurl{S-Q1-N768} & $768\!\times\!2880$ & 1 & 0.131 & 0.71 & 6 & 3008, 4512 & \\
\nolinkurl{S-Q1-N1024} & $1024\!\times\!3840$ & 1 & 0.145 & 0.72 & 6 & 2005, 4512 & \\
\hline
\multicolumn{8}{@{}l}{{\itshape inner boundary}: conductor reservoir varied, and the initial peak radius}\\
\nolinkurl{S-Q1-peakwall} & $384\!\times\!1440$ & 1 & 0.135 & 0.71 & 6 & 1504, 3008 & peak at $\rin$, $(\rho,T)_{\rm BC}=(1.1,4.5)$\\
\nolinkurl{S-Q1-BC-r1T0.1} & $384\!\times\!1440$ & 1 & 0.14 & 0.71 & 6 & 1009, 3028 & $(\rho,T)_{\rm BC}=(1,0.1)$\\
\nolinkurl{S-Q1-BC-r1T0.25} & $384\!\times\!1440$ & 1 & 0.153 & 0.71 & 6 & 1512, 3024 & $(\rho,T)_{\rm BC}=(1,0.25)$\\
\nolinkurl{S-Q1-BC-r0.5T0.5} & $384\!\times\!1440$ & 1 & 0.146 & 0.66 & 6 & 1512, 3024 & $(\rho,T)_{\rm BC}=(0.5,0.5)$\\
\nolinkurl{S-Q1-BC-r2T0.5} & $384\!\times\!1440$ & 1 & 0.155 & 0.71 & 6 & 1003, 3008 & $(\rho,T)_{\rm BC}=(2,0.5)$\\
\nolinkurl{S-Q1-BC-r0.25T1} & $384\!\times\!1440$ & 1 & 0.131 & 0.68 & 6 & 1764, 3024 & $(\rho,T)_{\rm BC}=(0.25,1)$\\
\nolinkurl{S-Q1-BC-r0.25T0.25} & $384\!\times\!1440$ & 1 & 0.122 & 0.67 & 6 & 1514, 3029 & $(\rho,T)_{\rm BC}=(0.25,0.25)$\\
\nolinkurl{S-Q1-BC-r8T0.125} & $384\!\times\!1440$ & 1 & 0.187 & 0.85 & 6 & 1003, 3008 & $(\rho,T)_{\rm BC}=(8,0.125)$\\
\hline
\multicolumn{8}{@{}l}{{\itshape fixed-rate fuelling}: mass controller replaced by a prescribed constant rate}\\
\nolinkurl{S-Q1-F0.25} & $384\!\times\!1440$ & 1 & 0.132 & 0.67 & 6 & 9024, 11029 & $\dot M=0.25\dot M_{\rm ref}$\\
\nolinkurl{S-Q1-F1} & $384\!\times\!1440$ & 1 & 0.154 & 0.70 & 6 & 2005, 4512 & $\dot M=\dot M_{\rm ref}$\\
\nolinkurl{S-Q1-F4} & $384\!\times\!1440$ & 1 & 0.166 & 0.78 & 6 & 13035, 14539 & $\dot M=4\dot M_{\rm ref}$\\
\nolinkurl{S-Q20-F1} & $384\!\times\!1440$ & 20 & 0.906 & 0.77 & 6 & 1504, 3008 & $\dot M=\dot M_{\rm ref}$\\
\hline
\multicolumn{8}{@{}l}{{\itshape passive-label extraction}: edge sink applied to label 0 alone}\\
\nolinkurl{S-Q1-Anone} & $384\!\times\!1440$ & 1 & 0.145 & 0.68 & 6 & 6517, 9024 & $\tau_{\rm sink}=\infty$\\
\nolinkurl{S-Q1-Aweak} & $384\!\times\!1440$ & 1 & 0.135 & 0.67 & 6 & 2005, 4512 & $\tau_{\rm sink}/\tau_{A0}=16.4$\\
\nolinkurl{S-Q1-Amidweak} & $384\!\times\!1440$ & 1 & 0.143 & 0.71 & 6 & 2005, 4512 & $\tau_{\rm sink}/\tau_{A0}=5.17$\\
\nolinkurl{S-Q1-Amedium} & $384\!\times\!1440$ & 1 & 0.149 & 0.71 & 6 & 2005, 4512 & $\tau_{\rm sink}/\tau_{A0}=1.64$\\
\nolinkurl{S-Q1-Amidstrong} & $384\!\times\!1440$ & 1 & 0.147 & 0.73 & 6 & 4011, 4512 & $\tau_{\rm sink}/\tau_{A0}=0.517$\\
\nolinkurl{S-Q1-Astrong} & $384\!\times\!1440$ & 1 & 0.148 & 0.69 & 6 & 2005, 4512 & $\tau_{\rm sink}/\tau_{A0}=0.164$\\
\nolinkurl{S-Q1-Asuperstrong} & $384\!\times\!1440$ & 1 & 0.135 & 0.68 & 6 & 2005, 4512 & $\tau_{\rm sink}/\tau_{A0}=0.0517$\\
\hline
\multicolumn{8}{@{}l}{{\itshape edge particle sink}: core fuelled, with material exhaust diverted to the outer edge}\\
\nolinkurl{S-Q1-CF-PSw} & $384\!\times\!1440$ & 1 & 0.137 & 0.78 & 6 & 7019, 9525 & $\tau_{\rm sink}/\tau_{A0}=132$\\
\nolinkurl{S-Q1-CF-PSm} & $384\!\times\!1440$ & 1 & 0.13 & 0.78 & 6 & 7019, 9525 & $\tau_{\rm sink}/\tau_{A0}=66.2$\\
\nolinkurl{S-Q1-CF-PSs} & $384\!\times\!1440$ & 1 & 0.127 & 0.79 & 6 & 7019, 9525 & $\tau_{\rm sink}/\tau_{A0}=31.9$\\
\hline
\multicolumn{8}{@{}l}{{\itshape axial length}: $L_z$ varied, with the extensive rates scaled with it}\\
\nolinkurl{S-Lz/12} & $384\!\times\!120$ & 0.0833 & 1.65 & 0.28 & 6 & 2632, 3008 & $L_z=L_{z0}/12$\\
\nolinkurl{S-Lz/6} & $384\!\times\!240$ & 0.167 & 0.219 & 0.42 & 6 & 1003, 1504 & $L_z=L_{z0}/6$\\
\nolinkurl{S-Lz/3} & $384\!\times\!480$ & 0.333 & 0.169 & 0.65 & 6 & 1253, 1504 & $L_z=L_{z0}/3$\\
\nolinkurl{S-2Lz/3} & $384\!\times\!960$ & 0.667 & 0.174 & 0.67 & 6 & 1003, 1504 & $L_z=2L_{z0}/3$\\
\nolinkurl{S-4Lz/3} & $384\!\times\!1920$ & 1.33 & 0.146 & 0.73 & 6 & 1003, 1504 & $L_z=4L_{z0}/3$\\
\nolinkurl{S-2Lz} & $384\!\times\!2880$ & 2 & 0.157 & 0.68 & 6 & 1003, 1504 & $L_z=2L_{z0}$\\
\hline
\multicolumn{8}{@{}l}{{\itshape radial domain}: annulus widened, energy sink moved to the new outer edge}\\
\nolinkurl{S-Q1-R9} & $576\!\times\!1440$ & 1 & 0.339 & 0.70 & 9 & 2005, 3008 & energy sink at $r_{\rm out}$\\
\nolinkurl{S-Q1-R12} & $768\!\times\!1440$ & 1 & 0.542 & 0.52 & 12 & 5264, 6016 & energy sink at $r_{\rm out}$\\
\hline
\multicolumn{8}{@{}l}{{\itshape three-dimensional continuation}: extruded two-dimensional states, continued in $\theta$}\\
\nolinkurl{3D-S-Q1} & $192\!\times\!720\!\times\!192$ & 1 & 0.135 & -- & 6 & 0, 1003 & 3-D survey\\
\nolinkurl{3D-S-Q4} & $192\!\times\!720\!\times\!192$ & 4 & 0.331 & -- & 6 & 0, 1003 & 3-D survey\\
\nolinkurl{3D-S-Q4-HR} & $384\!\times\!1440\!\times\!384$ & 4 & 0.331 & -- & 6 & 0, 501 & high-resolution 3-D\\
\nolinkurl{3D-S-Q10} & $192\!\times\!720\!\times\!192$ & 10 & 0.51 & -- & 6 & 0, 1504 & 3-D survey\\
\nolinkurl{3D-S-Q20} & $192\!\times\!720\!\times\!192$ & 20 & 0.893 & -- & 6 & 0, 1003 & 3-D survey\\
\nolinkurl{3D-S-Q20-HR} & $384\!\times\!1440\!\times\!384$ & 20 & 0.893 & -- & 6 & 0, 501 & high-resolution 3-D\\
\nolinkurl{3D-S-Q40} & $192\!\times\!720\!\times\!192$ & 40 & 1.37 & -- & 6 & 0, 1003 & 3-D survey\\
\nolinkurl{3D-S-Q100} & $192\!\times\!720\!\times\!192$ & 100 & 2.38 & -- & 6 & 0, 501 & 3-D survey\\
\hline
\end{longtable}
\endgroup

\section{Short axial boxes and the zonal-flow regime}\label{app:zonal}

Unlike the Z-pinch, a true dipole has no single axial period.  The pinch coordinate $z$ stands in for the
toroidal angle, whose physical length $2\pi R$ varies both along a field line
and between flux surfaces (\cref{app:dipole}), so a box length can only  represent
the  field-line average of the toroidal circumference.  We therefore loosely 
associate the ``physical'' axial length with the 
circumference around the middle of the device, motivating $L_{z0}=2\pi r_{\rm mid}$  
for the default parameter set.  This appendix explores the dependence on $\Lz$, 
which is important both for understanding the dipole to pinch correspondence and 
for the general physics of the pinch. We will see a strong dependence on axial 
length below a critical threshold, due to the formation of zonal flows.

\begin{figure}
  \centerline{\includegraphics[width=\textwidth]{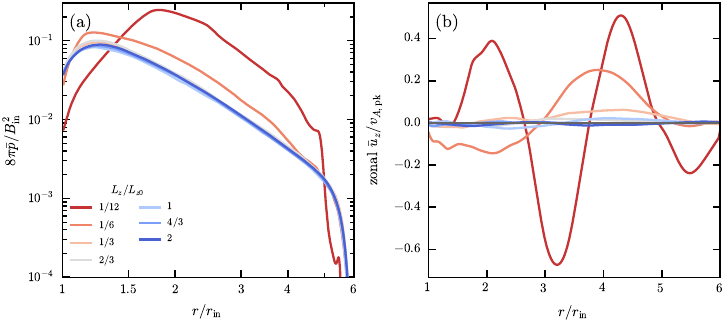}}
  \caption{The short-box zonal-flow regime, explored via  the axial-length scan \texttt{S-Lz/12} to
  \texttt{S-2Lz}.  (\textit{a}) Saturated pressure profiles and (\textit{b}) axially averaged zonal  flow
  normalised to $v_{A,{\rm pk}}$.  Boxes at $\Lz\geq L_{z0}/3$ share one
  pressure profile and carry little mean flow, while shorter ones develop
  strong banded flow, a steeper profile and a maximum that has stalled well
  outside the hard core.  The corresponding axial spectra are in
  \cref{fig:globality}(\textit{b}).}
  \label{fig:zonal}
\end{figure}

\Cref{fig:zonal} shows the relaxed pressure and the axially averaged $u_{z}$ over
the box-length scan from \texttt{S-Lz/12} to \texttt{S-2Lz}.  For $\Lz\geq L_{z0}/3$ the profiles collapse onto
one another, with near-identical pressure profiles and a mean flow of a
few per cent of $v_{A,{\rm pk}}$.  Shorter boxes reach a different state.  The
mean flow becomes strong and  banded, suppressing transport and
causing the  pressure profile to steepen significantly. Likewise, 
 the pressure maximum stalls further out from the wall, with significantly higher  stored
 thermal energy 
per unit volume compared to longer boxes.

The threshold length, below which the behaviour changes, is set by the axial correlation scale.  The spectra of
\cref{fig:globality}(\textit{b}) place the energy-containing wavelength at
$5$--$8\rin$ in the long boxes, and $L_{z0}/3=6.3\rin$ is the shortest box
in the scan that can capture this scale.  Below this
the radially extended overturning motions of \cref{fig:state} cannot fit, and
the energy condenses into axially averaged sheared flow instead.  The turbulence
is then broken up  by that shear and not by the quasi-global radial mixing, which significantly 
suppresses transport, thereby improving  confinement and increasing the pressure. 
This likewise affects various other measures explored in the main text 
such as the Mach-number scaling (\cref{fig:closures}\textit{b}) and the physics of the marginality front (\cref{fig:front}\textit{b}).

The most straightforward conclusion is that zonal-flow physics should  be treated with care in local domains.  Flux-tube gyrokinetic treatments of this instability
\citep[e.g.,][]{kobayashi10,Hallenbert2022,Hoffmann2023} generally resolve a binormal
extent that is a small
fraction of the toroidal circumference,  so care may be needed to ensure that the 
domain is not too short to hold the energy-containing binormal scale, thereby leading to a different saturated state.
More generally, the results show the potential relevance of zonal flows, which might arise due to edge  or kinetic physics absent here, for regulating transport and 
the migration of the pressure peak.

\section{Axially nonuniform heating and fuelling}\label{app:zmod}

A dipole reactor may heat or fuel through a few ports rather than uniformly
around the torus.  Its field lines close in the poloidal plane, so parallel
transport cannot spread that input toroidally as it can along a tokamak's
helical field.  We test whether interchange turbulence can do so by
modifying the default fractional-sink run \texttt{S-Q1}: its heat or particle source
is multiplied by $1+\cos(2\pi z/\Lz)$, with $\Lz=L_{z0}=6\pi\rin$.
The source ranges from zero to twice its axial mean, while its radial shape
and total heating power are unchanged and the mass controller still sets
the total particle input.  No mean rotation is imposed.

\begin{figure}
  \centerline{\includegraphics[width=\textwidth]{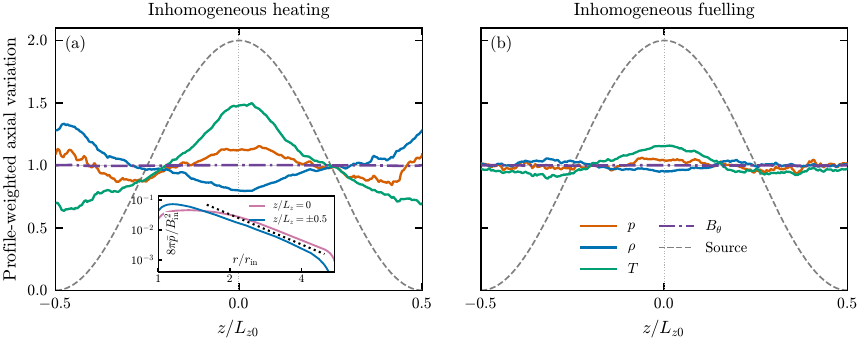}}
  \caption{Axial response to (\textit{a}) nonuniform heating and
  (\textit{b}) nonuniform fuelling, each imposed on \texttt{S-Q1}.  For each field
  $f$, the curves show $\int\langle f\rangle_t\,\mathrm{d}r\big/
  \big(\Lz^{-1}\!\int\langle f\rangle_t\,\mathrm{d}r\,\mathrm{d}z\big)$,
  with the $r$ integrals over the full radial domain,
  equivalent to weighting its fractional axial variation by its relaxed
  radial profile.  Every curve has unit axial mean.  The dashed curve is the
  modulated source, peaking at $z=0$.  The inset in (\textit{a}) shows the pressure averaged over
  bands of width $\Lz/4$ centred on the heating maximum ($z/\Lz=0$) and
  minimum ($z/\Lz=\pm0.5$, identified periodically), with a common normalisation.  The dotted black line shows a  
 low-$\beta$  $r^{-10/3}$ guide.}
  \label{fig:zmod}
\end{figure}

\Cref{fig:zmod} compares heating-only and fuelling-only modulation, averaging
each field's fractional axial variation over the full radius with its relaxed
radial profile as the weight.  Heating leaves a temperature modulation of
about $35$ per cent in the lowest harmonic, while the density falls near the
heated sector and partly offsets the pressure increase.  Fuelling leaves
much less variation: the density harmonic is about $2$ per cent and the
temperature harmonic about $9$ per cent.  The inset shows that pressure
profiles near the heating maximum and minimum retain a similar radial form,
close to the $r^{-10/3}$ marginal slope, although curiously,  the heated-sector peak has a lower maximum because it
is flatter, which seems to result from averaging over  larger radial excursions.
Overall, however, there is a modest degradation in performance, with the stored energy falling by ${\simeq}25$\%
for the modulated-heating case (there is little change with modulated fuelling).
Modulating both sources together, either in phase or on opposite sides of the box,
gives similar results (not shown).

We also show the magnetic field, which is  almost perfectly uniform and  does not balance the axial pressure
variation.  This implies that the mean profiles
are not in static axial equilibrium.  Instead, mean circulation and
fluctuation stresses make comparable contributions to the balance by carrying
axial momentum both radially and axially.  

The picture that emerges is a balance between the forcing, which
continually re-imposes the axial pattern, and interchange mixing, which
erases it.  If a harmonic is replaced on a time $\tau_f$ and mixed on a time $\tau_{\rm mix}$, a source harmonic of fractional
amplitude $a_f$ will leave the response $A_f\simeq a_f/(1+\tau_f/\tau_{\rm mix})$:
short-lived material retains the source pattern, while long-lived material is
mixed out.  The $n=1$ wavelength, $L_{z0}\simeq19\rin$, is several times the
energy-containing eddy scale (\cref{fig:globality}\textit{b}), so we estimate
its mixing as diffusive, with the mixing-length diffusivity
$D\sim u_{r,{\rm rms}}\ell_r$ from smaller-scale, roughly isotropic eddies in the $r$--$z$
plane.  The bulk values in \texttt{S-Q1} then give
$\tau_{\rm mix}\simeq(L_{z0}/2\pi)^2/D\simeq40\,\tau_{A,{\rm pk}}
\simeq0.2\,\tauE$.  With $\tau_f=\tauE$ and $a_f=1$ for the heat, this
predicts a thermal harmonic of about $15$ per cent, compared with the measured
$35$ per cent, which is reasonable agreement given the order-unity
uncertainty in $D$.

The same estimate applies to the particles and to a short-lived passive
scalar.  Particles are replaced about five times more slowly than thermal
energy, with the diffusive estimate then predicting a density response of  ${\simeq}4$\%,
compared with the measured $2$\%.  To test the opposite limit, we apply
the same axial modulation only to the ash-label source of the strong
selective-exhaust run \texttt{S-Q1-Astrong} (\cref{sec:confinement}).  With
$\tau_{\rm ash}\simeq0.7\tauE$, the estimate yields ${\simeq}20$\% against 
a measured variation of approximately $50$\%.  Given that
the sources sit at different radii and that temperature is not a passive scalar, this
agreement to within a factor of about two in each case supports the simple
picture: short-lived material retains a substantial source imprint, while
long-lived material is mixed to be nearly uniform. 

For a dipole reactor, the substantial mixing of even the largest axial mode
suggests that localised injection will likely not greatly change the relaxed
profiles, even without toroidal  rotation.  Similar  
arguments should carry over to cases with less extreme source asymmetry (e.g., from two or more toroidally spaced antennae):
 a diffusion estimate, appropriate for asymmetry above the source scale, gives $\tau_{{\rm mix},n}\propto n^{-2}$, and once the spacing 
 falls below the axial/toroidal correlation scale,  individual
eddies  should mix adjacent sectors in approximately the turnover time.  
Overall, this is promising for the robustness of the dipole concept to asymmetries arising 
from engineering design considerations.

\section{Application of our results to dipole geometry}\label{app:dipole}

A hard-core pinch and a levitated dipole look very different.  The pinch has
constant-curvature, uniform field lines wrapped around a cylinder, whereas the field
lines of a dipole are strongly varying with a field strength that can vary by orders
of magnitude between the coil and the outer equator.  For flute-like
interchange motion, however, the two are much more similar than this naive comparison suggests.
In both, the motion carries whole flux tubes together with their particle and
entropy contents, and the geometry enters only through a few functions of the
flux label.  The purpose of the first part of this appendix (\ref{app:dipoleflute}) is to make this similarity precise.
This provides the basis for our expectation that the general results of the
2-D simulations --- namely the self-organisation onto the marginal
family, its insensitivity to the sources, and the inward migration of the pressure
maximum --- should  carry over qualitatively to
a dipole. In contrast, as explored in \cref{app:dipoleballooning}, the geometry
has a strong effect on field-line-dependent (bending) dynamics, where the
dipole differs significantly from the pinch.

Our starting point is the  reduced flute equations for 
general axisymmetric, shearless geometry from \citet{pastukhov05b}, which extended the
cylindrical model of \citet{pastukhov01}.   The new result is a map, via an effective radius coordinate, of these general-geometry equations
onto those of an effective hard-core pinch.  This shows that locally the geometry
changes only the units of angle and time and the strength of the interchange
drive, in a similar way to what occurs at finite $\beta$ in the pinch, isolating the few ways in
which a global dipole differs from the pinch.  

Throughout, the reference field is
${\bm B}_0=\boldsymbol{\nabla}\psi\mathbin{\times}
\boldsymbol{\nabla}\varphi$, where $\varphi$ is the toroidal angle,
$\psi$ labels nested flux surfaces, and $\ell$ measures distance around a
closed poloidal field line.  We orient $\psi$ to increase outward.  In the
pinch, $z$ takes the place of $\varphi$, and ${\rm d}\psi=B_0\,{\rm d}r$ where $B_{0}$ is the background field strength ($B_{0}=B_{\theta}$ in the main text).  Fuelling, heating and transport change the
axisymmetric equilibrium, and with it the metric, $\bm{B}_{0}$, and the flux labels, only on
a slow transport time.  During the faster interchange dynamics considered here,
${\bm B}_0$ and the coordinates are therefore treated as frozen, but they may
be updated between saturated states, a two-time construction that is made
explicit in Pastukhov's equations. It implies no low-$\beta$
ordering because near-marginal interchange dynamics remain approximately electrostatic.

\subsection{The field-line-averaged system}\label{app:dipoleflute}

\subsubsection{Density and entropy contents}

The volume element of the geometry just described is
${\rm d}V={\rm d}\ell\,{\rm d}\psi\,{\rm d}\varphi/B_0$.  Define
\begin{equation}
 \mathcal{V}(\psi)\equiv\oint\frac{{\rm d}\ell}{B_0},\qquad
 \langle X\rangle_\parallel\equiv
 \frac{1}{\mathcal{V}}\oint X\frac{{\rm d}\ell}{B_0}.                  \label{eq:tubeaverage}
\end{equation}
Here $\mathcal{V}$ is the differential flux-tube volume per unit poloidal
flux and toroidal angle.

The main text showed that in the pinch the tube contents $\rho\mathcal V$ and
$p^{1/\gamma}\mathcal V$ are materially conserved, \cref{eq:tubeinvariants}.
The same argument applies to a closed flux tube of any shape, provided it is
applied to the content of the whole tube rather than pointwise.  Let
${\rm D}/{\rm D}t\equiv\partial/\partial t+{\bm u}\boldsymbol\cdot\boldsymbol\nabla$
follow the fluid, and consider a fluid element of length $\delta\ell$ along a
field line and cross-section $\delta A$.  Ideal flux freezing conserves the
flux $B_0\,\delta A$ through it, while its volume $\delta\ell\,\delta A$
changes at the rate $\boldsymbol\nabla\mathbin\cdot{\bm u}$, so its volume per
unit flux obeys
${\rm D}(\delta\ell/B_0)/{\rm D}t=(\delta\ell/B_0)\,\boldsymbol\nabla\mathbin\cdot{\bm u}$.
Flux freezing also keeps elements that start on one closed line on a single
line, so summing around it gives
\begin{equation}
 \frac{{\rm D}\mathcal V}{{\rm D}t}
 =\oint \boldsymbol\nabla\mathbin\cdot{\bm u}\,\frac{{\rm d}\ell}{B_0}
 =\mathcal V\langle\boldsymbol\nabla\mathbin\cdot{\bm u}\rangle_\parallel .  \label{eq:generaltubelaw}
\end{equation}
Now consider the continuity and the adiabatic
pressure equation, which, away from  dissipation and sources,  have the common form
${\rm D}C/{\rm D}t=-C\,\boldsymbol\nabla\mathbin\cdot{\bm u}$, with
$C=\rho$ or $p^{1/\gamma}$.  Averaging this, using
the commutator from the element law,
${\rm D}\langle C\rangle_\parallel/{\rm D}t
=\langle{\rm D}C/{\rm D}t\rangle_\parallel
+\langle C\,\boldsymbol\nabla\mathbin\cdot{\bm u}\rangle_\parallel
-\langle C\rangle_\parallel\langle\boldsymbol\nabla\mathbin\cdot{\bm u}\rangle_\parallel$, one finds
\begin{equation}
 \frac{{\rm D}\langle C\rangle_\parallel}{{\rm D}t}
 =-\langle C\rangle_\parallel\langle\boldsymbol\nabla\mathbin\cdot{\bm u}\rangle_\parallel ,   \label{eq:averagedC}
\end{equation}
which, added to
\cref{eq:generaltubelaw} in logarithmic form, gives the
conservation law
\begin{equation}
 \frac{{\rm D}}{{\rm D}t}\left(\mathcal V\langle C\rangle_\parallel\right)=0,
 \qquad C=\rho,\ p^{1/\gamma}.                                      \label{eq:tubecontentlaw}
\end{equation}
This holds for any ideal motion on closed lines, with $B_0$ then the
instantaneous field.  Thus the pinch argument carries over exactly for
the line-integrated tube contents
\begin{equation}
 \hat\rho\equiv\mathcal{V}\langle\rho\rangle_\parallel,
 \qquad
 G\equiv\mathcal{V}\langle p^{1/\gamma}\rangle_\parallel .          \label{eq:tubecontents}
\end{equation}
As in the pinch, a flute displacement interchanges whole tubes, each carrying
its own $\hat\rho$ and $G$, without changing the field, so each tube keeps its
flux and takes the $\mathcal V$ of its new position.  Random
interchange mixing therefore drives both towards uniformity in $\psi$,
whatever the shape of the field line.  Field-line-bending motions
(\cref{app:dipoleballooning}) also conserve the contents but deform the
tubes, so this conclusion does not carry over directly to them.  Uniform $G$ is interchange marginality
(see below), and uniform $\hat\rho$ is the general form of turbulent
equipartition.

\subsubsection{Pastukhov's reduced equations}

\Citet{pastukhov05b} derives the flute equations by ordering in the departure
from marginality.  If the entropy content departs from uniformity by a small
fraction $\varepsilon^2$, the interchange motion has frequency
$\omega\sim\varepsilon k_\perp c_s$, slow compared with magnetosonic waves,
shear-Alfv\'en waves, and sound waves along the line.  These are removed by
restricting the motion to displacements that neither perturb the field nor
make the pressure vary along a line,
$\boldsymbol\nabla\mathbin\times(\bm\xi_\perp\mathbin\times{\bm B}_0)=0$ and
$\boldsymbol\nabla\mathbin\cdot(\bm\xi/\mathcal V)=0$ (his eq.~9).  This
requires $\beta$ to lie below the field-line-bending threshold
(\cref{app:dipoleballooning}), but not $\beta\ll1$.  

The perpendicular flow is
then an E$\times$B drift in a potential $\Phi(\psi,\varphi)$, constant along
field lines, and the velocity is (his eq.~11)
\begin{equation}
 {\bm u}=\frac{{\bm B}_0\mathbin{\times}\boldsymbol{\nabla}\Phi}{B_0^2}
 +\lambda{\bm B}_0\,\frac{\partial\Phi}{\partial\varphi},\qquad
 u^\psi\equiv{\bm u}\boldsymbol\cdot\boldsymbol\nabla\psi
 =-\frac{\partial\Phi}{\partial\varphi},\qquad
 u^\varphi\equiv{\bm u}\boldsymbol\cdot\boldsymbol\nabla\varphi
 =\frac{\partial\Phi}{\partial\psi} ,                                              \label{eq:flutekinematics}
\end{equation}
where the parallel-flow coefficient $\lambda$ satisfies
\begin{equation}
 B_0\,\frac{\partial\lambda}{\partial\ell}
 =\boldsymbol\nabla\mathbin\cdot{\bm e}_\psi
 -\langle\boldsymbol\nabla\mathbin\cdot{\bm e}_\psi\rangle_\parallel,
 \qquad
 \langle B_0^2\lambda\rangle_\parallel=0 ,                             \label{eq:slavedflow}
\end{equation}
with ${\bm e}_\psi\equiv\boldsymbol\nabla\psi/|\boldsymbol\nabla\psi|^2$ the
displacement per unit flux.\footnote{Pastukhov's eq.~10 gives $\lambda$
explicitly for a non-orthogonal poloidal angle, with the free constant chosen
so that $\lambda$ vanishes at the equator; on an up--down-symmetric line this
is equivalent to the second condition in \cref{eq:slavedflow}.  }  Omitting dissipation and sources, the reduced system (his eqs.~21--24)
is
\begin{gather}
 \frac{\partial\hat\rho}{\partial t}+\{\Phi,\hat\rho\}=0,\qquad
 \frac{\partial G}{\partial t}+\{\Phi,G\}=0,\qquad
 p=G^\gamma\mathcal{V}^{-\gamma},                                    \label{eq:contentdynamics}\\
 \frac{\partial\hat w}{\partial t}+\{\Phi,\hat w\}
 -\big\{\hat\rho,\tfrac12\langle u^2\rangle_\parallel\big\}
 =-\frac{1}{\mathcal V^\gamma}\frac{\partial\mathcal V}{\partial\psi}\,
 \frac{\partial G^\gamma}{\partial\varphi},                                         \label{eq:exactflutevorticity}\\
 \hat w\equiv
 \frac{\partial}{\partial\psi}\!\left(\hat\rho\,\langle R^2\rangle_\parallel\,
 \frac{\partial\Phi}{\partial\psi}\right)
 +\frac{\partial}{\partial\varphi}\!\left(\hat\rho\,
 \left\langle\frac{1}{R^2B_0^2}+\lambda^2B_0^2\right\rangle_\parallel
 \frac{\partial\Phi}{\partial\varphi}\right),                                      \label{eq:generalvorticity}
\end{gather}
where $R$ is the distance from the symmetry axis,
$\{A,C\}\equiv(\partial A/\partial\psi)(\partial C/\partial\varphi)
-(\partial A/\partial\varphi)(\partial C/\partial\psi)$, and
\begin{equation}
 \langle u^2\rangle_\parallel
 =\langle R^2\rangle_\parallel\left(\frac{\partial\Phi}{\partial\psi}\right)^2
 +\left\langle\frac{1}{R^2B_0^2}+\lambda^2B_0^2\right\rangle_\parallel
 \left(\frac{\partial\Phi}{\partial\varphi}\right)^2 ,                               \label{eq:kineticenergy}
\end{equation}
so that $\tfrac12\langle u^2\rangle_\parallel$ is the line-averaged kinetic
energy per unit mass.  This is a closed two-dimensional system for $\hat\rho$, $G$ and
$\Phi$, in which the magnetic geometry enters only through $\mathcal V$ and the
two inertia weights in the vorticity $\hat w$. Note that all of the line-averaged coefficients  and $\lambda$ can 
be precomputed once the field is specified. 

Pastukhov's complete model adds dissipation, sources and a slow subsystem, in
which the toroidally averaged contents evolve by transport on a time longer
by $\varepsilon^{-2}$, with the mean field following from the Grad--Shafranov
equation.  

\emph{Parallel flow.}  Since $u^\psi$ and $u^\varphi$ are constant along each
line, the E$\times$B flow moves whole flux tubes, but it is compressible, with
divergence $u^\psi\,\boldsymbol\nabla\mathbin\cdot{\bm e}_\psi$.  In the pinch, with uniformity along field lines so that $\lambda=0$,  
there is no parallel flow.  In a dipole, a flute
displacement compresses the plasma most strongly near the outer equator, where
the flux surfaces are widely spaced and strongly curved, which would drive
a pressure differential along the line.  The parallel flow removes this by carrying
plasma from where the E$\times$B flow converges to where it diverges, so that
$\boldsymbol\nabla\mathbin\cdot{\bm u}=u^\psi\,\partial\ln\mathcal V/\partial\psi
={\rm D}\ln\mathcal V/{\rm D}t$ is uniform along the line.  Density and
pressure that start uniform along a line therefore remain so, and
$\hat\rho=\rho\mathcal V$ and $G=p^{1/\gamma}\mathcal V$ hold at leading order
in any closed-line geometry.  Radial motion must drag this flow with it, which
adds the $\langle\lambda^2B_0^2\rangle_\parallel$ factor to the radial inertia in
$\hat w$, effectively increasing the mass of tubes.  This added mass is the closed-line analogue of how
the Pfirsch--Schl\"uter return flow raises the inertia of
poloidal rotation in a tokamak \citep{hassam78}.

\emph{Advection of the tube contents.}  Because $u^\psi$ and $u^\varphi$ are
constant along the line, the tube derivative in \cref{eq:tubecontentlaw} is
$\partial/\partial t+u^\psi\,\partial/\partial\psi+u^\varphi\,\partial/\partial\varphi$.
Written in terms of $\Phi$, the  content laws thus become \cref{eq:contentdynamics}: $\hat\rho$ and $G$ are
advected by an incompressible flow in the $(\psi,\varphi)$ plane.  

\emph{Non-Boussinesq inversion.}  \Cref{eq:exactflutevorticity,eq:generalvorticity}
describe two-dimensional buoyancy-driven flow.  The right-hand side of
\cref{eq:exactflutevorticity} is the interchange drive: a displaced tube keeps
its $G$, so its pressure differs from that of its new surroundings unless $G$
is uniform.  Likewise, linearising about $G(\psi)$ shows that the squared growth rate is proportional to $-\mathcal V'\,\partial G/\partial\psi$, so interchange marginality is $\partial G/\partial\psi=0$.
The flow is recovered from $\hat w$ by inverting an elliptic
operator whose coefficients are the tube mass $\hat\rho$ times the
line-averaged inertia of toroidal ($\langle R^2\rangle_\parallel$) and radial
($\langle R^{-2}B_0^{-2}+\lambda^2B_0^2\rangle_\parallel$) motion.  Because
$\hat\rho$ varies from tube to tube, this inversion does not commute with
advection, giving the kinetic-energy correction (third term on the left of \cref{eq:exactflutevorticity}).  The ideal system
conserves the energy
$\int\big[\tfrac12\hat\rho\langle u^2\rangle_\parallel
+\mathcal Vp/(\gamma-1)\big]\,{\rm d}\psi\,{\rm d}\varphi$.

\emph{Hard-core pinch.}  In the cylindrical case field lines are 
circles ${\bm B}_0=B_\theta(r)\hat{\bm\theta}$, $\varphi$ becomes
$z$, ${\rm d}\psi=B_\theta\,{\rm d}r$ so that $\mathcal V=2\pi r/B_\theta$, 
$\lambda=0$, and the two inertia weights are
$\langle R^2\rangle_\parallel=1$ and
$\langle R^{-2}B_0^{-2}+\lambda^2B_0^2\rangle_\parallel= B_\theta^{-2}$.  Written in $(r,z)$, with
$\{A,C\}_{r,z}\equiv(\partial A/\partial r)(\partial C/\partial z)-
(\partial A/\partial z)(\partial C/\partial r)$,
\cref{eq:contentdynamics,eq:exactflutevorticity,eq:generalvorticity} become
\begin{gather}
 \frac{\partial\hat\rho}{\partial t}+\frac{1}{B_\theta}\{\Phi,\hat\rho\}_{r,z}=0,\qquad
 \frac{\partial G}{\partial t}+\frac{1}{B_\theta}\{\Phi,G\}_{r,z}=0,\qquad
 p=G^\gamma\mathcal{V}^{-\gamma},                                    \label{eq:pinchcontent}\\
 \frac{\partial\hat w}{\partial t}+\frac{1}{B_\theta}\{\Phi,\hat w\}_{r,z}
 -\frac{1}{B_\theta}\big\{\hat\rho,\tfrac12u^2\big\}_{r,z}
 =-\frac{1}{B_\theta\mathcal V^\gamma}\frac{{\rm d}\mathcal V}{{\rm d}r}\,
 \frac{\partial G^\gamma}{\partial z},                              \label{eq:pinchvorticityeq}\\
 \hat w=\frac{1}{B_\theta}\frac{\partial}{\partial r}\!\left(
   \frac{\hat\rho}{B_\theta}\frac{\partial\Phi}{\partial r}\right)
 +\frac{1}{B_\theta^{2}}\frac{\partial}{\partial z}\!\left(
   \hat\rho\,\frac{\partial\Phi}{\partial z}\right),                \label{eq:pinchvorticity}
\end{gather}
with $u^2=B_\theta^{-2}|\boldsymbol\nabla\Phi|^2$.  This is the reduced model
of the hard-core pinch simulated by \citet{pastukhov01,pastukhov05}, the only version 
that has (to our knowledge) been studied in detail.  It
should capture the near-marginal interchange dynamics of the saturated states
in the main text, though not effects related to order-unity Mach-number compressibility and the evolution of $B_{\theta}$.

\subsubsection{Relation to other flux-tube-averaged dipole models}

The flux-tube-averaged MHD system solved by \citet{li21} and the electrostatic
two-fluid model of \citet{ou20}, which has an additional entropy mode, have the
same structure as
\cref{eq:contentdynamics,eq:generalvorticity,eq:exactflutevorticity}.  Their
prescribed dielectric functions correspond to the two inertia weights in
$\hat w$.  They differ from Pastukhov's equations in three ways.  First, the
nonlinear advective polarisation current is approximated by the convective
derivative of the generalised vorticity, which drops
$\{\hat\rho,\tfrac12\langle u^2\rangle_\parallel\}$ from
\cref{eq:exactflutevorticity}.  Second, the dielectric functions omit the
dependence of the inertia on the evolving $\hat\rho$.  Third, because they
are built from the perpendicular polarisation current alone, they contain
$\langle R^{-2}B_0^{-2}\rangle_\parallel$ but not the parallel-flow inertia
$\langle\lambda^2B_0^2\rangle_\parallel$.  More recently, \citet{Jiang2025}
recovered the same flux-tube system and perpendicular point-dipole inertia
weights from a bounce-averaged kinetic derivation for an isotropic
warm-electron, cold-ion plasma.  The extra kinetic physics adds magnetic-drift
and curvature-heat-flux terms to their density and entropy equations, which
generate the entropy-mode sector absent from \cref{eq:contentdynamics}.

The first and second approximations are less severe than it may appear, because the dropped
term vanishes when $\hat\rho$ is uniform.  Likewise, if $\ell_r$ is the
radial eddy scale and $L_{\hat\rho}$ the gradient scale of $\hat\rho$, the kinetic energy term is
smaller than the retained vorticity advection by roughly $\ell_r/L_{\hat\rho}$.
These models are therefore 
well controlled for $\ell_r\ll L_{\hat\rho}$, or where turbulent equipartition has
flattened $\hat\rho$, but the differences become of order unity for
large-scale turbulent cells with a strong gradient of $\hat\rho$ (e.g., 
during relaxation from an arbitrary initial state).   The third effect (the parallel-flow inertia) has a
more uniform effect, which we quantify for a dipole below.

\subsubsection{Vacuum dipole geometry}

To concretely compare geometries, we normalise the two inertia weights by their values at
the outer equator, denoted by the subscript ``eq'':
\begin{equation}
 Y_\varphi\equiv
 \frac{\langle R^2\rangle_\parallel}{R_{\rm eq}^2},\qquad
 Y_\psi\equiv R_{\rm eq}^2B_{\rm eq}^2
 \left\langle\frac{1}{R^2B_0^2}+\lambda^2B_0^2\right\rangle_\parallel .   \label{eq:Ygeometry}
\end{equation}
Both are purely geometrical, because $\rho$ is uniform along each line, and
$Y_\varphi=Y_\psi=1$ in a pinch.  The geometrical numbers used below come either
from an ideal point dipole or from the vacuum field of a physical circular current loop of radius
$R_c$.\footnote{In cylindrical coordinates
$(R,Z)$, the latter is
$\boldsymbol B_{\rm vac}=\boldsymbol\nabla\psi_{\rm vac}\mathbin\times
\boldsymbol\nabla\varphi$, with
$\psi_{\rm vac}\propto\sqrt{RR_c}\big[(2-k^2)\mathrm{K}(k)-2\mathrm{E}(k)\big]/k$ and
$k^2=4RR_c/[(R+R_c)^2+Z^2]$, where $\mathrm{K}$ and $\mathrm{E}$ are the complete
elliptic integrals of the first and second kind
\citep[e.g.][]{jackson99}.
For $d\equiv(R^2+Z^2)^{1/2}\gg R_c$, this approaches
$\psi_{\rm vac}\propto R^2/d^3$, the point-dipole limit.}
Defining $R_{{\rm eq},0}$ as a
reference outer-equatorial radius,  direct evaluation of \cref{eq:Ygeometry} for a point dipole, with $R_{\rm eq}B_{\rm eq}\propto R_{\rm eq}^{-2}$, gives
\begin{equation}
 \frac{Y_\psi R_{{\rm eq},0}^2B_{{\rm eq},0}^2}{R_{\rm eq}^2B_{\rm eq}^2}
 =0.81\left(\frac{R_{\rm eq}}{R_{{\rm eq},0}}\right)^4,\qquad
 \frac{Y_\varphi R_{\rm eq}^2}{R_{{\rm eq},0}^2}
 =0.75\left(\frac{R_{\rm eq}}{R_{{\rm eq},0}}\right)^2.
 \label{eq:pointdipoleweights}
\end{equation}
The coefficients are the 
counterparts of the $0.66$ and $0.77$ dipole factors used by \citet{ou20} (or 
$0.6$ and $0.7$ in  \citealt{li21}), but 
here the parallel flow contribution ($\lambda^{2}B_{0}^{2}$)  raises the radial
weight by a further ${\simeq}26$\% ($0.64$ to $0.81$). 
This has the effect of  lowering the local growth rate of
interchange modes, whose inertia is mainly radial, by ${\simeq}10$\%.

  On the vacuum ring dipole, shown in  \cref{fig:flutegeometry},  the coefficients and the parallel-flow contribution vary with position.  
The fractional effect of the latter vanishes at the coil, where the field lines become
circles, peaks at $\langle\lambda^2B_0^2\rangle_\parallel/\langle
R^{-2}B_0^{-2}\rangle_\parallel\simeq0.35$ for $R_{\rm eq}\simeq1.5$--$2R_c$, and
approaches the point-dipole value of $0.26$ further out.  Across the traced
domain, $Y_\psi=0.77$--$0.83$.

\subsubsection{Exact equivalence for radially local models}

At any chosen surface, all equilibrium and metric factors may be treated as
constants.  We can then evaluate the two inertia weights and compare with a straight
hard-core pinch whose local field is $B_{\rm p}$ by rescaling the coordinates, defining
\begin{equation}
 \begin{gathered}
 a\equiv\langle R^2\rangle_\parallel=Y_\varphi R_{\rm eq}^2,\qquad
 b\equiv\left\langle\frac{1}{R^2B_0^2}+\lambda^2B_0^2\right\rangle_\parallel
 =\frac{Y_\psi}{R_{\rm eq}^2B_{\rm eq}^2},\\
 \sigma_z\equiv\frac{\sqrt a}{B_{\rm p}\sqrt b},\qquad
 z=\sigma_z\varphi,\qquad t=\sigma_t t_{\rm p},\qquad
 \Phi_{\rm p}=\sigma_t\sigma_z\Phi,\qquad
 w_{\rm p}=\frac{\sigma_t\sigma_z}{a}\hat w ,
  \end{gathered}                                                     \label{eq:localmap}
\end{equation}
where the clock $\sigma_t$ is, for now, an arbitrary constant.  With the
bracket now taken in $(\psi,z)=(\psi,\sigma_{z}\varphi)$,
\cref{eq:contentdynamics,eq:generalvorticity,eq:exactflutevorticity} become
\begin{gather}
 \frac{\partial\hat\rho}{\partial t_{\rm p}}+\{\Phi_{\rm p},\hat\rho\}_{\psi,z}=0,\qquad
 \frac{\partial G}{\partial t_{\rm p}}+\{\Phi_{\rm p},G\}_{\psi,z}=0,       \label{eq:stretchedcontent}\\
 \frac{\partial w_{\rm p}}{\partial t_{\rm p}}+\{\Phi_{\rm p},w_{\rm p}\}_{\psi,z}
 -\big\{\hat\rho,\tfrac12u_{\rm p}^2\big\}_{\psi,z}
 =-\frac{\sigma_t^2}{B_{\rm p}^2b}\,
 \frac{1}{\mathcal V^{\gamma}}\frac{\partial\mathcal V}{\partial\psi}
 \frac{\partial G^\gamma}{\partial z},                                   \label{eq:stretchedvorticityeq}\\
 w_{\rm p}=\frac{\partial}{\partial\psi}\!\left(\hat\rho\,\frac{\partial\Phi_{\rm p}}{\partial\psi}\right)
 +\frac{1}{B_{\rm p}^{2}}\frac{\partial}{\partial z}\!\left(\hat\rho\,\frac{\partial\Phi_{\rm p}}{\partial z}\right),
                                                                         \label{eq:stretchedvorticity}
\end{gather}
where $u_{\rm p}^2=(\partial\Phi_{\rm p}/\partial\psi)^2+B_{\rm p}^{-2}(\partial\Phi_{\rm p}/\partial z)^2$.
Apart from the factor multiplying the drive, this is exactly the pinch system
\cref{eq:pinchcontent,eq:pinchvorticityeq,eq:pinchvorticity}, written in
$(\psi,z)$ at fixed field $B_\theta=B_{\rm p}$: both dipole inertia weights have disappeared, without
linearisation and including the nonlinear terms.  The only local geometrical
number inherited from the closed-line field is the drive
$\mathcal V^{-\gamma}\partial\mathcal V/\partial\psi$, and it appears only in
combination with the free clock.  A pinch, whose own drive is
$\mathcal V_{\rm p}^{-\gamma}\partial\mathcal V_{\rm p}/\partial\psi$, is
therefore reproduced exactly by choosing
\begin{equation}
 \sigma_t^2=B_{\rm p}^2b\,
 \frac{\mathcal V_{\rm p}^{-\gamma}\partial\mathcal V_{\rm p}/\partial\psi}
 {\mathcal V^{-\gamma}\partial\mathcal V/\partial\psi} .            \label{eq:clockchoice}
\end{equation}
Any radially local closed-line flute system can therefore be mapped onto any
pinch, including the vacuum (low-$\beta$) field
$B_\theta\propto r^{-1}$.  Thus, a radially local flute model is insensitive
to the shape of its closed field line, which sets only the units of time and
toroidal length.  

The same argument relates pinches with different $\beta$.  Compare a
finite-$\beta$ pinch with a vacuum pinch of the same $r$, $B_\theta$ and
$\mathcal V$, taking $B_{\rm p}=B_\theta$.  Both have the pinch weights
$a=1$ and $b=B_\theta^{-2}$, so $\sigma_z=1$, and they differ only in their
drives.  The vacuum pinch has $\mathcal V_{\rm p}\propto r^2$, so
$\partial\mathcal V_{\rm p}/\partial\psi=2\mathcal V/(rB_\theta)$, and
\cref{eq:clockchoice} gives
\begin{equation}
 \sigma_t^2=\frac{2\mathcal V}{rB_\theta\,\partial\mathcal V/\partial\psi}
 =2\left(\frac{{\rm d}\ln\mathcal V}{{\rm d}\ln r}\right)^{-1}
 =1+\frac{\gamma\beta}{2},                                         \label{eq:betaclock}
\end{equation}
where the last form uses the marginal slope
${\rm d}\ln\mathcal V/{\rm d}\ln r=4/(2+\gamma\beta)$ from
\cref{eq:vlogslope,eq:kdef}.  This is the factor by which finite $\beta$
reduces $N^2$ in \cref{sec:kadomtsev} at a given gradient of entropy content,
since $K=(1+\gamma\beta/2)^{-1}{\rm d}\ln G/{\rm d}\ln r$.  Within the flute
model, a high-$\beta$ pinch is thus locally a low-$\beta$ pinch whose dynamics
run more slowly, by the (modest) factor $(1+\gamma\beta/2)^{1/2}$.  This
offers a nonlinear, though radially local, justification for why the
self-organisation dynamics studied in the main text behave similarly at low
and high $\beta$; globally, $\beta$ also flattens the marginal profile, which
no single clock can absorb.

\subsubsection{A  global dipole--pinch map}

The radially local mapping above relied on rescaling $\varphi$ and time 
to map arbitrary geometry onto an equivalent pinch.
A radially non-local version must  attempt to match the radial variation of the two
scale factors in \cref{eq:localmap}; to do so, it is helpful to define an equivalent 
pinch radial coordinate. Insofar as the coefficients in the global system stay nearly constant in this coordinate, any global dipole system can be mapped onto its equivalent pinch.

We preserve the differential tube volume.  For a pinch,
$\mathcal V_{\rm p}=2\pi r/B_\theta$ and
${\rm d}\psi=B_\theta\,{\rm d}r$, and hence
${\rm d}r^2/{\rm d}\psi=\mathcal V_{\rm p}/\pi$.  We therefore define
\begin{equation}
 r_{\rm eff}^2(\psi)=r_{{\rm eff},{\rm in}}^2
 +\frac{1}{\pi}\int_{\psi_{\rm in}}^\psi
 \mathcal V(\psi')\,{\rm d}\psi',\qquad
 B_{\rm eff}\equiv\frac{2\pi r_{\rm eff}}{\mathcal V},\qquad
 n_{\mathcal V}\equiv
 \frac{{\rm d}\ln\mathcal V}{{\rm d}\ln r_{\rm eff}}.             \label{eq:reff}
\end{equation}
Thus ${\rm d}r_{\rm eff}/{\rm d}\psi=B_{\rm eff}^{-1}$ and
${\rm d}\mathcal V/{\rm d}\psi=n_{\mathcal V}\mathcal V/
(r_{\rm eff}B_{\rm eff})$.  Since this pinch shares both $\mathcal V$ and
${\rm d}\mathcal V/{\rm d}\psi$, taking $B_{\rm p}=B_{\rm eff}$ and
$\sigma_t=B_{\rm p}\sqrt b$ in \cref{eq:localmap} gives the radius-dependent scale factors
\begin{equation}
 \sigma_t=B_{\rm eff}\sqrt b
 =\sqrt{Y_\psi}\,\frac{B_{\rm eff}}{R_{\rm eq}B_{\rm eq}},\qquad
 \sigma_z=\frac{\sqrt a}{B_{\rm eff}\sqrt b}
 =\sqrt{\frac{Y_\varphi}{Y_\psi}}\,
 \frac{R_{\rm eq}^2B_{\rm eq}}{B_{\rm eff}}.                      \label{eq:globalscales}
\end{equation}
A global comparison chooses one reference surface, denoted by a star, and
uses its constants everywhere:
\begin{equation}
 z=\sigma_{z*}\varphi,\qquad t=\sigma_{t*}t_{\rm p},\qquad
 \Phi_{\rm p}=\sigma_{t*}\sigma_{z*}\Phi,\qquad
 w_{\rm p}=\frac{\sigma_{t*}\sigma_{z*}}{a_*}\hat w .                  \label{eq:globalmap}
\end{equation}
Putting $r\equiv r_{\rm eff}$, the complete radially global system becomes
\begin{gather}
 \frac{\partial\hat\rho}{\partial t_{\rm p}}+\frac{1}{B_{\rm eff}}\{\Phi_{\rm p},\hat\rho\}_{r,z}=0,\qquad
 \frac{\partial G}{\partial t_{\rm p}}+\frac{1}{B_{\rm eff}}\{\Phi_{\rm p},G\}_{r,z}=0,\qquad
 p=G^\gamma\mathcal V^{-\gamma},                                     \label{eq:mappedcontent}\\
 \frac{\partial w_{\rm p}}{\partial t_{\rm p}}+\frac{1}{B_{\rm eff}}\{\Phi_{\rm p},w_{\rm p}\}_{r,z}
 -\frac{1}{B_{\rm eff}}\big\{\hat\rho,\tfrac12u_{\rm p}^2\big\}_{r,z}
 =-\frac{n_{\mathcal V}\mathcal V^{1-\gamma}}
 {rB_{\rm eff}}\,\frac{\partial G^\gamma}{\partial z},              \label{eq:mappedvorticityeq}\\
 w_{\rm p}=\frac{1}{B_{\rm eff}}\frac{\partial}{\partial r}\!\left(
   \frac{\hat\rho\,Y_r}{B_{\rm eff}}\frac{\partial\Phi_{\rm p}}{\partial r}\right)
 +\frac{1}{B_{\rm eff}^{2}}\frac{\partial}{\partial z}\!\left(
   \hat\rho\,Y_z\frac{\partial\Phi_{\rm p}}{\partial z}\right),     \label{eq:mappedpinch}
\end{gather}
where $u_{\rm p}^2=B_{\rm eff}^{-2}\big[Y_r(\partial\Phi_{\rm p}/\partial r)^2
+Y_z(\partial\Phi_{\rm p}/\partial z)^2\big]$ and
\begin{equation}
 Y_z(r)\equiv\left(\frac{\sigma_t}{\sigma_{t*}}\right)^2,
 \qquad
 Y_r(r)\equiv
 \left(\frac{\sigma_t\sigma_z}
 {\sigma_{t*}\sigma_{z*}}\right)^2 .                              \label{eq:globalweights}
\end{equation}
 By
construction,
$\mathcal V_{\rm p}(r_{\rm eff})=2\pi r_{\rm eff}/B_{\rm eff}
=\mathcal V(\psi(r_{\rm eff}))$, so the original geometry and its effective
pinch share $\mathcal V(r_{\rm eff})$, $B_{\rm eff}(r_{\rm eff})$ and, since
$n_{\mathcal V}\mathcal V/r_{\rm eff}={\rm d}\mathcal V/{\rm d}r_{\rm eff}$,
the drive.  Setting $Y_r=Y_z=1$ in
\cref{eq:mappedcontent,eq:mappedvorticityeq,eq:mappedpinch} therefore gives
exactly the pinch system
\cref{eq:pinchcontent,eq:pinchvorticityeq,eq:pinchvorticity} with
$B_\theta=B_{\rm eff}$, nonlinear terms included.  The explicit $B_{\rm eff}$
factors are not absorbed into the clock but rather capture the global
field-profile physics, while $Y_r$ and $Y_z$ are residual metric factors from
the geometry of the line.

\begin{figure}
  \centering
  \includegraphics[width=0.98\textwidth]{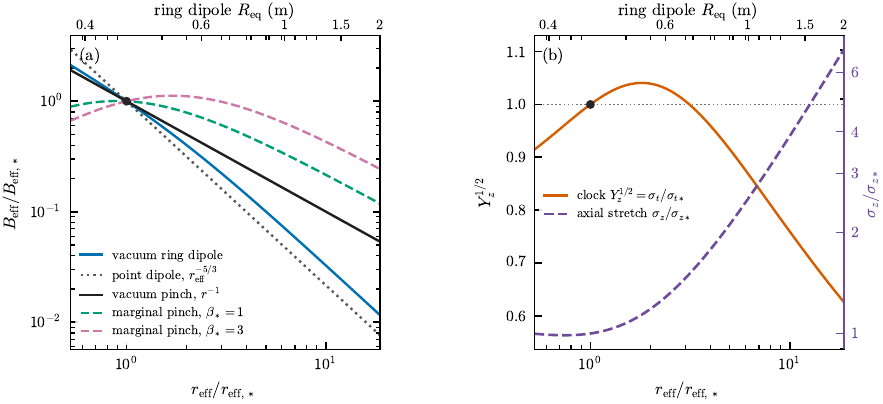}
  \caption{The global dipole--pinch comparison.  (\textit{a}) Effective field
  profiles $B_{\rm eff}(r_{\rm eff})$, normalised at the reference surface
  $R_{{\rm eq},*}=1.30R_c$, where $R_{\rm eq}$ is the dipole's
  outer-equatorial radius and $R_c=0.34\,{\rm m}$ is the LDX current-centroid
  radius \citep{boxer10}.  The vacuum circular-loop
  dipole is shown with its point-dipole limit, the vacuum pinch, and two
  finite-$\beta$ marginal pinch profiles, where
  $\beta_*=\beta(r_{{\rm eff},*})$.  For the vacuum ring dipole, the traced domain extends to
  $R_{\rm eq}=2\,{\rm m}$.  A finite-$\beta$ dipole is expected to be flatter
  than the vacuum ring-dipole curve but is not plotted because it requires a
  self-consistent equilibrium.  (\textit{b}) The residual geometry of the
  vacuum ring dipole's map onto its equivalent pinch: the local clock factor
  $Y_z^{1/2}$, \cref{eq:localclocks}, by which its interchange dynamics run
  slower than those of the pure pinch, and (purple, right axis) the
  angle-to-length stretch $\sigma_z/\sigma_{z*}$ of the axial domain, both
  relative to the reference surface.  Upper
  axes give the corresponding LDX outer-equatorial radius.  The line averages
  assume uniform density along each field line and include the inertia of the
  parallel flow.}
  \label{fig:flutegeometry}
\end{figure}
\FloatBarrier

The specific $B_{\rm eff}(r_{\rm eff})$ that arises is fixed by the geometry. 
For instance, a point dipole maps to
$B_{\rm eff}\propto r_{\rm eff}^{-5/3}$ and thus $n_{\mathcal V}=8/3$, while, as discussed 
above, a  vacuum pinch has $B_\theta\propto r^{-1}$ ($n_{\mathcal V}=2$) and 
a finite-$\beta$ pinch is yet flatter still ($n_{\mathcal V}=4/(2+\gamma\beta)$).
Thus, as far as its $B_{\rm eff}$ is concerned, a vacuum  dipole will behave like a 
pinch with a modestly steeper radial $B$ profile. Interestingly, a self-consistent finite-$\beta$ dipole equilibrium 
is likely closer to a vacuum pinch profile (through the same diamagnetic modification of its vacuum field that flattens the pinch),
although computing its actual $B_{\rm eff}(r_{\rm eff})$ profile
 requires a Grad--Shafranov equilibrium. 
 \Cref{fig:flutegeometry}\emph{(a)} plots $B_{\rm eff}(r_{\rm eff})$ for various relevant fields, including 
 the physical ring dipole of LDX \citep{boxer10}.

The global factors that remain after fixing $B_{\rm eff}$ are the geometry weights
$Y_r$ and $Y_z$, which are unity in the pure equivalent pinch.  To understand their effect, 
we consider the rescaling that would leave them constant.  Since
\cref{eq:globalweights} gives $Y_r=Y_z(\sigma_z/\sigma_{z*})^2$, stretching the
axial coordinate to $z'=(\sigma_z/\sigma_{z*})z=\sigma_z\varphi$ would make the
inversion isotropic, $Y_r\partial_r^2+Y_z\partial_z^2\to
Y_z(\sigma_z/\sigma_{z*})^2(\partial_r^2+\partial_{z'}^2)$, while rescaling
$t_{\rm p}$ by $Y_z^{1/2}$ (with $\Phi_{\rm p}$ and $w_{\rm p}$ rescaled
accordingly) then removes $Y_z$ as well.  Locally, therefore,
\cref{eq:mappedcontent,eq:mappedvorticityeq,eq:mappedpinch} are exactly the
pure-pinch equations ($Y_r=Y_z=1$), posed in a box of local length
$2\pi\sigma_z(r_{\rm eff})$ rather than $2\pi\sigma_{z*}$ and running slower by the radially varying clock
factor
\begin{equation}
 Y_z^{1/2}=\frac{\sigma_t}{\sigma_{t*}}
 =\frac{\sqrt{Y_\psi}\,B_{\rm eff}/(R_{\rm eq}B_{\rm eq})}
 {\big[\sqrt{Y_\psi}\,B_{\rm eff}/(R_{\rm eq}B_{\rm eq})\big]_*}. \label{eq:localclocks}
\end{equation}
For instance, a linear mode with radial wavenumber $k_r$ and toroidal number
$n$ grows at $Y_z^{-1/2}$ times the rate of the pure-pinch mode with the same
$k_r$ and $n$.  The influence of metric variation thus splits
into a clock error, $Y_z^{1/2}$, and an axial stretch,
$\sigma_z/\sigma_{z*}$, at each radius.

Both are shown in \cref{fig:flutegeometry}(\textit{b}) for the vacuum LDX ring
dipole.  Over this wide range,
$r_{\rm eff}/r_{{\rm eff},*}=0.53$--$18.65$, across which $B_{\rm eff}$ falls
by a factor of 180 (panel \textit{a}), the clock stays between $0.63$ and
$1.04$, so a global pinch calculation on one clock reproduces the dipole's
local interchange times to within 40\%.  The larger residual is the stretch,
$\sigma_z/\sigma_{z*}=0.99$--$7.0$: a fixed toroidal angle thus
corresponds to an equivalent-pinch length that grows outward, so a pure pinch
would need a box that expands with radius to match the dipole.  This is probably less consequential
dynamically than the raw factor suggests, because the axial spectra in
\cref{fig:globality}(\textit{b}) are insensitive to box length once the box is
long enough to contain the energy-carrying modes.

We thus see that the dominant geometry-dependent input is likely the effective field profile $B_{\rm eff}$ in
\cref{fig:flutegeometry}(\textit{a}), whose effect the $\beta$ scan of the
main text already probes (albeit for profiles flatter, rather than steeper, than the vacuum pinch); the clock varies only
 modestly, while the axial stretch is likely unimportant to the dynamics. This map is exact
within Pastukhov's ordering but does not include physics outside this ordering, which our full MHD 
simulations retain: compressibility at order-unity turbulent Mach number and
the self-consistent evolution of the background field. 
\FloatBarrier

\subsection{Linear field-line bending}\label{app:dipoleballooning}

We now use the high-$n$ ideal-MHD formalism of \citet{garnier99} to ask a
single linear question: is the field-line-dependent transition observed in
\cref{sec:3d} easier or harder to excite in a dipole?  
The results show that the ballooning-like $\beta$ limit of $0.4$, explored nonlinearly in 
\cref{sec:3d}, is much larger in a dipole than in the pinch, because
a dipole perturbation gets pinned to zero at the
outer equator, where the curvature drive is strongest, and must bend along
the rest of the line where the field is stiffer.  The results 
suggest that dipoles are likely immune to such instabilities  for 
useful operating regimes.

\subsubsection{Garnier's equation and the odd branch}

As above, $R$ is the cylindrical radius,
$\hat{\bm{b}}={\boldsymbol B}_0/B_0$, and
\begin{equation}
 \boldsymbol{\kappa}\equiv
 \hat{\bm{b}}\boldsymbol{\cdot}\boldsymbol{\nabla}\hat{\bm{b}}
 =\kappa_\psi\boldsymbol{\nabla}\psi,
 \qquad p_\psi\equiv\frac{{\rm d}p}{{\rm d}\psi},\qquad
 \beta(\ell)\equiv\frac{8\pi p}{B_0^2(\ell)} .                         \label{eq:balloondefs}
\end{equation}
 In Garnier's eikonal notation, the perpendicular
displacement is
\begin{equation}
 \boldsymbol{\xi}_\perp=\mathop{\rm Re}\!
 \left[\boldsymbol{\eta}_{\perp0}\exp({\rm i}S)\right],\qquad
 \boldsymbol{k}_\perp\equiv\boldsymbol{\nabla}S,\qquad
 {\boldsymbol B}_0\boldsymbol{\cdot}\boldsymbol{\nabla}S=0,\qquad
 \boldsymbol{\eta}_{\perp0}
 =\frac{X(\ell)}{B_0}\hat{\bm{b}}\mathbin{\times}\boldsymbol k_\perp ,
                                                                        \label{eq:garnierdisplacement}
\end{equation}
where $X$ is the arbitrarily normalised field-line envelope of the perturbation.  In an
axisymmetric shearless field,
$S=n\varphi+S_\psi(\psi)$ and minimising the positive bending energy over
${\rm d}S_\psi/{\rm d}\psi$ gives $\boldsymbol k_\perp=n\boldsymbol\nabla\varphi$ at $n\gg1$.
Substitution of \cref{eq:garnierdisplacement} into the ideal-MHD energy
principle, minimisation over magnetic and plasma compression, and variation
with respect to $X$ give the full closed-line ballooning equation of
\citet{garnier99}:
\begin{equation}
 B_0\frac{{\rm d}}{{\rm d}\ell}\!\left(
   \frac{1}{B_0R^2}\frac{{\rm d}X}{{\rm d}\ell}\right)
 +8\pi p_\psi\kappa_\psi X
 -\frac{16\pi\gamma p\kappa_\psi
 \langle X\kappa_\psi\rangle_\parallel}
 {1+\gamma\langle\beta\rangle_\parallel/2}=0 .                         \label{eq:garnierfull}
\end{equation}
Both $X$ and ${\rm d}X/{\rm d}\ell$ are periodic and $p(\psi)$ is flute-like, so averaging
\cref{eq:garnierfull} gives
\begin{equation}
 \langle X\kappa_\psi\rangle_\parallel\,
 \mathcal I=0,\qquad
 \mathcal I\equiv
 \frac{2\gamma p\langle\kappa_\psi\rangle_\parallel}
 {1+\gamma\langle\beta\rangle_\parallel/2}-p_\psi .                 \label{eq:garnierconstraint}
\end{equation}
Interchange modes, with flute-like $X$, require
$\langle\kappa_\psi\rangle_\parallel\mathcal I\geq0$ for stability.\footnote{This marginality condition is the same
 as $\partial G/\partial\psi=0$ given in
\cref{app:dipoleflute}.  The equilibrium identity
${\rm d}\ln\mathcal V/{\rm d}\psi=-2\langle\kappa_\psi\rangle_\parallel+
4\pi p_\psi\langle B_0^{-2}\rangle_\parallel$, together with
$G^\gamma=p\mathcal V^\gamma$, converts either form into the other.}  In the pinch,
$\kappa_\psi=-1/(rB_\theta)$ and $\mathcal I=-\gamma pK/(rB_\theta)$, with $K$
from \cref{eq:kdef}, so this  criterion becomes $K\geq0$.

The final term in \cref{eq:garnierfull} is stabilising despite its minus
sign, which can be seen  by multiplying by $X/B_0$ and integrating around the field line,
yielding a quadratic form that contains the positive-semidefinite term
$16\pi\gamma p\mathcal V\langle X\kappa_\psi\rangle_\parallel^2/
(1+\gamma\langle\beta\rangle_\parallel/2)$.  An even-parity mode with a
non-zero curvature average therefore pays an extra compression energy; this
term cannot lower its stability eigenvalue.  An odd $X$ on a
reflection-symmetric line avoids it because $B_0$, $R$ and $\kappa_\psi$ are
even.  
Thus, the first
distinct eigenfunction is odd, with a node at each of the two equatorial
crossings \citep{simakov00}; it obeys
\begin{equation}
 B_0\frac{{\rm d}}{{\rm d}\ell}\!\left(
   \frac{1}{B_0R^2}\frac{{\rm d}X}{{\rm d}\ell}\right)
 +8\pi p_\psi\kappa_\psi X=0,\qquad
 X(0)=X(\ell_h)=0 .                                                  \label{eq:dipoleballooning}
\end{equation}
$\ell=0$ and $\ell=\ell_h$ are the outer and inner equatorial crossings,
respectively.

For the uniform-line Z-pinch, inserting $X\propto\exp({\rm i}m\theta)$ in
\cref{eq:dipoleballooning} and using ${\rm d}\ell=r\,{\rm d}\theta$,
$\hat{\bm{b}}=\hat{\bm e}_\theta$,
$\boldsymbol\kappa=-\hat{\bm e}_r/r$ and
${\rm d}\psi=B_\theta\,{\rm d}r$ gives
$m^2=-(8\pi r/B_\theta^2)\,{\rm d}p/{\rm d}r=D$; thus $m=1$ is marginal at
$\beta=2/(3\gamma)=0.4$, recovering the $k_r\ll k_z$ limit of
\cref{eq:bendingcriterion} in the main text.

In true  LDX geometry, which is up--down asymmetric, \citet{garnier99} allow the node to move and, as
expected,  find a branch with approximate odd-like symmetry to be the least stable.
To handle their specified finite-$\beta$ equilibrium, \citet{garnier99} multiply the
pressure-gradient term by $\alpha$ while holding the equilibrium fixed.
The physical value is $\alpha=1$, so a marginal $\alpha>1$ denotes
stability; on their representative $\beta=5$ line,
$\alpha\simeq70$ for the odd-like branch and $119$ for the even branch,
suggesting the LDX geometry remains stable at even extreme values of $\beta$.
 \citet{Testa2026} find examples of less stable equilibria.

\subsubsection{Vacuum-ring estimate}

To provide a direct comparison to the Z-pinch results, we consider 
a vacuum circular-loop field (as in \cref{fig:flutegeometry}) and vary the pressure
 in the stability equation to test the effect of $\beta$.  As in \cref{app:dipoleflute},    let $R_c$ be the coil radius, with
$R_{\rm eq}$ and $B_{\rm eq}=B_0(\ell=0)$ the outer-equatorial radius and field.
For each field line, define the
dimensionless geometrical scalar $s$ and drive shape
$\mathcal C_\psi(\ell)$ by
\begin{equation}
\begin{aligned}
 s&\equiv\frac{\gamma}{2}B_{\rm eq}^2
 \langle B_0^{-2}\rangle_\parallel,\\
 \mathcal C_\psi(\ell)&\equiv
 2\gamma B_{\rm eq}^2\kappa_\psi(\ell)
 \langle\kappa_\psi\rangle_\parallel .
\end{aligned}                                                       \label{eq:vacuumgeometry}
\end{equation}
Following \citet{garnier99}, and motivated by the results of the main text, we choose the steepest interchange-stable
profile by setting $\mathcal I=0$.  This fixes the pressure gradient and,
using \cref{eq:vacuumgeometry}, gives
\begin{equation}
\begin{aligned}
 p_\psi&=\frac{2\gamma p\langle\kappa_\psi\rangle_\parallel}
 {1+\gamma\langle\beta\rangle_\parallel/2},\\
 8\pi p_\psi\kappa_\psi
 &=\frac{\beta_{\rm eq}}{1+s\beta_{\rm eq}}\,
 \mathcal C_\psi(\ell).
\end{aligned}                                                       \label{eq:marginaldrive}
\end{equation}
Substitution into \cref{eq:dipoleballooning} gives 
\begin{equation}
 B_0\frac{{\rm d}}{{\rm d}\ell}\!\left(
 \frac{1}{B_0R^2}\frac{{\rm d}X}{{\rm d}\ell}\right)
 +\frac{\beta_{\rm eq}}{1+s\beta_{\rm eq}}\,
 \mathcal C_\psi(\ell)X=0,\qquad X(0)=X(\ell_h)=0 .                 \label{eq:ringballooning}
\end{equation}
Here $\beta_{\rm eq}$ is directly the stability parameter; unlike \citet{garnier99}, no
auxiliary multiplier is needed because the field geometry is not self-consistent with changing $\beta$.
  We define
$\beta_{\rm eq}^*$ as its smallest positive value for which
\cref{eq:ringballooning} has a non-zero solution. 

\begin{figure}
  \centerline{\includegraphics[width=0.94\textwidth]{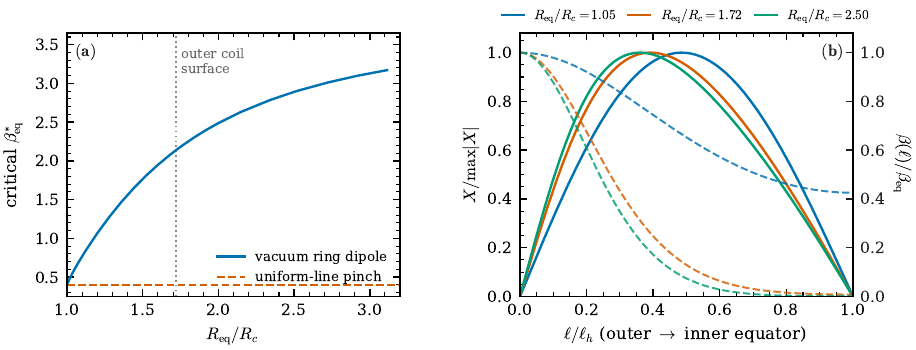}}
  \caption{Ballooning-mode stability on vacuum circular-loop (physical dipole) field lines
  for $\gamma=5/3$.  (\textit{a}) Critical outer-equatorial
  $\beta_{\rm eq}^{*}$ for an interchange-marginal pressure profile.  The
  dashed line is the uniform-line pinch limit $2/(3\gamma)$, which the
  computed curve reaches only very close to the idealised coil; the vertical line marks the
  outer radial surface of the LDX floating-coil structure, by which point $\beta_{\rm eq}^{*}$ has already risen substantially.  Values above
  $\beta\sim0.5$ are only  indicative because the geometry does not include the plasma pressure at finite $\beta$.
  (\textit{b}) First odd half-line eigenfunctions (solid, left axis) and the
  corresponding $\beta(\ell)/\beta_{\rm eq}$ (dashed, right axis), showing
  how the odd parity forces the mode to be localised in lower-$\beta$ regions, explaining 
  the enhanced stability compared to the pinch.}
  \label{fig:dipole}
\end{figure}

We trace each vacuum line, integrate
\eqref{eq:ringballooning} outward from $X(0)=0$, bisect on ${\beta_{\rm eq}}/({1+s\beta_{\rm eq}})$
until $X(\ell_h)=0$, and use this to solve for $\beta_{\rm eq}^{*}$.  As shown in \cref{fig:dipole}(\emph{a}), the threshold reaches the pinch value
$2/(3\gamma)$ as $R_{\rm eq}\rightarrow R_c$, but rises rapidly once the line samples appreciable variation of
$B_0$.  For instance, the outer radial surface of the LDX
floating-coil structure is at $R=0.585\,{\rm m}$, or
$R_{\rm eq}/R_c=1.72$ for $R_c=0.34\,{\rm m}$ \citep{garnier99,boxer10}, and the vacuum
calculation already gives $\beta_{\rm eq}^*\simeq2.1$ there.  \Cref{fig:dipole}(\emph{b}) shows the
eigenfunctions, illustrating how two things stabilise the dipole.  First, the odd displacement is
pinned to zero at the outer equator, which is exactly where $\kappa_\psi$
peaks, so the drive integral $\langle\kappa_\psi X^2\rangle_\parallel$ is
weighted away from its own maximum.  Second, as forced by its odd parity, the mode is localised
away from the high-$\beta$ outer midplane, seeing mostly much lower $\beta$ field regions on average. 

The vacuum curve is clearly idealised and should not be interpreted as a lower bound: pressure
changes $B_0$, $R$, $\kappa_\psi$ and $p_\psi$ together, and likely give equilibrium-dependent results. For example, 
 \citet{garnier99} find self-consistent Grad--Shafranov
equilibria that are yet more stable than seen here, while
the finite-pressure point-dipole equilibria of
\citet{Krasheninnikov1999} were  shown by \citet{simakov00}
to be   stable at arbitrary $\beta$.  However, other self-consistent equilibria have some
ballooning-unstable surfaces at peak flux-surface-averaged $\beta\simeq2$, comparable to the scale found here \citep{Testa2026}.  Nonetheless, all calculations 
of which we are aware
support the expectation of substantially greater dipole stability compared to the pinch,
meaning the destructive transitions in
\cref{fig:render,fig:3d} are likely  significantly harder to reach in a
dipole, including for  plasma fairly close to the coil.

\section{Linear theory and code validation}\label{app:numerics}

In this appendix, we outline the linear stability theory of the Z-pinch, as discussed in
\cref{sec:kadomtsev,sec:3dtheory}, and use this as a test for the code.  Both
criteria used in the main text are limits of a single eigenvalue problem, so we
state that problem first, reduce it, and then compare it with \textsc{Athena++} results.

Linearising ideal MHD about a static cylindrical equilibrium, taking
perturbations $\propto\exp({\rm i}m\theta+{\rm i}k_zz-{\rm i}\omega t)$ and
eliminating every component of the displacement in favour of $U=r\xi_r$ leaves
the Hain--L\"ust equation \citep{hain58},
\begin{equation}
 \frac{{\rm d}}{{\rm d}r}\left(A\frac{{\rm d}U}{{\rm d}r}\right)-CU=0 ,
 \label{eq:hl}
\end{equation}
where, for a pinch with no axial field
\citep[][\S11.5]{freidberg14}\footnote{The final term of
\citet{freidberg14}'s (11.134) appears to carry a factor of two less than follows from his
(11.131) and (11.132); we use the latter, which returns
\cref{eq:m0dispersion} and hence the standard interchange criterion.}
\begin{align}
 A&=\frac{\rho\left(c_s^{2}+v_{\rm A}^{2}\right)
   \left(\omega^{2}-\omega_{\rm A}^{2}\right)
   \left(\omega^{2}-\omega_h^{2}\right)}{r\,\mathcal{D}} ,
 \label{eq:hlA}\\
 C&=-\frac{\rho}{r}\left(\omega^{2}-\omega_{\rm A}^{2}\right)
   +\frac{B_\theta^{2}}{4\pi r^{3}}
    \frac{4k_z^{2}v_{\rm A}^{2}\left(\omega^{2}-\omega_g^{2}\right)}{\mathcal{D}}
   \notag\\
  &\quad+\frac{{\rm d}}{{\rm d}r}\left\{\frac{B_\theta^{2}}{4\pi r^{2}}
    \left[1+\frac{2k_z^{2}\left(c_s^{2}+v_{\rm A}^{2}\right)
     \left(\omega^{2}-\omega_h^{2}\right)}{\mathcal{D}}\right]\right\} ,
 \label{eq:hlC}
\end{align}
with
\begin{gather}
 \omega_{\rm A}^{2}=\frac{m^{2}v_{\rm A}^{2}}{r^{2}},\qquad
 \omega_h^{2}=\frac{c_s^{2}\omega_{\rm A}^{2}}{c_s^{2}+v_{\rm A}^{2}},\qquad
 \omega_g^{2}=\frac{c_s^{2}\omega_{\rm A}^{2}}{v_{\rm A}^{2}},\qquad
\mathcal{D}=\left(\omega^{2}-\omega_f^{2}\right)
  \left(\omega^{2}-\omega_s^{2}\right),
 \label{eq:hlfreqs}\\
  k_0^{2}=k_z^{2}+\frac{m^{2}}{r^{2}},\qquad
 \omega_{f,s}^{2}=\frac{k_0^{2}\left(c_s^{2}+v_{\rm A}^{2}\right)}{2}
  \left[1\pm\left(1-\frac{4c_s^{2}\omega_{\rm A}^{2}}
   {k_0^{2}\left(c_s^{2}+v_{\rm A}^{2}\right)^{2}}\right)^{1/2}\right] .
 \label{eq:hlfast}
\end{gather}
\noindent Here $\omega_{\rm A}$ is the Alfv\'en frequency of the azimuthal field and
$\omega_{f,s}$ are the local fast and slow magnetosonic frequencies.  The eigenvalue $\omega$ appears throughout, so \cref{eq:hl} is not a
Sturm--Liouville problem.

For $k_\perp\gg k_\parallel$, which here means $k_zr\gg m$, and for
$|\omega^{2}|\ll k_0^{2}(c_s^{2}+v_{\rm A}^{2})$, the fast branch decouples,
$\mathcal{D}\to-k_z^{2}(c_s^{2}+v_{\rm A}^{2})(\omega^{2}-\omega_h^{2})$, and the
coefficients simplify significantly,
\begin{equation}
 A=-\frac{\rho\left(\omega^{2}-\omega_{\rm A}^{2}\right)}{k_z^{2}r},\qquad
 C=-\frac{\rho}{r}\left(\omega^{2}-\omega_{\rm A}^{2}\right)
  -\frac{\rho v_{\rm A}^{2}}{r^{3}}
   \frac{4\left(v_{\rm A}^{2}\omega^{2}-c_s^{2}\omega_{\rm A}^{2}\right)}
   {\left(c_s^{2}+v_{\rm A}^{2}\right)\omega^{2}-c_s^{2}\omega_{\rm A}^{2}}
  -\frac{{\rm d}}{{\rm d}r}\frac{B_\theta^{2}}{4\pi r^{2}} .
 \label{eq:hlreduced}
\end{equation}
Writing $U\propto\exp({\rm i}\!\int\!k_r\,{\rm d}r)$ with $k_r$ slowly varying, to compute a local WKB dispersion relation,
turns \cref{eq:hl} into $Ak_r^{2}+C=0$, and force balance 
\eqref{eq:blogslope} then gives
\begin{equation}
 \left(\omega^{2}-\omega_{\rm A}^{2}\right)
 \left(1+\frac{k_r^{2}}{k_z^{2}}\right)
 =\frac{v_{\rm A}^{2}}{r^{2}}\left[4-\Dr
 -\frac{4\left(v_{\rm A}^{2}\omega^{2}-c_s^{2}\omega_{\rm A}^{2}\right)}
 {\left(c_s^{2}+v_{\rm A}^{2}\right)\omega^{2}-c_s^{2}\omega_{\rm A}^{2}}\right] ,
 \label{eq:dispersion}
\end{equation}
with $\Dr$ the bending drive of \cref{eq:bendingdrive}.

Two limits of \cref{eq:dispersion} are used in the paper.  At $m=0$, $\omega_{\rm A}=0$ so that
\begin{equation}
 \omega^{2}\left(1+\frac{k_r^{2}}{k_z^{2}}\right)=\frac{2c_s^{2}}{r^{2}}K ,
 \label{eq:m0dispersion}
\end{equation}
with $K$ as defined in \cref{eq:kdef}.  The mode is unstable wherever $K<0$,
which is the criterion of \citet{kadomtsev66}, and at $k_r\ll k_z$ its growth
rate is $\gamma_{\rm int}=(c_s/r)\sqrt{2|K|}$ as used in the main text.  For $m\geq1$ the term $\omega_{\rm A}^{2}$ dominates the left-hand
side instead, and $\omega^{2}=0$ leaves
\begin{equation}
 \Dr=m^{2}\left(1+\frac{k_r^{2}}{k_z^{2}}\right) ,
 \label{eq:m1marginal}
\end{equation}
which is the marginal condition used in \cref{eq:bendingcriterion}. The same results 
can be derived via the energy principle
\citep{kadomtsev66,freidberg14}.

\subsection{Measured growth rates}\label{app:linearvalidation}

We test  \textsc{Athena++} against the global linear Hain--L\"ust  solutions by 
initialising an unstable equilibrium and studying the perturbations that grow exponentially on top of this. 
For simplicity, we use the analytic family of \citet[][\S11.4.4]{freidberg14},
whose hard-core radius $r_c$ is where the pressure vanishes.  With
$x=r^{2}/r_c^{2}$ and quantities normalised to $B_*$,  this is
\begin{equation}
 \frac{8\pi p}{B_*^{2}}=\frac{3\beta_F}{4}\frac{x-1}{x^{5/2}},\qquad
 \frac{B_\theta^{2}}{B_*^{2}}=\frac{1}{x}
  -\frac{\beta_F}{4}\frac{9x-5}{x^{5/2}},\qquad
 \frac{\rho}{\rho_0}=\frac{2}{1+x^{2}} ,
 \label{eq:freidbergfamily}
\end{equation}
which satisfies the equilibrium \eqref{eq:forcebalance} exactly.  The parameter $\beta_F$ is a
convenient parameter to normalise the system  and the density is a free
choice, since interchange marginality does not involve it.  The pressure
vanishes at the core and peaks at $x=5/3$, and the family reaches an
equilibrium limit near $\beta_F=0.86$, beyond which $B_\theta$  vanishes in the
interior.  We take $\beta_F=0.70$, $0.80$ and $0.85$, which give local
$\beta_{\rm pk}=8\pi p/B_\theta^{2}=0.87$, $2.64$ and $15.94$ at the maximum and lie
above the $m=0$ marginal value $\beta_F=0.628$ of this family.  The
computational annulus is the production one, $\rin\leq r\leq6\rin$.  Its core
$r_c$ is placed just inside the domain's inner boundary at $r_c=0.8\rin$, so that
$p$, $\rho$ and $B_\theta^{2}$ remain positive in the ghost zones at $r<\rin$;
the pressure maximum then falls at $1.03\rin$.

\Cref{eq:hl} is solved on a uniform radial grid with $U=0$ at both walls.  For
a trial $\omega^{2}<0$ the operator $({\rm d}/{\rm d}r)(A\,{\rm d}/{\rm
d}r)-C$ is discretised as a symmetric tridiagonal matrix, and because $A$ and
$C$ themselves depend on $\omega^{2}$ the eigenvalue is found by scanning that
trial value and locating where the largest eigenvalue of the matrix crosses
zero.  This selects the fundamental unstable branch, the one with no interior
node.
\begin{figure}
  \centerline{\includegraphics[width=\textwidth]{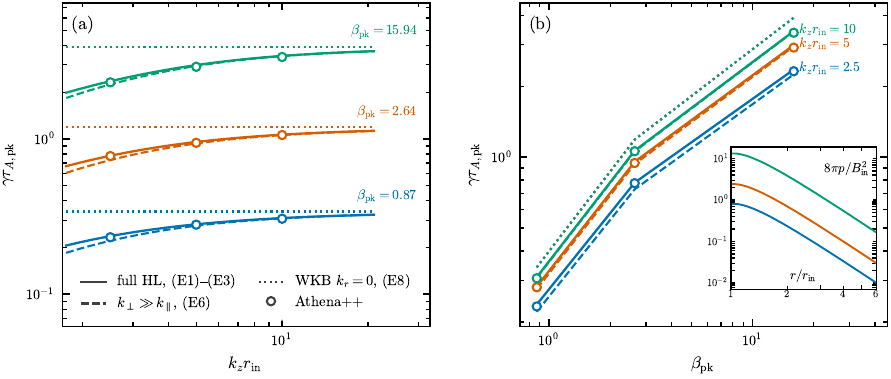}}
  \caption{Measured $m=0$ growth rates against the Hain--L\"ust eigenvalue for
  the source-free validation equilibria  \eqref{eq:freidbergfamily}.
  (\textit{a}) Against axial wavenumber at three values of the local $\beta$ at
  the pressure maximum and (\textit{b}) against that $\beta$ at three axial
  wavenumbers.  Solid curves solve \cref{eq:hl} with the full coefficients
  \cref{eq:hlA,eq:hlC}, dashed curves the $k_\perp\gg k_\parallel$ reduction
  \cref{eq:hlreduced}, and dotted lines the eikonal result
  \cref{eq:m0dispersion} at $k_r=0$, which is the large-$k_z$ asymptote.  Open
  circles are the measured rates at $N_r=96$.  The inset shows the three
  pressure profiles in matching colours.  Rates use the initial peak Alfv\'en
  curvature time of each equilibrium.}
  \label{fig:validation}
\end{figure}

\textsc{Athena++} is run without the sources and sinks used in the main text, initialised on
\cref{eq:freidbergfamily}.  One axial
period is retained by seeding a single $k_z$ of amplitude $10^{-6}$, and we run the 
system in 2-D to test  the $m=0$ interchange
mode (with two identical periodic $\theta$ cells, the minimum
\textsc{Athena++} allows with active $z$), on an $N_r\times N_z=96\times48$
grid.  We match the boundary condition of the eigenvalue problem ---
$\xi_r=0$ at both radii --- with an equilibrium-preserving, free-slip rigid
wall: the ghost zones hold the equilibrium \eqref{eq:freidbergfamily} continued analytically,
plus the perturbation of the mirror-image interior cell, taken odd for $u_r$
and $B_r$ (so these vanish on the wall) and even for all other variables.
Growth rates are then measured from the seeded axial Fourier component
projected onto the theoretical radial eigenfunction, and fitted once the
exponential growth  has stabilised but before the perturbation becomes
nonlinear.

\Cref{fig:validation} shows the outcome at $k_z\rin=2.5$, $5$ and $10$.  At
$N_r=96$ the measured rates lie within $0.4$--$2.4$ per cent of the
eigenvalue, with a mean absolute error of $1.3$ per cent, and the radial mode
shapes correlate with the theoretical eigenfunctions at $0.91$--$0.999$.  A
resolution series (not shown) scans from $N_r=64$, where the error is $4.8$\%, to
$N_r=192$, where it is $1.0$\%.  We also consider and plot two approximations to the full linear solution.   The $k_\perp\gg k_\parallel$ reduction \eqref{eq:hlreduced}, 
which eliminates fast-mode physics, 
 is low by $6.6$ per cent at $k_z\rin=2.5$ and by $0.3$ per
cent at $k_z\rin=10$, so the full coefficients are needed at the smaller
wavenumbers. The eikonal result \eqref{eq:m0dispersion} at $k_r=0$ is the
$k_z\rin\to\infty$ asymptote and still remains  ${\simeq}5$\% above the
true eigenvalue at the largest wavenumber shown.  
Overall, the comparison shows that \textsc{Athena++} and the setup we use are able to accurately capture 
interchange dynamics in the cylinder. 

\bibliographystyle{jpp}
\bibliography{zpinch_selforg_v5}

\begin{thebibliography}{61}
\expandafter\ifx\csname natexlab\endcsname\relax\def\natexlab#1{#1}\fi
\def\au#1{#1} \def\ed#1{#1} \def\yr#1{#1}\def\at#1{#1}\def\jt#1{\textit{#1}}
  \def\bt#1{#1}\def\bvol#1{\textbf{#1}} \def\vol#1{#1} \def\pg#1{#1}
  \def\publ#1{#1}\def\arxiv#1{#1}\def\org#1{#1}\def\st#1{\textit{#1}}

\bibitem[{Biskamp} \& {Welter}(1989)]{biskampwelter89}
{\sc \au{{Biskamp}, D.} \& \au{{Welter}, H.}} \yr{1989}  \at{{Dynamics of
  decaying two-dimensional magnetohydrodynamic turbulence}}.  \jt{Physics of
  Fluids B}  \bvol{1}~(10),  \pg{1964--1979}.

\bibitem[{Boxer} {\em et~al.\/}(2010){Boxer}, {Bergmann}, {Ellsworth},
  {Garnier}, {Kesner}, {Mauel} \& {Woskov}]{boxer10}
{\sc \au{{Boxer}, A.~C.}, \au{{Bergmann}, R.}, \au{{Ellsworth}, J.~L.},
  \au{{Garnier}, D.~T.}, \au{{Kesner}, J.}, \au{{Mauel}, M.~E.} \&
  \au{{Woskov}, P.}} \yr{2010}  \at{{Turbulent inward pinch of plasma confined
  by a levitated dipole magnet}}.  \jt{Nature Physics}  \bvol{6}~(3),
  \pg{207--212}.

\bibitem[{Chisholm} {\em et~al.\/}(2026){Chisholm}, {Berry}, {Garnier},
  {Badcock}, {Bioletti}, {Bouloukakis}, {Brewerton}, {Buchanan}, {Burt},
  {Chambers}, {Chappell}, {Coulson}, {Davidson}, {Ellingham}, {Geursen},
  {Hamilton}, {Hu}, {Hunter}, {Jones}, {Kusay}, {Lazi{\'c}}, {Leuw}, {Lynch},
  {Mataira}, {McCrohon}, {Meadows}, {Morris}, {Nowacki}, {Purvis}, {Rice},
  {Rutten}, {Schimanski}, {Sharma}, {Siamaki}, {Simpson}, {Simpson}, {Smith},
  {Stiers}, {Swanson-Dobbs}, {Todd}, {Treacher}, {Tyler}, {Venuturumilli},
  {Weijers}, {Wordsworth} \& {Zhou}]{Chisholm2026}
{\sc \au{{Chisholm}, C.~S.}, \au{{Berry}, T.}, \au{{Garnier}, D.~T.},
  \au{{Badcock}, R.~A.}, \au{{Bioletti}, G.}, \au{{Bouloukakis}, K.},
  \au{{Brewerton}, E.}, \au{{Buchanan}, M.~A.}, \au{{Burt}, P.~J.},
  \au{{Chambers}, E.~V.~W.}, \au{{Chappell}, K.~B.}, \au{{Coulson}, P.},
  \au{{Davidson}, R.~J.}, \au{{Ellingham}, J.~P.~M.}, \au{{Geursen}, P.},
  \au{{Hamilton}, K.}, \au{{Hu}, R.}, \au{{Hunter}, E.}, \au{{Jones}, J.~P.},
  \au{{Kusay}, P.}, \au{{Lazi{\'c}}, Z.}, \au{{Leuw}, B.}, \au{{Lynch}, M.},
  \au{{Mataira}, R.}, \au{{McCrohon}, M.}, \au{{Meadows}, L.}, \au{{Morris},
  J.~R.}, \au{{Nowacki}, R.}, \au{{Purvis}, J.~V.}, \au{{Rice}, J.~H.~P.},
  \au{{Rutten}, M.}, \au{{Schimanski}, S.}, \au{{Sharma}, A.}, \au{{Siamaki},
  M.}, \au{{Simpson}, A.}, \au{{Simpson}, T.}, \au{{Smith}, B.}, \au{{Stiers},
  E.}, \au{{Swanson-Dobbs}, E.}, \au{{Todd}, J.}, \au{{Treacher}, E.~O.~P.},
  \au{{Tyler}, J.~D.}, \au{{Venuturumilli}, S.}, \au{{Weijers}, H.~W.},
  \au{{Wordsworth}, T.} \& \au{{Zhou}, N.}} \yr{2026}  \at{{Design and initial
  results from the ``Junior'' Levitated Dipole Experiment}}.  \jt{Fusion
  Engineering and Design}  \bvol{223},  \pg{115551}.

\bibitem[{Crews} {\em et~al.\/}(2024){Crews}, {Datta}, {Meier} \&
  {Shumlak}]{Crews2024}
{\sc \au{{Crews}, Daniel~W.}, \au{{Datta}, Iman A.~M.}, \au{{Meier}, Eric~T.}
  \& \au{{Shumlak}, Uri}} \yr{2024}  \at{{The Kadomtsev Pinch Revisited for
  Sheared-Flow-Stabilized Z-Pinch Modeling}}.  \jt{IEEE Transactions on Plasma
  Science}  \bvol{52}~(10),  \pg{4804--4816},  \arxiv{arXiv: 2404.06636}.

\bibitem[{Davis} {\em et~al.\/}(2014){Davis}, {Mauel}, {Garnier} \&
  {Kesner}]{Davis2014}
{\sc \au{{Davis}, Matthew~S.}, \au{{Mauel}, M.~E.}, \au{{Garnier}, Darren~T.}
  \& \au{{Kesner}, Jay}} \yr{2014}  \at{{Pressure profiles of plasmas confined
  in the field of a magnetic dipole}}.  \jt{Plasma Physics and Controlled
  Fusion}  \bvol{56}~(9),  \pg{095021}.

\bibitem[{Freidberg}(2014)]{freidberg14}
{\sc \au{{Freidberg}, J.~P.}} \yr{2014} {\em {Ideal MHD}\/}.  \publ{Cambridge:
  Cambridge University Press}.

\bibitem[{Garnier} {\em et~al.\/}(2009){Garnier}, {Boxer}, {Ellsworth},
  {Kesner} \& {Mauel}]{Garnier2009}
{\sc \au{{Garnier}, D.~T.}, \au{{Boxer}, A.~C.}, \au{{Ellsworth}, J.~L.},
  \au{{Kesner}, J.} \& \au{{Mauel}, M.~E.}} \yr{2009}  \at{{Confinement
  improvement with magnetic levitation of a superconducting dipole}}.
  \jt{Nuclear Fusion}  \bvol{49}~(5),  \pg{055023}.

\bibitem[Garnier {\em et~al.\/}(2006)Garnier, Hansen, Mauel, Ortiz, Boxer,
  Ellsworth, Karim, Kesner, Mahar \& Roach]{Garnier2006}
{\sc \au{Garnier, D.~T.}, \au{Hansen, A.}, \au{Mauel, M.~E.}, \au{Ortiz, E.},
  \au{Boxer, A.~C.}, \au{Ellsworth, J.}, \au{Karim, I.}, \au{Kesner, J.},
  \au{Mahar, S.} \& \au{Roach, A.}} \yr{2006}  \at{Production and study of
  high-beta plasma confined by a superconducting dipole magnet}.  \jt{Physics
  of Plasmas}  \bvol{13}~(5),  \pg{056111}.

\bibitem[{Garnier} {\em et~al.\/}(1999){Garnier}, {Kesner} \&
  {Mauel}]{garnier99}
{\sc \au{{Garnier}, D.~T.}, \au{{Kesner}, J.} \& \au{{Mauel}, M.~E.}} \yr{1999}
   \at{{Magnetohydrodynamic stability in a levitated dipole}}.  \jt{Physics of
  Plasmas}  \bvol{6}~(9),  \pg{3431--3434}.

\bibitem[{Garnier} {\em et~al.\/}(2017){Garnier}, {Mauel}, {Roberts}, {Kesner}
  \& {Woskov}]{garnier17}
{\sc \au{{Garnier}, D.~T.}, \au{{Mauel}, M.~E.}, \au{{Roberts}, T.~M.},
  \au{{Kesner}, J.} \& \au{{Woskov}, P.~P.}} \yr{2017}  \at{{Turbulent
  fluctuations during pellet injection into a dipole confined plasma torus}}.
  \jt{Physics of Plasmas}  \bvol{24}~(1),  \pg{012506}.

\bibitem[{Goldreich} \& {Sridhar}(1995)]{goldreich95}
{\sc \au{{Goldreich}, P.} \& \au{{Sridhar}, S.}} \yr{1995}  \at{{Toward a
  theory of interstellar turbulence. II. Strong Alfv\'enic turbulence}}.
  \jt{The Astrophysical Journal}  \bvol{438},  \pg{763--775}.

\bibitem[{Grierson} {\em et~al.\/}(2010){Grierson}, {Mauel}, {Worstell} \&
  {Klassen}]{Grierson2010}
{\sc \au{{Grierson}, B.~A.}, \au{{Mauel}, M.~E.}, \au{{Worstell}, M.~W.} \&
  \au{{Klassen}, M.}} \yr{2010}  \at{{Transport Induced by Large Scale
  Convective Structures in a Dipole-Confined Plasma}}.  \jt{Physical Review
  Letters}  \bvol{105}~(20),  \pg{205004}.

\bibitem[{Grierson} {\em et~al.\/}(2009){Grierson}, {Worstell} \&
  {Mauel}]{Grierson2009}
{\sc \au{{Grierson}, B.~A.}, \au{{Worstell}, M.~W.} \& \au{{Mauel}, M.~E.}}
  \yr{2009}  \at{{Global and local characterization of turbulent and chaotic
  structures in a dipole-confined plasma}}.  \jt{Physics of Plasmas}
  \bvol{16}~(5),  \pg{055902}.

\bibitem[Guazzotto {\em et~al.\/}(2007)Guazzotto, Freidberg \&
  Kesner]{Guazzotto2007}
{\sc \au{Guazzotto, L.}, \au{Freidberg, J.~P.} \& \au{Kesner, J.}} \yr{2007}
  \at{Equilibrium beta limits in a dipole configuration}.  \jt{Physics of
  Plasmas}  \bvol{14}~(6),  \pg{062501}.

\bibitem[{Hain} \& {L\"ust}(1958)]{hain58}
{\sc \au{{Hain}, K.} \& \au{{L\"ust}, R.}} \yr{1958}  \at{{Zur Stabilit\"at
  zylindersymmetrischer Plasmakonfigurationen mit Volumenstr\"omen}}.
  \jt{Zeitschrift f\"ur Naturforschung A}  \bvol{13}~(11),  \pg{936--940}.

\bibitem[{Hallenbert} \& {Plunk}(2022)]{Hallenbert2022}
{\sc \au{{Hallenbert}, Axel} \& \au{{Plunk}, Gabriel~G.}} \yr{2022}
  \at{{Predicting the Z-pinch Dimits shift through gyrokinetic tertiary
  instability analysis of the entropy mode}}.  \jt{Journal of Plasma Physics}
  \bvol{88}~(4),  \pg{905880402},  \arxiv{arXiv: 2111.08121}.

\bibitem[{Hasegawa}(1987)]{hasegawa87}
{\sc \au{{Hasegawa}, A.}} \yr{1987}  \at{{A dipole field fusion reactor}}.
  \jt{Comments on Plasma Physics and Controlled Fusion}  \bvol{11}~(3),
  \pg{147--151}.

\bibitem[{Hasegawa} {\em et~al.\/}(1990){Hasegawa}, {Chen} \&
  {Mauel}]{Hasegawa1990}
{\sc \au{{Hasegawa}, Akira}, \au{{Chen}, Liu} \& \au{{Mauel}, M.~E.}} \yr{1990}
   \at{{A D--$^{3}$He fusion reactor based on a dipole magnetic field}}.
  \jt{Nuclear Fusion}  \bvol{30}~(11),  \pg{2405--2413}.

\bibitem[{Hasegawa} {\em et~al.\/}(1992){Hasegawa}, {Chen}, {Mauel}, {Warren}
  \& {Murakami}]{Hasegawa1992}
{\sc \au{{Hasegawa}, Akira}, \au{{Chen}, Liu}, \au{{Mauel}, Michael~E.},
  \au{{Warren}, Harry~H.} \& \au{{Murakami}, Sadayoshi}} \yr{1992}  \at{{A
  Description of a D--$^{3}$He Fusion Reactor Based on a Dipole Magnetic
  Field}}.  \jt{Fusion Technology}  \bvol{22}~(1),  \pg{27--34}.

\bibitem[{Hassam} \& {Kulsrud}(1978)]{hassam78}
{\sc \au{{Hassam}, A.~B.} \& \au{{Kulsrud}, R.~M.}} \yr{1978}  \at{{Time
  evolution of mass flows in a collisional tokamak}}.  \jt{Physics of Fluids}
  \bvol{21}~(12),  \pg{2271--2279}.

\bibitem[{Hoffmann} {\em et~al.\/}(2023){Hoffmann}, {Frei} \&
  {Ricci}]{Hoffmann2023}
{\sc \au{{Hoffmann}, A.~C.~D.}, \au{{Frei}, B.~J.} \& \au{{Ricci}, P.}}
  \yr{2023}  \at{{Gyrokinetic simulations of plasma turbulence in a Z-pinch
  using a moment-based approach and advanced collision operators}}.
  \jt{Journal of Plasma Physics}  \bvol{89}~(2),  \pg{905890214},
  \arxiv{arXiv: 2208.01346}.

\bibitem[{Ishizawa} \& {Hattori}(1998)]{ishizawa98}
{\sc \au{{Ishizawa}, A.} \& \au{{Hattori}, Y.}} \yr{1998}  \at{{Wavelet
  analysis of two-dimensional MHD turbulence}}.  \jt{Journal of the Physical
  Society of Japan}  \bvol{67}~(2),  \pg{441--450}.

\bibitem[{Isichenko} {\em et~al.\/}(1995){Isichenko}, {Gruzinov} \&
  {Diamond}]{isichenko95}
{\sc \au{{Isichenko}, M.~B.}, \au{{Gruzinov}, A.~V.} \& \au{{Diamond}, P.~H.}}
  \yr{1995}  \at{{Invariant Measure and Turbulent Pinch in Tokamaks}}.
  \jt{Physical Review Letters}  \bvol{74}~(22),  \pg{4436--4439}.

\bibitem[{Ivanov} {\em et~al.\/}(2025){Ivanov}, {Luhadiya}, {Adkins} \&
  {Schekochihin}]{ivanov25}
{\sc \au{{Ivanov}, P.~G.}, \au{{Luhadiya}, P.}, \au{{Adkins}, T.} \&
  \au{{Schekochihin}, A.~A.}} \yr{2025}  \at{{The gyrokinetic field invariant
  and electromagnetic temperature-gradient instabilities in `good-curvature'
  plasmas}}.  \jt{Journal of Plasma Physics}  \bvol{91}~(4),  \pg{E95}.

\bibitem[{Jackson}(1999)]{jackson99}
{\sc \au{{Jackson}, J.~D.}} \yr{1999} {\em {Classical Electrodynamics}\/}, 3rd
  edn.  \publ{New York: Wiley}.

\bibitem[{Jha} {\em et~al.\/}(2026){Jha}, {Verma}, {Nirgudkar} \&
  {Verma}]{jha26}
{\sc \au{{Jha}, A.~K.}, \au{{Verma}, M.}, \au{{Nirgudkar}, S.} \& \au{{Verma},
  M.~K.}} \yr{2026}  \at{{Spectral properties of anisotropic two-dimensional
  magnetohydrodynamic turbulence}}.  \jt{Physical Review E}  \bvol{114}~(1),
  \pg{015223}.

\bibitem[{Jiang} {\em et~al.\/}(2025){Jiang}, {Wang}, {Ou}, {Zhou}, {Li} \&
  {Mauel}]{Jiang2025}
{\sc \au{{Jiang}, Changzhi}, \au{{Wang}, Lipeng}, \au{{Ou}, Weike}, \au{{Zhou},
  Tianchun}, \au{{Li}, Bo} \& \au{{Mauel}, M.~E.}} \yr{2025}
  \at{{Bounce-averaged fluid equations for interchange dynamics in a
  dipole-confined plasma}}.  \jt{Physics of Plasmas}  \bvol{32}~(6),
  \pg{062302}.

\bibitem[{Kadomtsev}(1966)]{kadomtsev66}
{\sc \au{{Kadomtsev}, B.~B.}} \yr{1966}  \at{{Hydromagnetic stability of a
  plasma}}.  \bt{In {\em Reviews of Plasma Physics, vol.~2\/} (ed. \ed{M.~A.
  {Leontovich}})},  \pg{pp. 153--199}.  \publ{New York: Consultants Bureau}.

\bibitem[Kenmochi {\em et~al.\/}(2022)Kenmochi, Yokota, Nishiura, Saitoh, Sato,
  Nakamura, Mori, Ueda \& Yoshida]{Kenmochi2022}
{\sc \au{Kenmochi, N.}, \au{Yokota, Y.}, \au{Nishiura, M.}, \au{Saitoh, H.},
  \au{Sato, N.}, \au{Nakamura, K.}, \au{Mori, T.}, \au{Ueda, K.} \&
  \au{Yoshida, Z.}} \yr{2022}  \at{Inward diffusion driven by low frequency
  fluctuations in self-organizing magnetospheric plasma}.  \jt{Nuclear Fusion}
  \bvol{62}~(2),  \pg{026041}.

\bibitem[Kesner(2003)]{Kesner2003}
{\sc \au{Kesner, J.}} \yr{2003}  \at{Convective cell formation in a {Z} pinch}.
   \jt{Physics of Plasmas}  \bvol{10}~(3),  \pg{908--911}.

\bibitem[{Kesner} {\em et~al.\/}(2010){Kesner}, {Davis}, {Ellsworth},
  {Garnier}, {Kahn}, {Mauel}, {Michael}, {Wilson} \& {Woskov}]{Kesner2010}
{\sc \au{{Kesner}, J.}, \au{{Davis}, M.~S.}, \au{{Ellsworth}, J.~L.},
  \au{{Garnier}, D.~T.}, \au{{Kahn}, J.}, \au{{Mauel}, M.~E.}, \au{{Michael},
  P.}, \au{{Wilson}, B.} \& \au{{Woskov}, P.~P.}} \yr{2010}  \at{{Stationary
  density profiles in the Levitated Dipole Experiment: toward fusion without
  tritium fuel}}.  \jt{Plasma Physics and Controlled Fusion}  \bvol{52}~(12),
  \pg{124036}.

\bibitem[Kesner \& Garnier(2000)]{KesnerGarnier2000}
{\sc \au{Kesner, J.} \& \au{Garnier, D.~T.}} \yr{2000}  \at{Convective cell
  formation in a levitated dipole}.  \jt{Physics of Plasmas}  \bvol{7}~(6),
  \pg{2733--2737}.

\bibitem[{Kesner} {\em et~al.\/}(2004){Kesner}, {Garnier}, {Hansen}, {Mauel} \&
  {Bromberg}]{Kesner2004}
{\sc \au{{Kesner}, J.}, \au{{Garnier}, D.~T.}, \au{{Hansen}, A.}, \au{{Mauel},
  M.} \& \au{{Bromberg}, L.}} \yr{2004}  \at{{Helium catalysed D--D fusion in a
  levitated dipole}}.  \jt{Nuclear Fusion}  \bvol{44}~(1),  \pg{193--203}.

\bibitem[{Kesner} {\em et~al.\/}(2011){Kesner}, {Garnier} \&
  {Mauel}]{Kesner2011}
{\sc \au{{Kesner}, J.}, \au{{Garnier}, D.~T.} \& \au{{Mauel}, M.~E.}} \yr{2011}
   \at{{Fluctuation driven transport and stationary profiles}}.  \jt{Physics of
  Plasmas}  \bvol{18}~(5),  \pg{050703}.

\bibitem[{Kesner} \& {Mauel}(1997)]{Kesner1997}
{\sc \au{{Kesner}, J.} \& \au{{Mauel}, M.}} \yr{1997}  \at{{Plasma confinement
  in a levitated magnetic dipole}}.  \jt{Plasma Physics Reports}
  \bvol{23}~(9),  \pg{742--750}.

\bibitem[Kesner {\em et~al.\/}(2001)Kesner, Simakov, Garnier, Catto, Hastie,
  Krasheninnikov, Mauel, {Sunn Pedersen} \& Ramos]{Kesner2001}
{\sc \au{Kesner, J.}, \au{Simakov, A.~N.}, \au{Garnier, D.~T.}, \au{Catto,
  P.~J.}, \au{Hastie, R.~J.}, \au{Krasheninnikov, S.~I.}, \au{Mauel, M.~E.},
  \au{{Sunn Pedersen}, T.} \& \au{Ramos, J.~J.}} \yr{2001}  \at{Dipole
  equilibrium and stability}.  \jt{Nuclear Fusion}  \bvol{41}~(3),
  \pg{301--308}.

\bibitem[{Kobayashi} {\em et~al.\/}(2009){Kobayashi}, {Rogers} \&
  {Dorland}]{kobayashi09}
{\sc \au{{Kobayashi}, S.}, \au{{Rogers}, B.~N.} \& \au{{Dorland}, W.}}
  \yr{2009}  \at{{Gyrokinetic simulations of turbulent transport in a ring
  dipole plasma}}.  \jt{Physical Review Letters}  \bvol{103}~(5),  \pg{055003}.

\bibitem[{Kobayashi} {\em et~al.\/}(2010){Kobayashi}, {Rogers} \&
  {Dorland}]{kobayashi10}
{\sc \au{{Kobayashi}, S.}, \au{{Rogers}, B.~N.} \& \au{{Dorland}, W.}}
  \yr{2010}  \at{{Particle pinch in gyrokinetic simulations of closed
  field-line systems}}.  \jt{Physical Review Letters}  \bvol{105}~(23),
  \pg{235004}.

\bibitem[{Kouznetsov} {\em et~al.\/}(2007){Kouznetsov}, {Freidberg} \&
  {Kesner}]{kouznetsov07}
{\sc \au{{Kouznetsov}, A.}, \au{{Freidberg}, J.~P.} \& \au{{Kesner}, J.}}
  \yr{2007}  \at{{Quasilinear theory of interchange modes in a closed field
  line configuration}}.  \jt{Physics of Plasmas}  \bvol{14}~(10),  \pg{102501}.

\bibitem[{Krasheninnikov} {\em et~al.\/}(1999){Krasheninnikov}, {Catto} \&
  {Hazeltine}]{Krasheninnikov1999}
{\sc \au{{Krasheninnikov}, Sergei~I.}, \au{{Catto}, Peter~J.} \&
  \au{{Hazeltine}, R.~D.}} \yr{1999}  \at{{Magnetic Dipole Equilibrium Solution
  at Finite Plasma Pressure}}.  \jt{Physical Review Letters}  \bvol{82}~(13),
  \pg{2689--2692}.

\bibitem[{Li} {\em et~al.\/}(2016){Li}, {Liao}, {Sun}, {Ou}, {Liu}, {Gui} \&
  {Wang}]{Li2016}
{\sc \au{{Li}, B.}, \au{{Liao}, X.}, \au{{Sun}, C.~K.}, \au{{Ou}, W.},
  \au{{Liu}, D.}, \au{{Gui}, G.} \& \au{{Wang}, X.~G.}} \yr{2016}
  \at{{Magnetic curvature effects on plasma interchange turbulence}}.
  \jt{Physics of Plasmas}  \bvol{23}~(6),  \pg{062304}.

\bibitem[{Li} {\em et~al.\/}(2021){Li}, {Wang}, {Ou}, {Zhou}, {Li}, {Rogers} \&
  {Mauel}]{li21}
{\sc \au{{Li}, Bo}, \au{{Wang}, Lipeng}, \au{{Ou}, Weike}, \au{{Zhou}, Zhenyu},
  \au{{Li}, Zhuoyi}, \au{{Rogers}, Barrett~N.} \& \au{{Mauel}, M.~E.}}
  \yr{2021}  \at{{Self-organized confinement in whole-device modeling of
  laboratory magnetospheres}}.  \jt{Physics of Plasmas}  \bvol{28}~(10),
  \pg{102301}.

\bibitem[{Makhin} {\em et~al.\/}(2005){Makhin}, {Siemon}, {Bauer}, {Esaulov},
  {Lindemuth}, {Sotnikov}, {Ryutov} \& {Sheehey}]{Makhin2005}
{\sc \au{{Makhin}, V.}, \au{{Siemon}, R.~E.}, \au{{Bauer}, B.~S.},
  \au{{Esaulov}, A.}, \au{{Lindemuth}, I.~R.}, \au{{Sotnikov}, V.~I.},
  \au{{Ryutov}, D.~D.} \& \au{{Sheehey}, P.~T.}} \yr{2005}
  \at{{Self-organization observed in numerical simulations of a hard-core
  diffuse Z pinch}}.  \jt{Physics of Plasmas}  \bvol{12}~(4),  \pg{042312}.

\bibitem[Numata(2025)]{Numata2025}
{\sc \au{Numata, R.}} \yr{2025}  \at{Thermal equilibration in collisionless
  magnetospheric plasmas via entropy-mode turbulence}.  \jt{Monthly Notices of
  the Royal Astronomical Society: Letters}  \bvol{538}~(1),  \pg{L94--L99}.

\bibitem[{Ou} {\em et~al.\/}(2020){Ou}, {Wang}, {Li} \& {Rogers}]{ou20}
{\sc \au{{Ou}, Weike}, \au{{Wang}, Lipeng}, \au{{Li}, Bo} \& \au{{Rogers},
  B.~N.}} \yr{2020}  \at{{Turbulent pinch in whole-plasma simulations of a
  dipole-confined plasma}}.  \jt{Physical Review E}  \bvol{101}~(2),
  \pg{021201}.

\bibitem[{Pastukhov}(2005)]{pastukhov05b}
{\sc \au{{Pastukhov}, V.~P.}} \yr{2005}  \at{{Equations for nonlinear MHD
  convection in shearless magnetic systems}}.  \jt{Plasma Physics Reports}
  \bvol{31}~(7),  \pg{577--590}.

\bibitem[{Pastukhov} \& {Chudin}(2001)]{pastukhov01}
{\sc \au{{Pastukhov}, V.~P.} \& \au{{Chudin}, N.~V.}} \yr{2001}  \at{{Plasma
  convection near the threshold for MHD instability in nonparaxial magnetic
  confinement systems}}.  \jt{Plasma Physics Reports}  \bvol{27}~(11),
  \pg{907--921}.

\bibitem[{Pastukhov} \& {Chudin}(2005)]{pastukhov05}
{\sc \au{{Pastukhov}, V.~P.} \& \au{{Chudin}, N.~V.}} \yr{2005}
  \at{{Self-consistent turbulent convection in a magnetized plasma}}.  \jt{JETP
  Letters}  \bvol{82}~(6),  \pg{356--365}.

\bibitem[Ricci {\em et~al.\/}(2006)Ricci, Rogers \& Dorland]{Ricci2006PRL}
{\sc \au{Ricci, P.}, \au{Rogers, B.~N.} \& \au{Dorland, W.}} \yr{2006}
  \at{Small-scale turbulence in a closed-field-line geometry}.  \jt{Physical
  Review Letters}  \bvol{97}~(24),  \pg{245001}.

\bibitem[{Ricci} {\em et~al.\/}(2006){Ricci}, {Rogers}, {Dorland} \&
  {Barnes}]{Ricci2006}
{\sc \au{{Ricci}, Paolo}, \au{{Rogers}, B.~N.}, \au{{Dorland}, W.} \&
  \au{{Barnes}, M.}} \yr{2006}  \at{{Gyrokinetic linear theory of the entropy
  mode in a Z pinch}}.  \jt{Physics of Plasmas}  \bvol{13}~(6),  \pg{062102}.

\bibitem[Roberts {\em et~al.\/}(2015)Roberts, Mauel \& Worstell]{Roberts2015}
{\sc \au{Roberts, T.~M.}, \au{Mauel, M.~E.} \& \au{Worstell, M.~W.}} \yr{2015}
  \at{Local regulation of interchange turbulence in a dipole-confined plasma
  torus using current-collection feedback}.  \jt{Physics of Plasmas}
  \bvol{22}~(5),  \pg{055702}.

\bibitem[Saitoh {\em et~al.\/}(2026)Saitoh, Mimura, Aoyagi, Shirokoshi, Ueda,
  Nishiura, Kenmochi, Sato \& Yoshida]{Saitoh2026}
{\sc \au{Saitoh, H.}, \au{Mimura, H.}, \au{Aoyagi, S.}, \au{Shirokoshi, N.},
  \au{Ueda, K.}, \au{Nishiura, M.}, \au{Kenmochi, N.}, \au{Sato, N.} \&
  \au{Yoshida, Z.}} \yr{2026}  \at{Inward particle flux caused by low-frequency
  electric fluctuations in the edge region of dipole magnetic field
  configuration}.  \jt{Nuclear Fusion}  \bvol{66}~(9),  \pg{096013}.

\bibitem[Saitoh {\em et~al.\/}(2024)Saitoh, Nakagawa, Ueda, Mori, Nishiura,
  Kenmochi, Sato \& Yoshida]{Saitoh2024}
{\sc \au{Saitoh, H.}, \au{Nakagawa, R.}, \au{Ueda, K.}, \au{Mori, T.},
  \au{Nishiura, M.}, \au{Kenmochi, N.}, \au{Sato, N.} \& \au{Yoshida, Z.}}
  \yr{2024}  \at{Transition from electrostatic to electromagnetic modes of
  low-frequency fluctuations in {RT-1}}.  \jt{Nuclear Fusion}  \bvol{64}~(12),
  \pg{126011}.

\bibitem[{Saitoh} {\em et~al.\/}(2011){Saitoh}, {Yoshida}, {Morikawa}, {Yano},
  {Mizushima}, {Ogawa}, {Furukawa}, {Kawai}, {Harima}, {Kawazura}, {Kaneko},
  {Tadachi}, {Emoto}, {Kobayashi}, {Sugiura} \& {Vogel}]{Saitoh2011}
{\sc \au{{Saitoh}, H.}, \au{{Yoshida}, Z.}, \au{{Morikawa}, J.}, \au{{Yano},
  Y.}, \au{{Mizushima}, T.}, \au{{Ogawa}, Y.}, \au{{Furukawa}, M.},
  \au{{Kawai}, Y.}, \au{{Harima}, K.}, \au{{Kawazura}, Y.}, \au{{Kaneko}, Y.},
  \au{{Tadachi}, K.}, \au{{Emoto}, S.}, \au{{Kobayashi}, M.}, \au{{Sugiura},
  T.} \& \au{{Vogel}, G.}} \yr{2011}  \at{{High-beta plasma formation and
  observation of peaked density profile in RT-1}}.  \jt{Nuclear Fusion}
  \bvol{51}~(6),  \pg{063034}.

\bibitem[{Simakov} {\em et~al.\/}(2001){Simakov}, {Catto} \&
  {Hastie}]{simakov01}
{\sc \au{{Simakov}, A.~N.}, \au{{Catto}, P.~J.} \& \au{{Hastie}, R.~J.}}
  \yr{2001}  \at{{Kinetic stability of electrostatic plasma modes in a dipolar
  magnetic field}}.  \jt{Physics of Plasmas}  \bvol{8}~(10),  \pg{4414--4426}.

\bibitem[{Simakov} {\em et~al.\/}(2000){Simakov}, {Catto}, {Krasheninnikov} \&
  {Ramos}]{simakov00}
{\sc \au{{Simakov}, A.~N.}, \au{{Catto}, P.~J.}, \au{{Krasheninnikov}, S.~I.}
  \& \au{{Ramos}, J.~J.}} \yr{2000}  \at{{Ballooning stability of a point
  dipole equilibrium}}.  \jt{Physics of Plasmas}  \bvol{7}~(6),
  \pg{2526--2529}.

\bibitem[{Simpson} {\em et~al.\/}(2026){Simpson}, {Badcock}, {Berry},
  {Chisholm}, {Fimognari}, {Fisher}, {Garnier}, {Lenagh-Glue}, {Leuw},
  {Mataira}, {Meadows}, {McIntosh}, {Poata}, {Richardson}, {Smith}, {Simpson},
  {Tyler} \& {Wordsworth}]{Simpson2026}
{\sc \au{{Simpson}, T.~E.}, \au{{Badcock}, R.~A.}, \au{{Berry}, T.},
  \au{{Chisholm}, C.~S.}, \au{{Fimognari}, P.~J.}, \au{{Fisher}, P.},
  \au{{Garnier}, D.~T.}, \au{{Lenagh-Glue}, K.}, \au{{Leuw}, B.},
  \au{{Mataira}, R.}, \au{{Meadows}, L.}, \au{{McIntosh}, T.}, \au{{Poata},
  J.}, \au{{Richardson}, K.}, \au{{Smith}, B.}, \au{{Simpson}, A.},
  \au{{Tyler}, J.~D.} \& \au{{Wordsworth}, T.}} \yr{2026}
  \at{{Deuterium--tritium levitated dipole fusion power plants}}.  \jt{Fusion
  Engineering and Design}  \bvol{230},  \pg{115864}.

\bibitem[{Stone} {\em et~al.\/}(2020){Stone}, {Tomida}, {White} \&
  {Felker}]{stone20}
{\sc \au{{Stone}, J.~M.}, \au{{Tomida}, K.}, \au{{White}, C.~J.} \&
  \au{{Felker}, K.~G.}} \yr{2020}  \at{{The Athena++ adaptive mesh refinement
  framework: design and magnetohydrodynamic solvers}}.  \jt{The Astrophysical
  Journal Supplement Series}  \bvol{249}~(1),  \pg{4}.

\bibitem[{Testa} \& {Fitzpatrick}(2026)]{Testa2026}
{\sc \au{{Testa}, D.~M.} \& \au{{Fitzpatrick}, R.}} \yr{2026}  \at{{Revisiting
  ballooning modes in levitated dipole geometries}}.  \jt{Physics of Plasmas}
  \bvol{33}~(7),  \pg{072510}.

\bibitem[{Yan'kov}(1994)]{yankov94}
{\sc \au{{Yan'kov}, V.~V.}} \yr{1994}  \at{{The pinch effect explains turbulent
  transport in tokamaks}}.  \jt{JETP Letters}  \bvol{60}~(3),  \pg{171}.

\bibitem[{Yoshida} {\em et~al.\/}(2013){Yoshida}, {Saitoh}, {Yano}, {Mikami},
  {Kasaoka}, {Sakamoto}, {Morikawa}, {Furukawa} \& {Mahajan}]{Yoshida2013}
{\sc \au{{Yoshida}, Z.}, \au{{Saitoh}, H.}, \au{{Yano}, Y.}, \au{{Mikami}, H.},
  \au{{Kasaoka}, N.}, \au{{Sakamoto}, W.}, \au{{Morikawa}, J.}, \au{{Furukawa},
  M.} \& \au{{Mahajan}, S.~M.}} \yr{2013}  \at{{Self-organized confinement by
  magnetic dipole: recent results from RT-1 and theoretical modeling}}.
  \jt{Plasma Physics and Controlled Fusion}  \bvol{55}~(1),  \pg{014018}.

\end{thebibliography}

\end{document}